\documentclass[12pt,twoside,a4paper]{article}

\usepackage{amsmath}
\usepackage{a4wide}
\usepackage[dvips]{graphicx}
\usepackage{dsfont}
\usepackage{amsthm}
\usepackage{amssymb}
\usepackage{epsfig}
\usepackage{float}
\usepackage{url}

\newcommand{\heff}{{\hbar_{\mathrm{eff}}}}
\newcommand{\crp}{{\mathcal{N}}}

\newcommand{\vecq}{{\bf q}}
\newcommand{\vecp}{{\bf p}}
\newcommand{\vecz}{{\bf z}}
\newcommand{\vecx}{{\bf x}}

\newcommand{\ud}{\mathrm{d}}
\newcommand{\ui}{\mathrm{i}}
\newcommand{\ue}{\mathrm{e}}

\newcommand{\R}{\mathds{R}}
\newcommand{\Z}{\mathds{Z}}
\newcommand{\N}{\mathds{N}}

\newcommand{\M}{\mathcal{M}}

\newcommand{\Hcnf}{H_{\text{CNF}}}

\newcommand{\flux}{f}

\providecommand{\abs}[1]{\lvert#1\rvert}

\newcommand{\la}{\langle}
\newcommand{\ra}{\rangle}

\renewcommand{\Im}{\operatorname{Im}}
\renewcommand{\Re}{\operatorname{Re}}

\newcommand{\dof}{DoF\ }

\newcommand{\nhim}{{S_{\text{NHIM}}^{2d-3}(E)}}
\newcommand{\ts}{{S_{\text{ds}}^{2d-2}(E)}}

\newcommand{\tsf}{{B_{\text{ds,\,f}}^{2d-2}(E)}}
\newcommand{\tsb}{{B_{\text{ds,\,b}}^{2d-2}(E)}}

\newcommand{\sphere}{\mathbb{S}}

 \normalsize

\newcommand{\capsty}{\footnotesize}

\newcommand{\rem}[1]{}

\newtheorem{platz}{{\bf Fig.}} 

\rem{
\newcommand{\Capts}[1]{#1}
\newcommand{\FIGo}[3]{%
\marginpar{ \begin{platz} \label{#1} ~ \end{platz} \vspace*{1.5ex} }}

} 

\newcommand{\Capts}[1]{}
\newcommand{\FIGo}[3]{\begin{figure}%
#3
\caption[]{\capsty #2}%
\label{#1}%
\end{figure}}

\numberwithin{equation}{section}

\title{Quantum Theory of Reactive Scattering in Phase Space}
\author{Arseni Goussev$^1$\thanks{Arseni.Goussev@bristol.ac.uk}, Roman
  Schubert$^1$\thanks{Roman.Schubert@bristol.ac.uk}, Holger
  Waalkens$^{1,2}$\thanks{H.Waalkens@rug.nl}, and Stephen
  Wiggins$^1$\thanks{S.Wiggins@bristol.ac.uk} }

\begin{document}
\maketitle

\noindent
{\small $^1$ School of Mathematics, University of Bristol, University Walk, Bristol BS8 1TW, UK}\\
{\small $^2$ Johann Bernoulli Institute for Mathematics and Computer Sciences,
  University of Groningen, PO Box 407, 9700 AK Groningen, The
  Netherlands}

\abstract{We review recent results on quantum reactive scattering from
  a phase space perspective. The approach uses classical and quantum
  versions of Poincar\'e-Birkhoff normal form theory and the
  perspective of dynamical systems theory.  Over the past ten years
  the classical normal form theory has provided a method for realizing
  the phase space structures that are responsible for determining
  reactions in high dimensional Hamiltonian systems. This has led to
  the understanding that a new (to reaction dynamics) type of phase
  space structure, a {\em normally hyperbolic invariant manifold} (or,
  NHIM) is the ``anchor'' on which the phase space structures
  governing reaction dynamics are built, e.g. it is the classical
  analogue of the chemists notion of the ``activated complex'' and it
  is essential for the construction of a surface that divides
  reactants from products which has the ``no-recrossing'' property for
  trajectories and minimal flux. The quantum normal form theory
  provides a method for quantizing these phase space structures
  through the use of the Weyl quantization procedure. We show that
  this approach provides a solution of the time-independent
  Schr\"odinger equation leading to a (local) S-matrix in a
  neighborhood of the saddle point governing the reaction. These
  results can be obtained for any dimensional system for which an
  accurate normal form can be computed, and it does {\em not} require
  numerical solution of the Schr\"odinger equation or the generation
  of any classical trajectories. It follows easily that the
  quantization of the directional flux through the dividing surface
  with the properties noted above is a flux operator that can be
  expressed in a ``closed form''.  Moreover, from the local S-matrix
  we easily obtain an expression for the cumulative reaction
  probability (CRP), which is the essential ingredient for the
  computation of microcanonical reaction rates and thermal reaction
  rates. Significantly, the expression for the CRP can be evaluated
  without the need to compute classical trajectories. This is a by
  product of the quantization of classical phase space structures that
  govern ``exact'' classical dynamics. The quantization of the NHIM is
  shown to lead to the activated complex, and the lifetimes of quantum
  states initialized on the NHIM correspond to the Gamov-Siegert
  resonances.  We apply these results to the collinear nitrogen
  exchange reaction and a three degree-of-freedom system corresponding
  to an Eckart barrier coupled to two Morse oscillators. We end by
  describing some further challenges that are topics of current
  research, but where some preliminary results are known:
  corner-cutting tunneling, state-to-state reaction rates, the
  flux-flux autocorrelation function formalism and the convergence of
  the quantum normal form.  We emphasize that this dynamical systems,
  phase space approach to quantum reactive scattering through the
  quantum normal form provides a completely new approach to the
  computation of the relevant quantum scattering quantities (e.g. CRP,
  resonances) which shows promise in leading to computationally
  efficient methods for ``high dimensional'' systems.}

\tableofcontents


\section{Introduction}
\label{sec:introduction}

Over the past fifty years the computation of quantities describing quantum reactive scattering processes has been a topic of great interest in the chemistry and physics community, see the Perspective article of Miller \cite{Miller00} for background and history. The interest in this topic continues to grow as a result of the need to include quantum effects in order to understand the behavior of certain biomolecules and molecular materials (\cite{Miller08}).  These contemporary applications are concerned with  ``large systems'', and quantum mechanical calculations are notorious for their consumption of computational resources as the number of degrees-of-freedom grows. 
As noted in \cite{doe2}, the Nobel Prize winner Walter Kohn has identified the problem as the Òexponential wallÓ of difficulty when one tries to perform numerical calculations in the traditional manner using the rules of quantum mechanics.  In this review we discuss a new (phase space) approach to quantum reactive scattering that allows us to bypass the traditional computational difficulties for certain types of physically relevant quantities.

Our purpose here is not to review the entire field of quantum reactive scattering. That would be a  huge task that is beyond the scope of this review. Rather, we begin by describing the standard quantities that are computed to describe quantum reactive scattering processes and the issues associated with their computation. Our description follows the excellent review of Miller \cite{Miller98}.

The quantity that contains a great deal of detailed information about a particular quantum reactive scattering process is the {\bf S}-matrix.
The {\bf S}-matrix, $\{S_{n_p, \, n_r} (E, J) \}$ as a function of total energy $E$ and total angular momentum $J$, can be obtained by solving the Schr\"odinger equation with scattering boundary conditions.  From the {\bf S}-matrix state-to-state differential scattering cross-sections can be computed as follows:

\begin{equation}
\sigma_{n_p \leftarrow n_r} (\theta, E) = \vert (2 i k_{n_r})^{-1} \sum_{J} (2J + 1) d^J_{m_p, \,  m_r} (\theta) S_{n_p, \, n_r} (E, J) \vert^2,
\label{eq:ss_cs}
\end{equation}

\noindent
where $n_r$ ($n_p$) labels the reactant (product) rotational and 
vibrational states, $\theta$ is the scattering angle between the relative velocity vectors of reactants and products, $m_r$ ($m_p$) is the projection of total angular momentum onto the relative velocity vector of the reactants (products) and $d^J_{m \, m'} (\theta)$ is the Wigner rotation matrix.

For many chemical reactions a sufficient description is provided by the {\em rate constant}, either the canonical rate constant characterized by the temperature, $k(T)$, or the microcanonical rate constant characterized by the total energy, $k(E)$. These rate constants can be obtained using appropriate averages of the state-to-state differential cross sections.  This averaging  process yields the {\em cumulative reaction probability} (CRP):

\begin{equation}
\crp(E) = \sum_{J} (2J + 1) \sum_{n_p, \, n_r} \vert S_{n_p, \, n_r} (E, J) \vert^2,
\label{eq:crp}
\end{equation}

\noindent
which is used to compute the microcanonical and canonical rate constants, as follows

\begin{eqnarray}
k(E) & = &[ 2 \pi \hbar \rho_r(E) ]^{-1} \crp(E), \label{eq:microrate}\\
k(T) & = &  [ 2 \pi \hbar Q_r(T)]^{-1} \int_{-\infty}^{\infty} dE e^{-\beta E} \crp(E). \label{eq:canrate}
\end{eqnarray}

\noindent
where $\beta = (kT)^{-1}$, $\rho_r$ is the density of reactant states per unit energy, and $Q_r$ is the reactant partition function per unit volume.

Miller has pointed out  (\cite{Miller98}) that if one is interested only in the rate,  then  solving the Schr\"odinger equation to obtain the {\bf S}-matrix from which the state-to-state differential scattering cross section \eqref{eq:ss_cs} is computed, and then subsequently averaged to obtain the cumulative reaction probability \eqref{eq:crp},  is an extremely inefficient process in the sense that a great deal of the detailed information obtained through the computation of the {\bf S}-matrix is ``thrown away'' in the averaging process.  Instead, Miller \cite{Miller98} discusses a {\em direct} way to compute  $\crp(E)$, or $k(T)$, i.e. a method that avoids first computing the  {\bf S}-matrix, that is also {\em correct} in the sense that the expressions for $\crp(E)$, or $k(T)$ are exact, i.e. no approximations are involved.  These expressions are given as follows (\cite{Miller98}):

\begin{equation}
\crp(E) = 2 \pi \hbar \, {\rm Tr} \left(\delta(E- \hat{H}) \, \hat{F} \, P_r  \right)
\label{eq:microcan_crp}
\end{equation}

\noindent

\begin{equation}
k(T) = Q_r (T)^{-1} \, {\rm Tr} \left( e^{-\beta \hat{H}} \, \hat{F} \,  P_r \, \right)
\label{eq:canrate_trace}
\end{equation}

\noindent
where $\hat{H}$ is the Hamilton operator, $\hat{F}$ is a flux operator, $P_r$ is the long time limit of the quantum mechanically time evolved Heaviside function, and ${\rm Tr} ( \cdot )$ denotes  the trace operation. Schatz and Ratner \cite{schatz02} point out  that there are three possibilities for the use of \eqref{eq:microcan_crp} and \eqref{eq:canrate_trace} for computing rates.  The first is to evaluate the trace of the operators in a basis, the second is to develop semiclassical theories for the rate constants (\cite{Miller07, Miller08}), and the third is to use the expressions to develop a version of quantum transition state theory (\cite{Miller74}). An alternative approach is to express  \eqref{eq:microcan_crp} and \eqref{eq:canrate_trace} in  terms of a form of flux-flux autocorrelation function (\cite{Yamamoto60, MillerSchwartzTromp83}), and we discuss this approach in some detail in Section \ref{sec:flux-flux}. A recent survey of methods for computing reaction rates is given in \cite{manol08}.

Our approach us firmly rooted in phase space, both the classical and quantum theories. The classical phase space theory forms the ``skeleton'' on which our quantum theory of reactive scattering in phase space is built. The classical theory of reaction dynamics in phase space is reviewed in Section~\ref{sec:CNF}.  Certainly there has been earlier work on classical reaction dynamics in phase space. In particular, we note the seminal work of Pechukas, Pollak, Child, and McLafferty from the  1970's and early 1980's on two degree-of-freedom Hamiltonian systems (\cite{Pechukas1, PechukasPollak77, Pechukas2, PollakPechukas79, PechukasPollak79, Pechukas3, cp, PollakChild80}) where the notion of a periodic orbit dividing surface (PODS) was introduced.  The PODS bound a two dimensional dividing surface (in the three dimensional energy surface) having the ``no-recrossing'' property and minimal flux.  This work did not generalize to more than two degrees-of-freedom. The generalization has required a new concept--the notion of a {\em normally hyperbolic invariant manifold}, or NHIM. This was introduced in the study of phase space transport in Hamiltonian systems with three or more degrees-of-freedom in \cite{Wiggins90}, and was subsequently used in the study of chemical reaction dynamics in \cite{ge91}.  For a $d$ degree-of-freedom Hamiltonian system,  the NHIM has the structure of a $2d-3$ dimensional sphere in the  $2d-1$ dimensional energy surface (so for $d=2$, the NHIM has the structure of a one dimensional sphere, or periodic orbit). Conceptually, the NHIM is a fundamentally new dynamical object. It is a {\em manifold} of ``saddle type stability'' that exists in phase space. In Section \ref{sec:CNF} we discuss in more detail why this is the mathematical manifestation of the chemists notion of the {\em activated complex}.  Later, it was shown in \cite{WaalkensWiggins04} that the NHIM played the role of the PODS for systems with three or more degrees-of-freedom in the sense that the NHIM is the boundary of a dividing surface having the ``no-recrossing'' property and minimal flux.

This approach to phase space reaction dynamics for three or more degree-of-freedom Hamiltonian systems did not receive a great deal of applications throughout the 1990's.  One reason for this was that there was not a computational method for realizing NHIMs in specific Hamiltonian systems (with three or more degrees-of-freedom). This changed as a result of the work in \cite{wwju, ujpyw} where it was shown that the classical  Poincar\'e-Birkhoff normal form theory could be used to compute a coordinate system where the NHIM could be realized, along with a variety of associated phase space structures.  Software was developed that enabled the computation of the classical and quantum normal forms, and this allowed the realization of these phase space structures in specific examples\footnote{The software, along with documentation, is freely available for download at  \\ \texttt{http://lacms.maths.bris.ac.uk/publications/software/index.html}.}. This enabled the application of this 
approach to classical  phase space reaction dynamics  to HCN isomerization  (\cite{WaalkensBurbanksWiggins04}),  computation of the volume of points in an energy surface corresponding to reactive trajectories (the ``reactive volume'', \cite{WaalkensBurbanksWiggins05, WaalkensBurbanksWiggins05c}), and a realization of Thiele's \cite{thiele} notion of ``gap times'' and reaction rates (\cite{eww09}).

The generalization of this approach to the quantum setting has been carried out in \cite{SchubertWaalkensWiggins06,WaalkensSchubertWiggins08,gsww09}.  The essential tools that make this possible are a quantum mechanical generalization of the classical Poincar\'e-Birkhoff normal form (described in Appendix~\ref{sec:qnf_algorithm}) and the Weyl quantization procedure. This firmly establishes a phase space approach to quantum mechanics.  Our development and understanding of classical phase space structures associated with reaction dynamics is essential because one sees  precisely their quantum mechanical manifestations. For example, in the classical setting the normal form is  integrable in a  (phase space) neighborhood of a particular saddle point that is relevant to  the reaction of interest (in a way that we precisely describe in Section \ref{sec:CNF}). This property is manifested quantum mechanically by yielding a Hamilton operator in (quantum) normal form that factors into separate one-dimensional problems that can be solved exactly. This, in turn, gives rise to a block diagonal {\bf S}-matrix that can easily be computed and analyzed. This is described in some detail in Section~\ref{sec:smatrix}.  This is a significant result because the implication is if the relevant quantum normal form for a system can be computed, then {\em regardless of the size of the system}, the {\bf S}-matrix can be calculated. In other words, our approach leads to an expression for the {\bf S}-matrix that does {\em not} require a solution of the Schr\"odinger equation. In Section \ref{sec:smatrix} we show that the
classical flux through a dividing surface in phase space having the no-recrossing and minimal flux properties is  directly related to the quantum mechanical cumulative reaction probability.  We apply this approach to computing the cumulative reaction probability for the collinear nitrogen exchange reaction in Section~\ref{sec:nitrogen_exchange} and coupled Eckart-Morse-Morse oscillators in Section~\ref{sec:Eckart_Morse_Morse}.  In Section~\ref{sec:resonances} we show that quantum resonances are related to the time of decay of quantum states initialized on the activated complex. We conclude in Section~\ref{sec:challenges} by considering four developing areas of the theory: ``corner cutting'' tunneling trajectories, state-to-state reaction rates, the flux-flux autocorrelation formalism, and convergence of the quantum normal form. 



\section{Phase-Space Structures Underlying Reaction Dynamics}
\label{sec:CNF}

In this section we describe the dynamics and geometry associated with phase space structures governing reaction dynamics. While the emphasis in this review is on quantum mechanics, the phase space structure that we describe forms the classical mechanical ``skeleton'' on which the quantum mechanical theory is supported. Here we merely summarize the basic results, more details can be  found in the references given in the introduction. Our exposition here  follows  \cite{ww10}.

We consider a Hamiltonian system with $d$ degrees of freedom, phase
space coordinates $(\tilde{q}, \tilde{p})\in \R^d \times\R^d$ and
Hamiltonian function $H$.  We assume that $(\tilde{q}_0, \tilde{p}_0)$
is an equilibrium point of Hamilton's equations which is of
saddle-centre-$\ldots$-centre stability type.
\footnote{We will define this more precisely shortly. However, briefly, it means that the matrix associated with the linearization of Hamilton's equations about this equilibrium point has two real eigenvalues of equal magnitude, with one positive and one negative, and $d-1$ purely imaginary complex conjugate pairs of eigenvalues. We will assume that the eigenvalues satisfy a nonresonance condition that we will describe more fully in the following.}  
By adding a constant  term to the Hamiltonian function (which does not change the dynamics) we can, without loss of generality, assume that 
 $H(\tilde{q}_0, \tilde{p}_0)=0$. Moreover,  for simplicity of exposition,
we can assume that the coordinates have been suitably
translated so that the relevant equilibrium point $(\tilde{q}_0, \tilde{p}_0)$
is at the origin. For much of the discussion below, we will consider iso-energetic
geometrical structures belonging to a single positive energy surface
$\Sigma(E):=H^{-1}(E)$ for some constant $E>0$. However, we note that in \cite{ew09a} essentially the same ``picture'' of phase space structures governing reaction dynamics has been shown to occur for  constant temperature dynamics governed by the Hamiltonian isokinetic thermostat. 

We construct geometric structures in a neighborhood of the  equilibrium point. We emphasize that by ``neighborhood'' we mean a neighborhood in phase space, not just on a fixed energy surface. We denote this neighborhood of the equilibrium point by
 $\mathcal{L}$.  We will defer until later a discussion of the constraints on the size of this neighborhood; suffice it to say for now that the region is chosen so that a new set of coordinates can be constructed (the normal form coordinates) 
in which an accurate  nonlinear approximation of the Hamiltonian can be expressed (the normal form Hamiltonian) such that it provides an integrable approximation to the dynamics, as well as an algorithmic procedure for realizing phase space structures to within a given desired accuracy. 

Before describing the  dynamics  in normal form coordinates, as well as the  realization of the phase space structures in normal form coordinates that govern ``reaction'', i.e. trajectories in a phase space neighborhood of the saddle-centre-$\ldots$-centre equilibrium point (henceforth, referred to as a ``saddle'') we describe the relevant geometric structures that have been developed in the in the references given above. We emphasize that although the normal form procedure provides a specific method for realizing these geometric structures in a given coordinate system, their existence is independent of any specific set of coordinates. 

Locally, the $(2d-1)$-dimensional energy surface $\Sigma(E)$ has
the structure of $\sphere^{2d-2}\times\R$ in the
$2d$-dimensional phase space.  The energy surface $\Sigma(E)$ is
split locally into two components, ``reactants''  and
``products'' , by a $(2d-2)$-dimensional  ``dividing surface''
that is diffeomorphic to $\sphere^{2d-2}$ and which we therefore denote by $\ts$.  The
dividing surface that we construct has the following properties:-

\begin{itemize}

\item The only way that trajectories can evolve from reactants  to products  (and
vice-versa), without leaving the local region $\mathcal{L}$, is
through $\ts$. In other words, initial conditions (ICs) on  this dividing surface  specify all reacting trajectories.

\item The dividing surface that we construct is free of local
recrossings; any trajectory which crosses it must leave the
neighbourhood $\mathcal{L}$ before it might possibly cross again.

\item The dividing surface that we construct minimizes the
 flux, i.e. the directional flux through the dividing surface will increase upon a generic deformation of the dividing surface (see \cite{WaalkensWiggins04} for the details).

\end{itemize}

The fundamental phase space building block that allows the construction of a dividing surface with these properties is a particular \emph{Normally Hyperbolic Invariant Manifold} (NHIM) which, for a
fixed positive energy $E$, will be denoted $\nhim$.  The
NHIM is diffeomorphic to $\sphere^{2d-3}$ and forms the natural
\emph{dynamical equator} of the dividing surface: The dividing
surface is split by this equator into $(2d-2)$-dimensional
hemispheres, each diffeomorphic to the open $(2d-2)$-ball,
$B^{2d-2}$.  We will denote these hemispheres by
$\tsf$ and $\tsb$ and call them the
``forward reactive'' and ``backward reactive'' hemispheres,
respectively.  $\tsf$ is crossed by trajectories
representing ``forward'' reactions (from reactants to products),
while $\tsb$ is crossed by trajectories representing
``backward'' reactions (from products to reactants).

The $(2d-3)$-dimensional NHIM can be viewed as the energy surface of an 
(unstable) invariant subsystem which as mentioned above, in
chemistry terminology, corresponds to the ``activated complex'', which as an
oscillating ``supermolecule''  is located between reactants and
products. 

The NHIM is of saddle stability type, having $(2d-2)$-dimensional
stable and unstable manifolds $W^s(E)$ and $W^u(E)$ that are
diffeomorphic to $\sphere^{2d-3}\times\mathbb{R}$.  Being of
co-dimension
\footnote{Briefly, the co-dimension of a submanifold is the dimension of the space in which the submanifold exists, minus the dimension of the submanifold. The significance of a submanifold being ``co-dimension one'' is that it is one less dimension than the space in which it exists. Therefore it can ``divide'' the space and act as a separatrix, or barrier, to transport.} 
one with respect to the energy surface, these invariant manifolds act as separatrices, partitioning the energy
surface into ``reacting'' and ``non-reacting'' parts as will explain in detail in Sec.~\ref{sec:structures_explicit}.

\subsection{Dynamics and phase space structures near the saddle in normal form coordinates} \label{sec:normal_form}

As mentioned in the previous section, reaction type dynamics are
induced by equilibrium points of
saddle$\times$centre$\times\ldots\times$centre stability type. These
are equilibria for which the matrix associated with the linearisation
of Hamilton's equations have eigenvalues which consist of a pair of
real eigenvalues of equal magnitude and opposite sign,
$(+\lambda,-\lambda)$, $\lambda\in\R$, and $(d-1)$ pairs of complex
conjugate purely imaginary eigenvalues, $(+i\omega_k,-i\omega_k)$,
$\omega_k\in\R$, for $k=2,\ldots,d$.

The phase space structures  near equilibria of this type exist independently of a specific coordinate system. However, in order to carry out specific calculations we will need to be able to express these phase space structures in coordinates. 
This is where Poincar\'e-Birkhoff normal form theory is used.This is a well-known theory and has been the subject of many
review papers and books, see,  e.g.,  \cite{Dep69,Meye74,DragtFinn76,akn88,MeYe91,MeHa92,Murdock03}.  
For our purposes it provides an algorithm whereby the phase space structures described in the previous section can be realised
for a particular system by means of the normal form transformation
which involves making a nonlinear symplectic change of variables,

\begin{equation}\label{eq:NFtransform}
 (q, p) =T(\tilde{q}, \tilde{p}),
\end{equation}

\noindent into \emph{normal form coordinates}, $(q,p)$ which, in a local
neighbourhood $\mathcal{L}$ of the equilibrium point, ``decouples'' the
dynamics into a ``reaction coordinate'' and ``bath modes.'' The coordinate transformation $T$ is obtained from imposing conditions on the form of $H$ expressed the new coordinates, $(q, p)$,

\begin{equation}
\Hcnf(q, p) = H\big(T^{-1}(q,p)\big)=H(\tilde{q}, \tilde{p}).
\end{equation}

\noindent These conditions are chosen such that  $\Hcnf$ and the resulting equations of motions assume a simple form in which the reaction coordinate and bath modes ``decouple''.  
This decoupling  is one way of understanding how we are able to construct the phase space structures, in the normal form coordinates, that govern the dynamics of reaction.

In fact, we will assume that a (generic)
non-resonance condition holds between the eigenvalues, namely that

\begin{equation}
k_2\omega_2+\dots+k_n\omega_d \ne 0 
\end{equation}

\noindent for all integer vectors $(k_2,\ldots,k_d)\in\Z^{d-1}$. 
\footnote{We note that the
inclusion of $\pm\lambda$ in a non-resonance condition would be
vacuous; one cannot have a resonance of this kind between a real
eigenvalue, $\pm\lambda$, and purely imaginary eigenvalues,
$\pm i\omega_k$, $k=2,\ldots,d$.}
When such a condition
holds, the classical normal form procedure, see Appendix~\ref{sec:cnf_algorithm}, yields an explicit expression for the
normalised Hamiltonian $\Hcnf$ as a function of $d$ 
integrals of motion:

\begin{align}
  \Hcnf^{(N)} &= K_{\mathrm{CNF}}^{(N)}(I,
  J_2,
  J_3, \ldots, J_d) \nonumber\\
  &= E_0 + \lambda I + \sum_{k=2}^d \omega_k  J_k +
  \sum_{n=2}^{\lfloor N/2 \rfloor} \! \sum_{|\alpha| = n} \!\!
  \kappa_{n,\alpha} I^{\alpha_1} J_2^{\alpha_2} \ldots
  J_d^{\alpha_d}  \, .
\label{nf_int}
\end{align}


\noindent Note that the normal form is expressed as a polynomial in the action integrals $I$ and $J_k$, $k=2,\ldots,d$, corresponding to a ``reaction coordinate'' and $(d-1)$ ``bath modes'' respectively. In Eq.~\eqref{nf_int} $E_0$ denotes the energy of the system at the equilibrium point, and the expansion coefficient $\kappa_{n,\alpha}$ are obtained by the classical normal form algorithm presented in Appendix~\ref{sec:QNF_algorithm}; here, $\alpha = (\alpha_1, \ldots, \alpha_d)$ with $\alpha_k \in \N_0$ for $k=1,\ldots,d$, and $\abs{\alpha} = \sum_k \alpha_k$. In Appendix \ref{sec:QNF_algorithm} we show that the normal form algorithm generates a polynomial at each step, where the normalization algorithm at a certain order does not modify the normalized terms at lower order. However, in practice it is necessary to stop the algorithm at a given order. In this sense the normal form is an approximation to the original Hamiltonian in a neighborhood of the saddle, which we have denoted by $\cal L$.  Certainly, it is essential that the approximation is accurate enough to yield useful information. Numerous examples have show this to be the case, and we discuss this issue more fully in Section \ref{sec:trunc_error}.

The integral, $I$, corresponds to a ``reaction
coordinate'' (saddle-type DoF):

\begin{equation}\label{eq:def_I}
I = q_1p_1.
\end{equation}

\noindent We note that there is an equivalent form of the reaction
coordinate:- making the linear symplectic change of variables
$q_1=(P_1-Q_1)/\sqrt{2}$ and $p_1=(P_1+Q_1)/\sqrt{2}$, transforms the
above into the following form, which may be more familiar to many
readers,

\begin{equation}\label{eq:def_I_tilde}
I = q_1p_1 = \frac{1}{2}\left(P_1^2-Q_1^2\right).
\end{equation}

\noindent Geometrically speaking, one can move freely between
these two representations by considering the plane $(q_1,p_1)$ and
rotating it by angle $\pi/4$, to give $(Q, P)$.

The integrals $J_k$, for $k=2,\ldots,d$, correspond to ``bath
modes'' (centre-type \dof)\footnote{Throughout our work we use, somewhat interchangeably, terminology from both chemical reaction dynamics and dynamical systems theory. This is most noticable in our reference to the integrals of motion. $I$ is the integral related to reaction, and in the context of dynamical systems theory it is related to hyperbolic behaviour. The term ``reactive mode'' might also be used to describe the dynamics associated with this integral. The integrals $J_2, \ldots , J_d$ describe the dynamics associated with ``bath modes''. In the context of dynamical systems theory, the dynamics associated with these integrals is referred to as ``center type dynamics'' or ``center modes''. A key point here is that integrals of the motion provide us with the natural way of defining and describing the physical notion of a ``mode''. The nature of the mode is defined in the context of the specific application (i.e. chemical reactions) or, in the context of dynamical systems theory, through its stability properties (i.e. hyperbolic or centre).}:

\begin{equation}\label{eq:def_J}
J_k = \frac{1}{2}\left(p_k^2 + q_k^2\right),\quad k=2,\ldots,d\,.
\end{equation}

\noindent In the new coordinates, Hamilton's equations have a
particularly simple form:

\begin{equation} \label{eq:NF_equations_motion_gen}
\begin{split}
\dot{q}_1 = \hphantom{-}\frac{\partial{\Hcnf}}{\partial p_1}
&= \hphantom{-}\Lambda_1(I,J_2,\ldots,J_d)q_1,\\
\dot{p}_1 = {-}\frac{\partial{\Hcnf}}{\partial q_1}
&= {-}\Lambda(I,J_2,\ldots,J_d)p_1,\\
\dot{q}_k = \hphantom{-}\frac{\partial{\Hcnf}}{\partial p_k}
&= \hphantom{-}\Omega_k(I,J_2,\ldots,J_d)p_k\\
\dot{p}_k = {-}\frac{\partial{\Hcnf}}{\partial q_k} &=
{-}\Omega_k(I,J_2,\ldots,J_d)q_k,
\end{split}
\end{equation}

\noindent for $k=2,\ldots,d$, where we denote

\begin{eqnarray}
\Lambda(I,J_2,\ldots,J_d)  &:=&\frac{\partial{K_\text{CNF}(I,J_2,\ldots,J_d)}}{\partial
I}\\
\Omega_k(I,J_2,\ldots,J_d) &:=&\frac{\partial{K_\text{CNF}(I,J_2,\ldots,J_d)}}{\partial
J_k},\quad k=2,\ldots,d.
\end{eqnarray}

The integrals provide a natural definition of the term ``mode'' that is appropriate in the context of reaction, and they are a consequence of the (local) integrability in a neighborhood of the equilibrium point of saddle-centre-$\ldots$-centre stability type. Moreover, the expression of the normal form Hamiltonian in terms of the integrals provides us a way to partition the ``energy'' between the different modes. We will provide examples of how this can be done in the following.

The normal form transformation $T$ in \eqref{eq:NFtransform} can be computed in an algorithmic fashion. One can give explicit expression for the phase space structures discussed in the previous section in terms of the normal form coordinates, $(q,p)$. This way the phase space structures can be constructed in terms on the normal form coordinates, $(q,p)$, and for physical interpretation, transformed back to the original ``physical'' coordinates, $(\tilde{q}, \tilde{p})$, by the inverse of the transformation $T$. \footnote{The original coordinates $(\tilde{q}, \tilde{p})$ typically have an interpretation as configuration space coordinates and momentum coordinates. The normal form coordinates $(q, p)$, in general, do not have such a physical interpretation since both $q$ and $p$ are nonlinear functions of both $\tilde{q}$ and $\tilde{p}$.}

In summary, the ``output'' of the normal form algorithm is the following:

\begin{itemize}

\item A symplectic transformation  $T(\tilde{q}, \tilde{p}) = (q, p)$,  and its inverse $T^{-1}(q, p) = (\tilde{q}, \tilde{p})$, that relate the normal form coordinates $(q,p)$ to the original ``physical'' coordinates $(\tilde{q}, \tilde{p})$.

\item An expression for the normalized Hamiltonian: in the form, $\Hcnf$, in terms of the normal form coordinates $(q,p)$, and in the form $K_\text{CNF}$, in terms of the integrals $(I,J_2,\ldots,J_d)$.

\item Explicit expressions for the integrals of motion $I$ and $J_k$, $k=2, \ldots, d$, in terms of the original ``physical'' coordinates by means of the symplectic transformation $T(\tilde{q}, \tilde{p}) = (q, p)$.

\end{itemize}

\subsection{Explicit definition and construction of the phase space structures in the
normal form coordinates} \label{sec:structures_explicit}

As indicated in the previous section it is straightforward to
construct the local phase space objects governing ``reaction'' in the
normal form coordinates, $(q, p)$.  In this section, we will define
the various structures in the normal form coordinates and discuss
briefly the consequences for the original dynamical system.

\begin{figure}[htb!]
\centerline{\includegraphics[width=12.0cm]{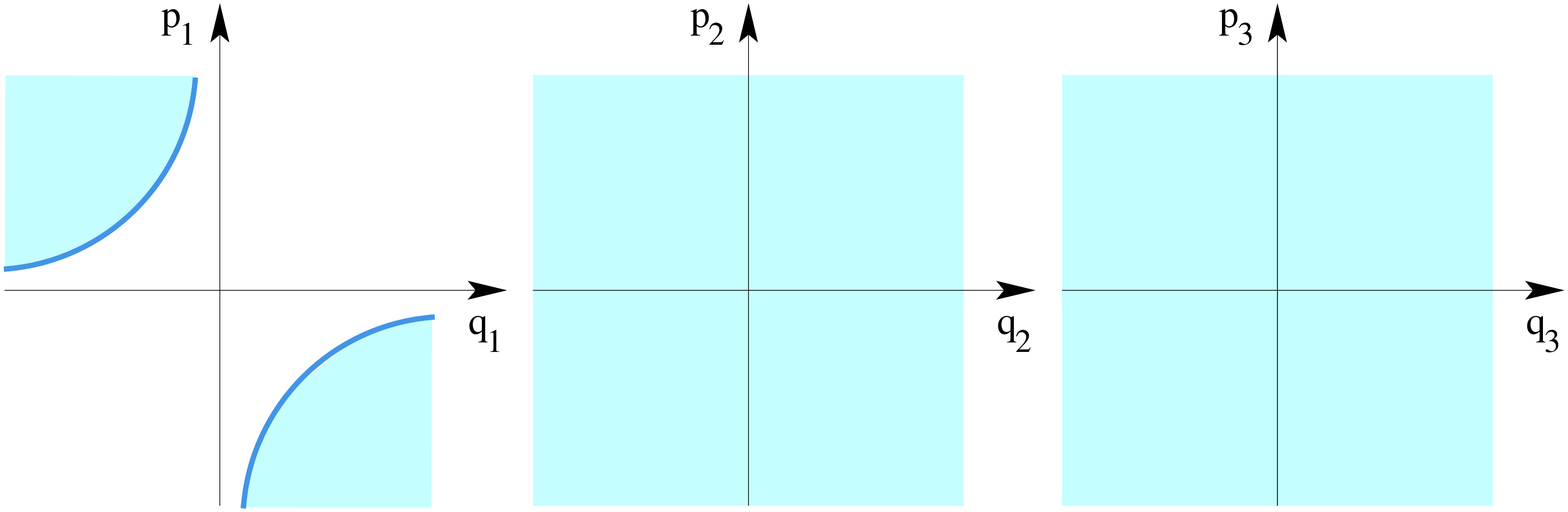}}
\centerline{\includegraphics[width=12.0cm]{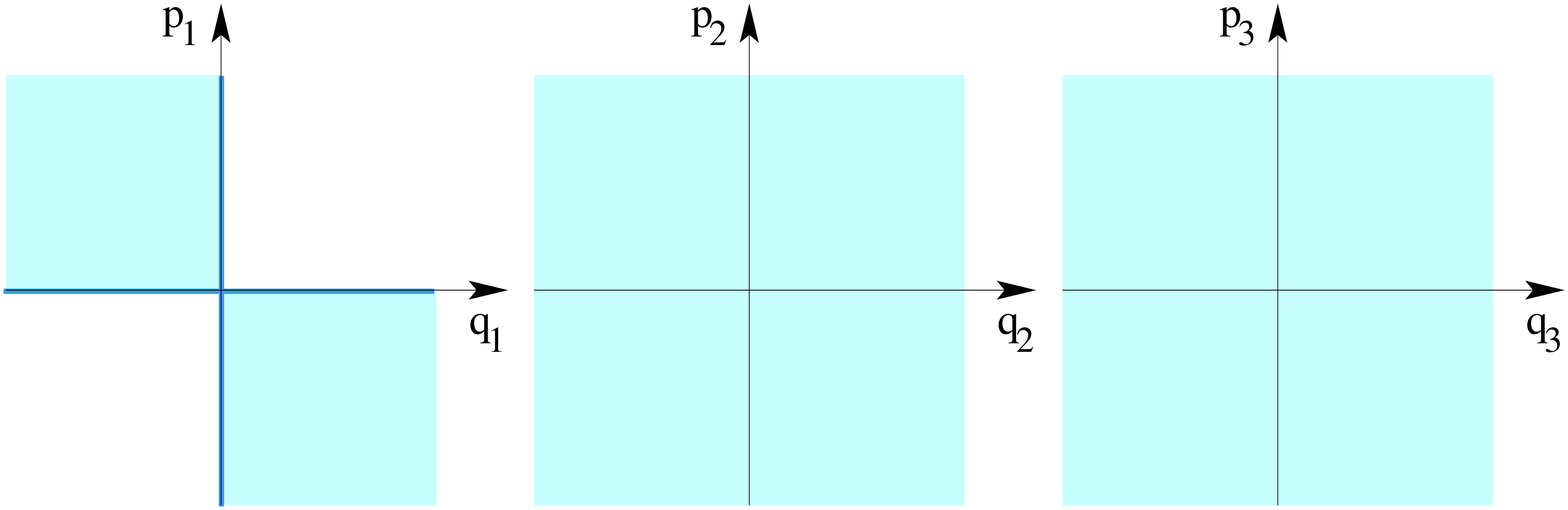}}
\centerline{\includegraphics[width=12.0cm]{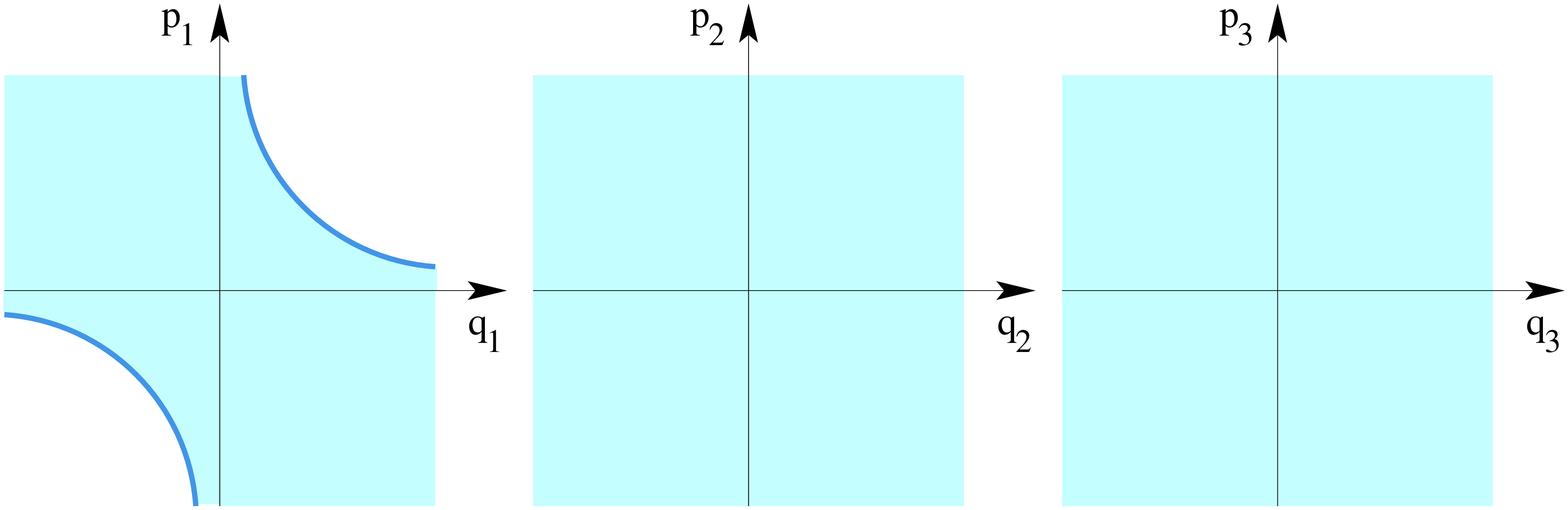}}
\caption{Projection of energy surfaces (turquoise regions) to the
  planes of the normal form coordinates.  The energy surface have
  energy $E<0$ (top panel), $E=0$ (middle panel), $E>0$ (bottom
  panel).}
\label{fig:esurfs}
\end{figure}

\vspace*{0.3cm} \noindent {\bf The structure of an energy surface
near a saddle point: } For $E<0$, the energy surface consists of
two disjoint components.  The two components correspond to
``reactants'' and ``products.''  The top panel of
Fig.~\ref{fig:esurfs} shows how the two components project to the
various planes of the normal form coordinates.  The projection to
the plane of the saddle coordinates $(q_1,p_1)$ is bounded away
from the origin by the two branches of the hyperbola, $q_1p_1=I<0$, where $I$ is
given implicitly by the energy equation with the centre actions
$J_k$, $k=2,\dots,d$, set equal to zero:
$K_{\text{CNF}}(I,0,\ldots,0)=E<0$.  The projections to the
planes of the centre coordinates, $(q_k,p_k)$, $k=2,\dots,d$, are
unbounded.

At $E=0$, the formerly disconnected components merge (the energy
surface bifurcates), and for $E>0$, the energy surface has
locally the structure of a spherical cylinder, $\sphere^{2d-2}\times
\R$.  Its projection to the plane of the saddle coordinates now
includes the origin.  In the first and third quadrants it is
bounded by the two branches of the  hyperbola, $q_1p_1=I>0$, where $I$ is again given
implicitly by the energy equation with all centre actions equal to
zero, but now with an energy greater than zero:
$K_{\text{CNF}}(I,0,\dots,0)=E>0$. The projections to the planes
of the centre coordinates are again unbounded. This is illustrated
in the bottom panel of Fig.~\ref{fig:esurfs}.

\begin{figure}[htb!]
\begin{center}
\includegraphics[width=12.0cm]{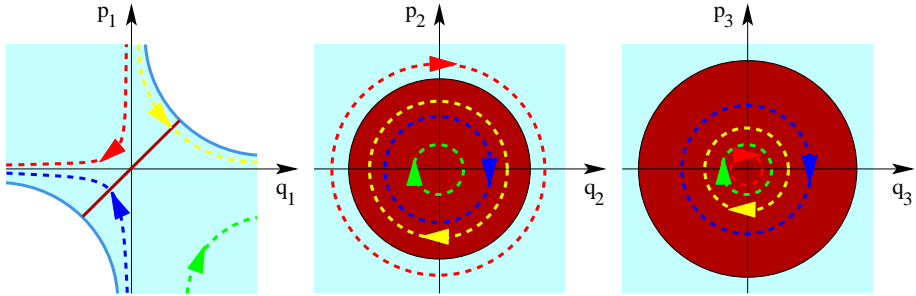}
\end{center}
\caption{Projection of the dividing surface and reacting and
  nonreacting trajectories to the planes of the normal form
  coordinates. In the plane of the saddle coordinates, the projection
  of the dividing surface is the dark red diagonal line segment, which
  has $q_1=p_1$.  In the planes of the centre coordinates, the
  projections of the dividing surface are the dark red discs.  Forward
  and backward reactive trajectories (yellow and blue) project to the
  first and third quadrant in the plane of the saddle coordinates,
  respectively, and pass through the dividing surface.  The red and
  green curves mark nonreactive trajectories on the reactant side
  ($p_1-q_1>0$), and on the product side ($p_1-q_1<0$), of the
  dividing surface, respectively.  The turquoise regions indicate the
  projections of the energy surface.}
\label{fig:tst}
\end{figure}

\vspace*{0.3cm}
\noindent
{\bf The dividing surface, and reacting and nonreacting trajectories:}
On an energy surface with $E>0$, we define the dividing surface by
$q_1=p_1$.  This gives a $(2d-2)$-sphere which we denote by $\ts$.
Its projection to the saddle coordinates simply gives a line segment
through the origin which joins the boundaries of the projection of the
energy surface, as shown in Fig.~\ref{fig:tst}.  The projections of the
dividing surface to the planes of the centre coordinates are bounded by
circles $(p_k^2+q_k^2)/2=J_k$, $k=2,\dots,d$, where $J_k$ is
determined by the energy equation with the other centre actions, $J_l$,
$l\ne k$, and the saddle integral, $I$, set equal to zero.  The
dividing surface divides the energy surface into two halves,
$p_1-q_1>0$ and $p_1-q_1<0$, corresponding to reactants and products.

As mentioned above, trajectories project to
hyperbolae in the plane of the saddle coordinates, and to circles
in the planes of the centre coordinates.  The sign of $I$
determines whether a trajectory is nonreacting or reacting, see
Fig.~\ref{fig:tst}.  Trajectories which have $I<0$ are nonreactive and for one branch of the hyperbola $q_1p_1=I$ they stay on the
reactants side and for the other branch they stay on the products side; trajectories with $I>0$ are reactive, and for one  branch of the hyperbola $q_1p_1=I$ they react
in the forward direction, i.e., from reactants to products, and for the other branch they react in the
backward direction, i.e., from products to reactants.  The
projections of reactive trajectories to the planes of the centre
coordinates are always contained in the projections of the dividing
surface. In this, and other ways, the geometry of the reaction is
highly constrained.  There is no analogous restriction on the
projections of nonreactive trajectories to the centre coordinates.

\vspace*{0.3cm} \noindent {\bf The normally hyperbolic invariant
  manifold (NHIM) and its relation to the ``activated complex'':} On
an energy surface with $E>0$, the NHIM is given by $q_1=p_1=0$.  The
NHIM has the structure of a $(2d-3)$-sphere, which we denote by
$\nhim$.  The NHIM is the equator of the dividing surface; it divides
it into two ``hemispheres'': the \emph{forward dividing surface},
which has $q_1=p_1>0$, and the \emph{backward dividing surface}, which
has $q_1=p_1<0$.  The forward and backward dividing surfaces have the
structure of $(2d-2)$-dimensional balls, which we denote by $\tsf$ and
$\tsb$, respectively.  All forward reactive trajectories cross $\tsf$;
all backward reactive trajectories cross $\tsb$.  Since $q_1=p_1=0$ in
the equations of motion \eqref{eq:NF_equations_motion_gen} implies
that $\dot{q}_1=\dot{p}_1=0$, the NHIM is an invariant manifold, i.e.,
trajectories started in the NHIM stay in the NHIM for all time.  The
system resulting from $q_1=p_1=0$ is an invariant subsystem with one
degree of freedom less than the full system.  In fact, $q_1=p_1=0$
defines the centre manifold associated with the
saddle-centre-$\cdots$-centre equilibrium point, and the NHIM at an
energy $E$ greater than the energy of the quilibrium point is given by
the intersection of the centre manifold with the energy surface of
this energy $E$ \cite{ujpyw,WaalkensWiggins04}.

This invariant subsystem with one degree of freedom less than the full
system is the ``activated complex'' (in phase space), located between
reactants and products.  The NHIM can be considered to be the energy
surface of the activated complex.  In particular, all trajectories in
the NHIM have $I=0$.

The equations of motion \eqref{eq:NF_equations_motion_gen} also show that
$\dot{p}_1-\dot{q}_1<0$ on the forward dividing surface $\tsf$, and
$\dot{p}_1-\dot{q}_1>0$ on the backward dividing surface $\tsb$.
Hence, except for the NHIM, which is is an invariant manifold, the
dividing surface is everywhere transverse to the Hamiltonian flow.
This means that a trajectory, after having crossed the forward or
backward dividing surface, $\tsf$ or $\tsb$, respectively, must leave
the neighbourhood of the dividing surface before it can possibly cross
it again. Indeed, such a trajectory must leave the local region in
which the normal form is valid before it can possibly cross the
dividing surface again.

The NHIM has a special structure: due to the conservation of the
centre actions, it is filled, or {\em foliated}, by invariant
$(d-1)$-dimensional tori, $\mathbb{T}^{d-1}$.  More precisely, for
$d=3$ degrees of freedom, each value of $J_2$ implicitly defines a
value of $J_3$ by the energy equation $K_{\text{CNF}}(0,J_2,J_3)=E$.
For three degrees of freedom, the NHIM is thus foliated by a
one-parameter family of invariant 2-tori.  The end points of the
parameterization interval correspond to $J_2=0$ (implying $q_2=p_2=0$)
and $J_3=0$ (implying $q_3=p_3=0$), respectively.  At the end points,
the 2-tori thus degenerate to periodic orbits, the so-called {\em
  Lyapunov periodic orbits}.

\begin{figure}[htb!]
\begin{center}
\includegraphics[width=12.0cm]{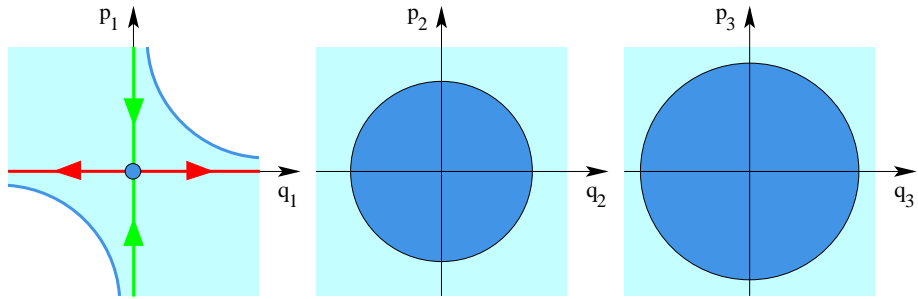}
\end{center}
\caption{The projection of the NHIM and the local parts of its stable
  and unstable manifolds, $W^{\text{s}}(E)$ and $W^u(E)$, to the
  planes of the normal form coordinates.  In the plane of the saddle
  coordinates, the projection of the NHIM is the origin marked by the
  blue bold point, and the projection of $W^\text{s}(E)$ and $W^u(E)$
  are the $p_1$-axis and $q_1$-axis, respectively.  $W^\text{s}(E)$
  consists of the forward and backward branches
  $W_{\text{f}}^{\text{s}}(E)$ and $W_{\text{b}}^{\text{s}}(E)$, which
  have $p_1>0$ and $p_1<0$, respectively; $W^{\text{u}}(E)$ consists
  of $W_{\text{f}}^{\text{u}}(E)$ and $W_{\text{b}}^{\text{u}}(E)$,
  which have $q_1>0$ and $q_1<0$, respectively.  In the plane of the
  centre coordinates, the projections of the NHIM, $W^{\text{s}}(E)$,
  and $W^{\text{u}}(E)$ (the blue circular discs) coincide with the
  projection of the dividing surface in Fig.~\ref{fig:tst}.  The
  turquoise regions mark the projections of the energy surface.}
\label{fig:nhim}
\end{figure}

\vspace*{0.3cm} \noindent {\bf The stable and unstable manifolds
of the NHIM forming the phase space conduits for reactions:} Since
the NHIM is of saddle stability type, it has stable and unstable
manifolds, $W^{\text{s}}(E)$ and $W^{\text{u}}(E)$.  The stable and unstable
manifolds have the structure of spherical cylinders,
$\sphere^{2d-3}\times\R$.  Each of them consists of two branches: the
``forward branches'', which we denote by $W_f^s(E)$ and
$W_{\text{f}}^{\text{u}}(E)$, and the ``backward branches'', which we denote by
$W_{\text{b}}^{\text{s}}(E)$ and $W_{\text{b}}^{\text{u}}(E)$.  In terms of the normal form
coordinates, $W_{\text{f}}^{\text{s}}(E)$ is given by $q_1=0$ with $p_1>0$,
$W_{\text{f}}^{\text{u}}(E)$ is given by $p_1=0$ with $q_1>0$, $W_b^s(E)$ is given
by $q_1=0$ with $p_1<0$, and $W_b^u(E)$ is given by $p_1=0$ with
$q_1<0$, see Fig.~\ref{fig:nhim}.  Trajectories on these manifolds
have $I=0$.

Since the stable and unstable manifolds of the NHIM are of one less
dimension than the energy surface, they enclose volumes of the energy
surface.  We call the union of the forward branches, $W_f^s(E)$ and
$W_{\text{f}}^{\text{u}}(E)$, the {\em forward reactive spherical cylinder} and denote it by
$W_{\text{f}}(E)$.  Similarly, we define the {\em backward reactive spherical
cylinder}, $W_{\text{b}}(E)$, as the union of the backward branches, $W_{\text{b}}^{\text{s}}(E)$ and
$W_{\text{b}}^{\text{u}}(E)$.

\begin{figure}[htb!]
\begin{center}
\includegraphics[width=12.0cm]{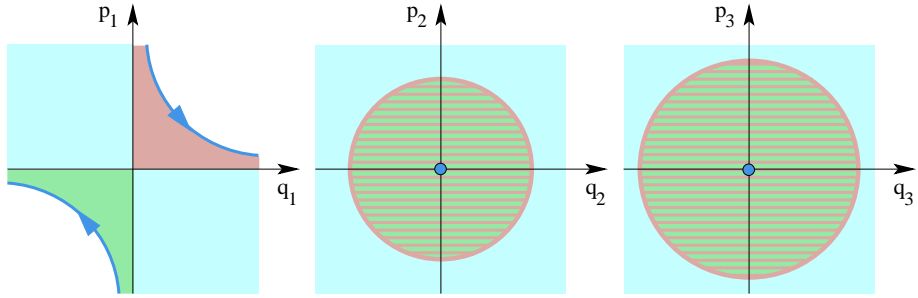}
\end{center}
\caption{Projections of the reactive volumes enclosed by the forward
  and backward reactive spherical cylinders, $W_{\text{f}}(E)$ and
  $W_{\text{b}}(E)$, and the forward and backward reactions paths, to
  the planes of the normal form coordinates.  The volumes enclosed by
  $W_{\text{f}}(E)$ and $W_{\text{b}}(E)$ project to the dark pink and
  green regions in the first and third quadrant in the plane of the
  saddle coordinates, respectively.  These volumes project to the dark
  green/dark pink brindled disks in the planes of the centre
  coordinates, where their projections coincide with the projection of
  the NHIM and the dividing surface in
  Figs.~\ref{fig:tst}~and~\ref{fig:nhim}.  The forward and backward
  reaction paths project to the two branches of a hyperbola marked
  blue in the first and third quadrant in the plane of the saddle
  coordinates, respectively, and to the origins (bold blue points) in
  the planes of the centre coordinates.  The turquoise regions mark
  the projections of the energy surface.}
\label{fig:volumes}
\end{figure}

The reactive volumes enclosed by $W_f(E)$ and $W_b(E)$ are shown
in Fig.~\ref{fig:volumes} as their projections to the normal form
coordinate planes.  In the plane of the saddle coordinates, the
reactive volume enclosed by $W_f(E)$ projects to the first
quadrant.  This projection is bounded by the corresponding
hyperbola $q_1 p_1= I$, with $I$ obtained from
$K_{\text{CNF}}(I,0,\dots,0)=E$.  Likewise, $W_b(E)$ projects to
the third quadrant in the $(q_1,p_1)$-plane.  $W_f(E)$ encloses
{\em all} forward reactive trajectories; $W_b(E)$ encloses {\em
all} backward reactive trajectories.  {\em All} nonreactive
trajectories are contained in the complement.


\subsection{The foliation of the reaction region by Lagrangian submanifolds}
\label{sec:Lagrange_foliation}

The existence of the $d$ integrals of motion, $(I,J_2,\ldots, J_d)$, induce phase space structures which lead to further constraints on the trajectories in addition to the ones described above. In order to describe these structures and the resulting constraints it is useful to introduce the so called \emph{momentum map}, ${\cal M}$ \cite{Gui94,MR99}
which maps a point
$(q_1,\ldots,q_n,p_1,\ldots,p_d)$ in the phase space $\R^{d}\times\R^d$ to
the $d$ integrals evaluated at this point:

\begin{equation} \label{eq:def_momentum_map}
{\cal M}:\R^d\times\R^d \to \R^d\,,\quad(q_1,\ldots,q_n,p_1,\ldots,p_d) \mapsto (I,J_2,\ldots,J_d)\,.
\end{equation}

\noindent
The preimage of a value for the constants of motion $(I,J_2,\ldots,J_d)$ under ${\cal M}$  is called a \emph{fibre}. A fibre thus corresponds to the common level set of the integrals in phase space.

A point $(q_1,\ldots,q_d,p_1,\ldots,p_d)$ is called a \emph{regular
point} of the momentum map if the linearisation of the momentum map, D$\M$, has full rank $d$
at this point, i.e., if the gradients of the $n$ integrals $I$, $J_k$,
$k=2,\ldots,d$, with respect to the
phase space coordinates $(q,p)$ are linearly independent at this point. If the rank of D$\M$ is less than $d$ then the point is called irregular.
A fibre is called regular if it
consists of regular points only. Otherwise, it is called an irregular fibre. 
In fact almost all fibres are regular.
They  are $d$-dimensional
manifolds given by the Cartesian product of an hyperbola $q_1 p_1 =
I$ in the saddle plane $(q_1,p_1)$ and $d-1$ circles
$\sphere^1$ in the centre planes $(q_k,p_k)$, $k=2,\ldots,d$.
Since the hyperbola $q_1 p_1 =
I$ consists of two branches
each of which have the topology of a line $\mathbb{R}$,  the
regular fibres consist of two disjoint \emph{toroidal cylinders}, $\mathbb{T}^{d-1}
\times \R $,
which are the Cartesian products of a
$(d-1)$-dimensional torus and a line.
We denote these toroidal cylinders by

\begin{equation} \label{eq:def_Lambda_plus}
\Lambda^+_{I,J_2,\ldots,J_d} =
\{(q,p)\in \R^{2d}\,:\,
  p_1q_1=I,\,\frac12\big(p_2^2+q_2^2\big)=J_2\,,\ldots\,,\frac12\big(p_d^2+q_d^2\big)=J_d\,,q_1>0 \}
\end{equation}

\noindent
and

\begin{equation}\label{eq:def_Lambda_minus}
\Lambda^-_{I,J_2,\ldots,J_d} =
\{(q,p)\in \R^{2d}\,:\,
  p_1q_1=I,\,\frac12\big(p_2^2+q_2^2\big)=J_2\,,\ldots\,,\frac12\big(p_d^2+q_d^2\big)=J_d\,,q_1<0 \}\,.
\end{equation}
$\Lambda^+_{I,J_2,\ldots,J_d}$ and $\Lambda^-_{I,J_2,\ldots,J_d}$ are
\emph{Lagrangian manifolds} \cite{Arnold78}. The Lagrangian manifolds consists of all trajectories which have the same constants of motion. In particular the Lagrangian manifolds are invariant, i.e. a trajectory with initial condition on a Lagrangian manifold will stay in the Lagrangian manifold for all time. 
For $I<0$, the Lagrangian manifolds $\Lambda^-_{I,J_2,\ldots,J_d}$ and $\Lambda^+_{I,J_2,\ldots,J_d}$
consist of nonreactive trajectories in the reactants resp. products components of the energy surface. For $I>0$, $\Lambda^+_{I,J_2,\ldots,J_d}$ consists of forward reactive trajectories, and $\Lambda^-_{I,J_2,\ldots,J_n}$ consists of backward reactive trajectories.

\subsection{The Directional Flux Through the Dividing Surface}
\label{sec:classicalrate}

\begin{figure}[htb!]
\begin{center}
\includegraphics[width=5.0cm]{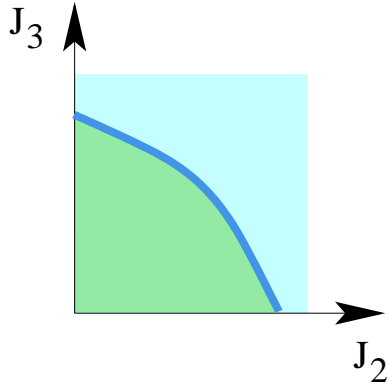}
\end{center}
\caption{Contour $K_{\text{CNF}}(0,J_2,\ldots,J_d)=E$ (blue line) in
  the space of the centre integrals $(J_2,\ldots,J_d)$ for $d=3$
  degrees of freedom. Up to the prefactor $(2\pi)^{d-1}$, the area
  ${\cal V}(E)$ of the enclosed region (marked green) gives the
  directional flux through the dividing surface, see
  Equation~\eqref{eq:fluxactions_classical}. The green region is the
  projection of the $I>0$ piece of the energy surface on the
  $(J_2,J_3)$-plane. \rem{The green region agrees with the projection
    of the piece of the image of the energy surface under the momentum
    map which has $I>0$ in Fig.~\ref{fig:momentummap} to the
    $(J_2,J_3)$-plane.} \rem{(This figure is from
    \cite{WaalkensSchubertWiggins08}.)}}
\label{fig:flux}
\end{figure}

A key ingredient of transition state theory and the classical
reaction rate is the directional flux through the dividing surface
defined in Sec.~\ref{sec:structures_explicit}. Given the Hamiltonian
function in normal form expressed as a function of the integrals
\eqref{nf_int}, and a fixed energy $E$ above the energy of the
saddle-centre-$\cdots$-centre,  $E_0$, it is shown in
\cite{WaalkensWiggins04} that the directional flux through the
dividing surface is given by

\begin{equation} \label{eq:fluxactions_classical}
\flux(E) = (2\pi)^{d-1} {\cal V}(E)\,,
\end{equation}

\noindent
where ${\cal V}(E)$ is the volume in the space of the actions
$(J_2,\dots,J_d)$ enclosed by the contour
$K_{\mathrm{CNF}}(0,J_2,\dots,J_d)=E$. This is a significant result
because it enables the computation of the directional flux {\em
  without computing trajectories}. Moreover, it directly connects the
directional flux to the NHIM, i.e. the activated complex. In
Fig.~\ref{fig:flux} we illustrate the volume ${\cal V}(E)$ for the
case of a 3 DoF systems: here ${\cal V}(E)$ is given by the area in
the $(J_2,J_3)$ plane enclosed by the light blue line corresponding to
the NHIM.

\rem{For the case of a system with three degrees of freedom for which
  we sketched the image of the energy surface in the space of the
  integrals in Fig.~\ref{fig:momentummap}, the volume ${\cal V}(E)$ is
  given by the area in the $(J_2,J_3)$ plane enclosed by the light
  blue line corresponding to the NHIM in
  Fig.~\ref{fig:momentummap}. For clarity we illustrate this area
  again in Fig.~\ref{fig:flux}. As we mentioned in
  Sec.~\ref{sec:nfcpss}. The NHIM can be considered as the energy
  surface of an invariant subsystem with one degree of freedom less
  than the full system which is referred to as the activated complex
  in the chemistry literature. Therefore the flux can be interpreted
  as the volume enclosed by the energy surface (given by the NHIM) in
  the phase space of this invariant subsystem. This gives a direct
  connection between the directional flux through the dividing surface
  and the activated complex.  In fact,}

We also note here that the dimensionless quantity

\begin{equation} \label{eq:NWeyl_def}
\crp_{\mathrm{Weyl}}(E) = \frac{\flux(E)}{(2\pi \hbar)^{d-1}} \,,
\end{equation}

\noindent
where $2\pi \hbar$ is Planck's constant, is Weyl's approximation of
the integrated density of states, or equivalently the mean number of
quantum states of the activated complex with energies less than or
equal to $E$ (see, e.g. \cite{Gutz90}). It is shown in
\cite{WaalkensSchubertWiggins08} that $\crp_{\mathrm{Weyl}}(E)$ can be
interpreted as the mean number of open quantum ``transition channels''
at energy $E$.

In the case where we only take into account the quadratic part of the
normal form, or equivalently, if we linearise Hamilton's equations, we
have $K_{\mathrm{CNF}}(I,J_2,\dots,J_d)=\lambda I+ \sum_{k=2}^d
\omega_k J_k$ and the energy surface
$K_{\mathrm{CNF}}(0,J_2,\dots,J_d)=E$ encloses a simplex in
$(J_2,\dots,J_d)$ whose volume leads to the well-known result
\cite{MacKay1990}
\begin{equation} \label{eq:linear_flux}
\flux(E) = \frac{ E^{d-1}}{(d-1)!} \prod_{k=2}^{d}
\frac{2\pi}{\omega_k}\,.
\end{equation}

\noindent
This shows, e.g, that the flux scales with $E^{d-1}$ for energies close to
the saddle energy. The key advantage of the normal form coordinates resulting 
from the normal form algorithm is that it allows one to include the
non-linear corrections to \eqref{eq:linear_flux} to any desired order.

\subsection{Effect of Truncation of the Normal Form Algorithm}
\label{sec:trunc_error}

The normalization procedure proceeds via
formal power series manipulations whose input is a Taylor
expansion of the original Hamiltonian, $H$, necessarily up to some
finite order, $N$, in homogeneous polynomials.  For a particular
application, this procedure naturally necessitates a suitable
choice of the order, $N$, for the normalization, after which one
must make a restriction to some local region, $\mathcal{L}$, about
the equilibrium point in which the resulting computations achieve
some desired accuracy. Hence, the accuracy of the normal form as a power series expansion truncated
at order $N$ in a neighborhood $\mathcal{L}$ is determined by comparing the dynamics associated with the normal form with the dynamics of the original system. There are several independent tests that can be carried out to verify accuracy of the normal form. Straightforward tests that we use are the following:

\begin{itemize}

\item Examine how the integrals associated with the normal form change on trajectories of the full Hamiltonian (the integrals will be constant on trajectories of the normal form). 

\item Check invariance of the different invariant manifolds (i.e. the NHIM and its stable and unstable manifolds) with respect to trajectories of the full Hamiltonian.

\end{itemize}

Both of these tests will require us to use the transformations between the original coordinates and the normal form coordinates. Specific examples where $N$, $\mathcal{L}$ and accuracy of the normal forms are considered can be found in \cite{WaalkensBurbanksWiggins04,WaalkensBurbanksWigginsb04,WaalkensBurbanksWiggins05,WaalkensBurbanksWiggins05b,WaalkensBurbanksWiggins05c}.

\subsection{``Globalizing'' the Geometrical Structures in the Reaction Region}
\label{sec:global}

As we have shown, the  normal form transformation to normal form
coordinates provides a method for providing a complete
understanding of the geometry of reaction dynamics in a
neighbourhood $\mathcal{L}$ (in phase space)  of the
saddle-centre-$\cdots$-centre equilibrium point of  Hamilton's
equations. By this, we mean that in the normal form coordinates we
can give an explicit equation for the surfaces and, as a result of
the ``simple''  structure of Hamilton's equations in the normal
form coordinates, we can describe precisely the influence of these
geometrical structures on trajectories of Hamilton's equations. In
Tab.~\ref{tab_surf} we summarize the results obtained this far by
providing a list of the different surfaces that control the
evolution of trajectories from reactants to products in the
neighbourhood $\mathcal{L}$  in Fig. \ref{tab_surf}.

\begin{table}[!htb]
\begin{center}
\begin{tabular}{|c|c|} \hline
{\bf Geometrical Structure} & {\bf Equation in Normal Form Coordinates} \\
\hline \hline dividing surface, $\ts$&  $q_1 = p_1$ \\
\hline forward reactive hemisphere, $\tsf$ & $q_1 = p_1 >0$ \\
\hline backward reactive hemisphere, $\tsb$ & $q_1 = p_1 <0$  \\
\hline NHIM, $\nhim$  & $q_1 = p_1 =0$\\
\hline stable manifold of the NHIM, $W^s(E)$& $q_1 =0$, $p_1\ne 0$  \\
\hline unstable manifold of the NHIM, $W^u(E)$& $p_1 =0$, $q_1\ne 0$  \\
\hline forward branch of $W^s(E)$, $W^s_f(E)$& $q_1 =0$, $p_1>0$  \\
\hline backward branch of $W^s(E)$, $W^s_b(E)$& $q_1 =0$, $p_1<0$  \\
\hline forward branch of $W^u(E)$, $W^u_f(E)$& $p_1 =0$, $q_1>0$  \\
\hline backward branch of $W^u(E)$, $W^u_b(E)$& $p_1 =0$, $q_1<0$  \\
\hline forward reactive spherical cylinder & $p_1 q_1 = 0$, $p_1,q_1\ge 0$, $q_1\ne p_1$ \\
 $W_f(E) \equiv W^s_f(E) \cup W^u_f(E)$ & \\
\hline backward reactive spherical cylinder & $p_1 q_1 = 0$, $p_1,q_1\le 0$, $q_1\ne p_1$ \\
$W_b (E) \equiv W^s_b(E) \cup W^u_b(E)$ & \\
\hline
\end{tabular}
\end{center}
\caption{Table of phase space surfaces influencing reaction
dynamics and their representations in normal form coordinates on an energy
surface of energy greater than the energy of the saddle equilibrium point.}
\label{tab_surf}
\end{table}

However, it must be  kept in mind that  all of these surfaces, and associated dynamical
phenomena, are only ``locally valid'' in the neighbourhood $\mathcal{L}$ . The
next step is to understand their influence on the dynamics outside
of $\mathcal{L}$ , i.e., their influence on the dynamics of reaction
throughout phase space in the original coordinates (as opposed to
the normal form coordinates). In order to do this we will need the
normal form transformation discussed in Appendix \ref{sec:QNF_algorithm}, to order
$N$ (where $N$ is determined according to the desired accuracy
following the discussion in Section \ref{sec:trunc_error}).

In Appendix \ref{sec:QNF_algorithm} we discuss the necessary transformations of the original 
physical coordinates required to transform the Hamiltonian into normal form.  In particular, we translate the
saddle-centre-$\cdots$-centre equilibrium point to the origin, we
``simplify'' the linear part of Hamilton's equations (what ''simplify'' precisely means is described in the Appendix), then we
iteratively construct a sequence of nonlinear coordinate
transformations that successively ``simplify'' the order 3, 4,
$\ldots$, N terms of the Hamiltonian according to the algorithm
described in Appendix  \ref{sec:QNF_algorithm}. We can invert each of these
transformations to return from the normal form coordinates to the
physical coordinates.

\paragraph{Computation of $W^u_b(E)$ and $W^u_f(E)$:}
Our approach to computing the stable and unstable manifolds of a
NHIM is, in principle, the same as for computing the stable and
unstable manifolds of a hyperbolic trajectory (however, the
practical implementation of the algorithm in higher dimensions is
a different matter and one that deserves much more investigation).

We describe the computation of $W^u_f(E)$ as follows.

\begin{itemize}

\item In the normal form coordinates, choose a distribution of
initial conditions on the NHIM and displace these initial
conditions ``slightly'' in the direction of the forward branch of
$W^u(E)$ ($p_1=0, \, q_1 = \varepsilon >0$, $\varepsilon$
``small'').

\item Map these  initial conditions back into the physical
coordinates using the inverse of the normal form  transformation.

\item Integrate the initial conditions forward in time using
  Hamilton's equations in the physical coordinates, for the desired
  length of time (typically determined by accuracy considerations)
  that will give the manifold of the desired ``size''. Since the
  initial conditions are in the unstable manifold they will leave the
  neighbourhood $\mathcal{L}$ in which the normal form transformation
  is valid (which is why we integrate them in the original coordinates
  with respect to the original equations of motion).

\end{itemize}

\noindent The backward branch of $W^u(E)$ can be computed in an
analogous manner by displacing the initial conditions on the NHIM
in the direction of the backward branch of $W^u(E)$ ($p_1=0, \,
q_1 = \varepsilon < 0$, $\varepsilon$ ``small'').

\paragraph{Computation of $W^s_b(E)$ and $W^s_f(E)$:} The forward
and backward branches of $W^s(E)$ can be computed in an analogous
fashion, except the initial conditions are integrated {\em
backward} in time.

\paragraph{Practical considerations:} By their very definition,
invariant manifolds consist of  trajectories, and the common way
of computing them, and visualizing them, that works well in low
dimensions is to integrate a distribution of initial conditions
located on the invariant manifold (hence, this illustrates the
value of the normal form coordinates and transformation for
locating appropriate initial conditions). In high dimensions there
are numerical and algorithmic issues that have yet to be fully
addressed. How does one choose a mesh on a $2d-3$ dimensional
sphere? As this mesh evolves in time, how does one ``refine'' the
mesh in such a way that the evolved mesh maintains the structure
of the invariant manifold? 

Examples where this ''globalization'' of phase space structures in the reaction region has been carried out can be found in \cite{WaalkensBurbanksWiggins05,WaalkensBurbanksWiggins05c,WaalkensBurbanksWiggins04,WaalkensBurbanksWigginsb04}. However, there is tremendous scope for future work in this direction, both from the point of view of mathematical and computational techniques, and applications to chemical dynamics. 



\section{Quantum Normal Form Representation of the Activated Complex}
\label{sec:activated_complex}

In this section we present an extension of the classical normal form
theory to quantum mechanics. Leaving all technical details for
Appendix~\ref{sec:QNF_algorithm} we only consider here key aspects and
results of the theory.

The quantum normal form (QNF) theory\footnote{See
  Ref.~\cite{WaalkensSchubertWiggins08} for a comprehensive review of
  the mathematical theory, or consult Appendix~\ref{sec:QNF_algorithm}
  for a concise summary.} provides an explicit algorithmic procedure
that allows one to approximate the Hamiltonian operator $\hat{H}$,
corresponding to a reactive system with a generally {\it
  nonintegrable} classical Hamiltonian, by a ``simpler'' Hamiltonian
operator $\hat{H}_{\mathrm{QNF}}$, having an {\it integrable}
classical counterpart. The operator $\hat{H}_{\mathrm{QNF}}$ takes the
form of a power series expansion in reactive and nonreactive
(bath-mode) action operators and the effective Planck's constant. The
(generally infinite) power series can then be terminated at any
desired order $N$ dictated by the desired accuracy of the
approximation. This leads to an $N$th order QNF approximation
$\hat{H}_{\mathrm{QNF}}^{(N)}$ of the original Hamiltonian $\hat{H}$:
\begin{align}
  \hat{H}_{\mathrm{QNF}}^{(N)} &= K_{\mathrm{QNF}}^{(N)}(\hat{I},
  \hat{J}_2,
  \hat{J}_3, \ldots, \hat{J}_d) \nonumber\\
  &= E_0 + \lambda \hat{I} + \sum_{k=2}^d \omega_k \hat{J}_k +
  \sum_{n=2}^{\lfloor N/2 \rfloor} \! \sum_{|\alpha|+j = n} \!\!\!\!
  \kappa_{n,\alpha,j} \hat{I}^{\alpha_1} \hat{J}_2^{\alpha_2} \ldots
  \hat{J}_d^{\alpha_d} \heff^j \, .
\label{eq:qnf_H}
\end{align}
This approximation holds locally, in the vicinity of a single {\it
  equilibrium} point of the Hamiltonian function of the corresponding
$d$-dimensional classical system (see Appendix~\ref{sec:QNF_algorithm}
for details). It is this equilibrium point whose phase-space
neighborhood is traversed by reactive trajectories on their way from
the valley of reactants to the valley of products. The equilibrium
point is considered to be of saddle-center-...-center stability type,
meaning that the $2d \times 2d$ matrix associated with the linearized
Hamilton's equations of motion has two real eigenvalues, $\pm\lambda$,
and $d-1$ purely imaginary complex conjugate pairs of eigenvalues,
$\pm i\omega_k$ with $k=2,\ldots,d$. The linear frequencies $\omega_k$
are further assumed to be \rem{(a) small compared to $\lambda$ and
  (b)} rationally independent, so that the condition $m_2\omega_2 +
\ldots + m_d \omega_d = 0$ implies $m_2 = \ldots = m_d = 0$ for all
integers $m_2,\ldots,m_d$. Other quantities entering
Eq.~(\ref{eq:qnf_H}) include (i) the energy $E_0$ of the system at the
equilibrium point, (ii) parameters
$\kappa_{n,\alpha_1,\ldots,\alpha_d,j}$ obtained from the QNF
expansion procedure, (iii) an effective (scaled) Planck's constant
$\heff$, which in atomic units is given by the inverse of the square
root of the reduced mass of the system and plays the role of a ``small
parameter'', (iv) an action integral operator
\begin{equation}
  \hat{I} = \frac{1}{2} \left( \hat{q}_1\hat{p}_1 + \hat{p}_1\hat{q}_1 \right)
\label{eq:I_op}
\end{equation}
corresponding to the reactive mode, and (v) action integral operators
\begin{equation}
  \hat{J}_k = \frac{1}{2} \left( \hat{q}_k^2 + \hat{p}_k^2 \right) \, ,
  \qquad k = 2,\ldots,d
\label{eq:J_op}
\end{equation}
of the bath modes. Here, the ``position'' and ``momentum'' operators,
$\hat{q}_k$ and $\hat{p}_k$ respectively, satisfy the commutation
relations
\begin{equation}
  [\hat{q}_k, \hat{q}_l] = [\hat{p}_k, \hat{p}_l] = 0 \, , \qquad
  [\hat{q}_k, \hat{p}_l] = i \heff \, \delta_{kl} \, ,
\label{eq:com_rel}
\end{equation}
where $k,l = 1,\ldots,d$, and $\delta_{kl}$ is the Kronecker's
delta. The operators $\hat{q}_1$, $\hat{p}_1$, $\ldots$, $\hat{q}_d$,
$\hat{p}_d$ are given by the Weyl quantization of classical
phase-space coordinates $q_1$, $p_1$, $\ldots$, $q_d$, $p_d$ that in
turn are obtained from the original ``physical'' phase-space
coordinates by way of a nonlinear canonical transformation
\cite{WaalkensSchubertWiggins08}. Thus, $\hat{q}_1$ and $\hat{p}_1$
correspond to the reaction coordinate, while $\hat{q}_k$ and
$\hat{p}_k$ to the $k$th bath mode with $k=2,\ldots,d$.

It sometimes proves convenient to work in a phase-space coordinate
basis ($Q_1$, $P_1$, $q_2$, $p_2$, $\ldots$, $q_d$, $p_d$) (which we
further refer as to ``QP-basis'') that is ``rotated'' at an angle
$\pi/4$ with respect to the basis ($q_1$, $p_1$, $q_2$, $p_2$,
$\ldots$, $q_d$, $p_d$) (further referred as to ``qp-basis''), i.e.,
\begin{equation}
  \hat{Q}_1 = \frac{1}{\sqrt{2}} \left( \hat{q}_1-\hat{p}_1 \right) \, , \qquad
  \hat{P}_1 = \frac{1}{\sqrt{2}} \left( \hat{q}_1+\hat{p}_1 \right) \, .
\label{eq:rot45}
\end{equation}
In the QP-basis, the action integral operator corresponding to the
reaction coordinate reads
\begin{equation}
  \hat{I} = \frac{1}{2} \left( \hat{P}_1^2 - \hat{Q}_1^2 \right) \, .
\label{eq:I_op_2}
\end{equation}

The main advantage of having the original Hamiltonian $\hat{H}$
approximated by a polynomial in the operators $\hat{I}$ and
$\hat{J}_k$, $k=2,\ldots,d$, is that the eigenstates of the QNF
operator $\hat{H}_{\mathrm{QNF}}^{(N)}$ can be chosen to be
simultaneously the eigenstates of the operators $\hat{I}$ and
$\hat{J}_k$, whose spectral properties are well known. Indeed,
\begin{equation}
  \hat{H}_{\mathrm{QNF}}^{(N)} |I,n_2,\ldots,n_d\rangle = E |I,n_2,\ldots,n_d\rangle
\label{eq:H_QNF_eignsys}
\end{equation}
with
\begin{equation}
  |I,n_2,\ldots,n_d\rangle = |\psi_I\rangle \otimes |\psi_{n_2}\rangle
  \otimes \ldots \otimes |\psi_{n_d}\rangle \, ,
\label{eq:H_QNF_eignfunc}
\end{equation}
where
\begin{subequations}
\begin{alignat}{3}
  &\hat{I} |\psi_I\rangle = I |\psi_I\rangle \, ,& \qquad  &I \in \R \, , \label{eq:I_J_eignsys_a}\\
  &\hat{J}_k |\psi_{n_k}\rangle = \heff (n_k + 1/2) |\psi_{n_k}\rangle
  \, ,& \qquad &n_k \in \N_0 \, , \label{eq:I_J_eignsys_b}
\end{alignat}
\label{eq:I_J_eignsys}
\end{subequations}
and
\begin{equation}
  E = K_{\mathrm{QNF}}^{(N)} \Big( I, \heff (n_2+1/2), \ldots, \heff (n_d+1/2) \Big) \, .
\label{eq:H_QNF_eignval}
\end{equation}
And, after having explicitly described the eigenstates
$|\psi_I\rangle$ and $|\psi_{n_k}\rangle$ of the reaction and bath
degrees of freedom respectively, one obtains a complete (approximate)
eigensystem for the original reactive scattering problem.

\begin{figure}[h]
\centerline{\epsfig{figure=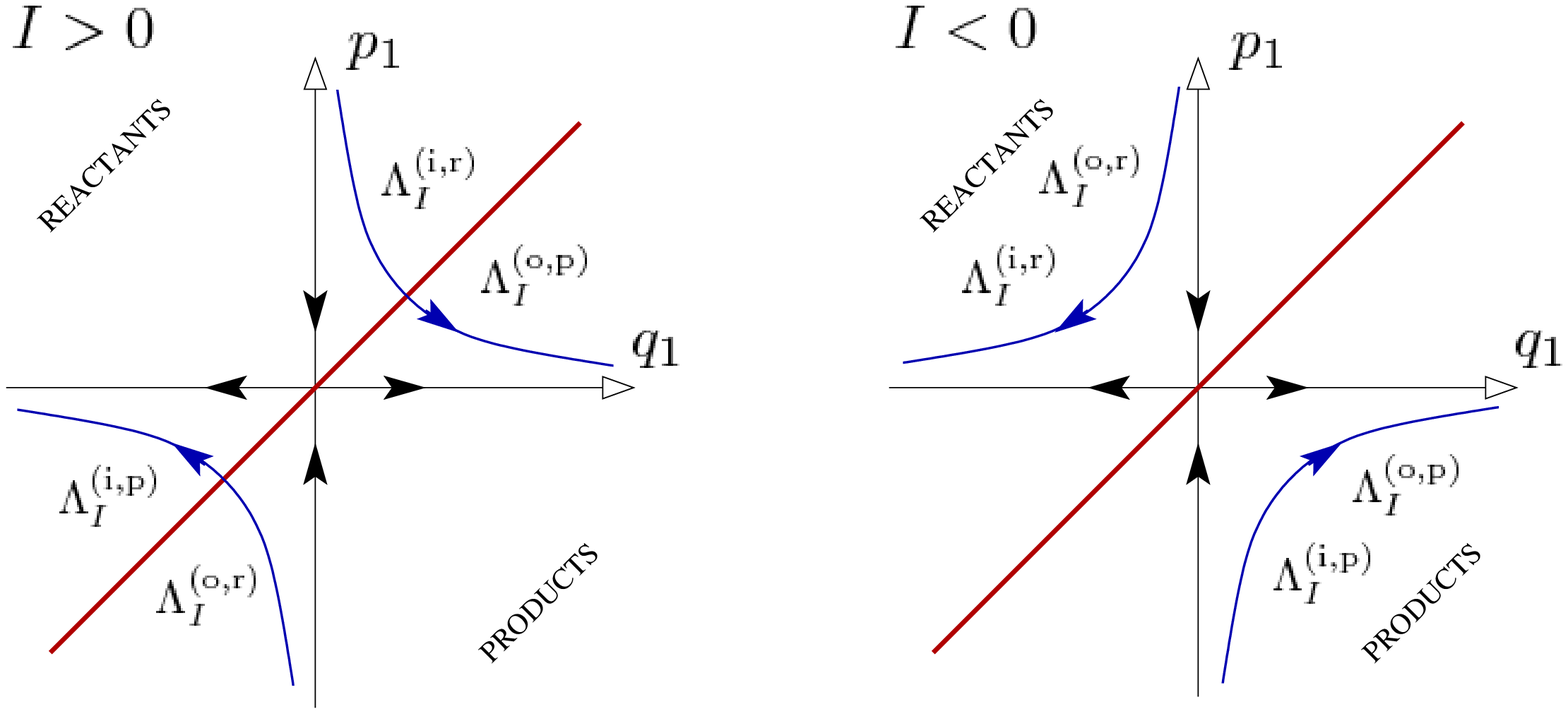,width=5in}}
\caption{The blue lines show the Lagrangian manifolds
  $\Lambda_I^{\mathrm{(i,r)}}$, $\Lambda_I^{\mathrm{(i,p)}}$,
  $\Lambda_I^{\mathrm{(o,r)}}$, and $\Lambda_I^{\mathrm{(o,p)}}$
  associated with the $\hat{I}$-eigenstates $| \psi_I^{\mathrm{(i,r)}}
  \rangle$, $| \psi_I^{\mathrm{(i,p)}} \rangle$, $|
  \psi_I^{\mathrm{(o,r)}} \rangle$, and $| \psi_I^{\mathrm{(o,p)}}
  \rangle$ respectively. The arrows indicate the classical Hamiltonian
  vector fields generated by $I=p_1q_1$. The red thick line
  corresponds to the dividing surface $s(q_1,p_1) = q_1-p_1$.}
\label{fig:lagrange_1}
\end{figure}

Two independent solution of Eq.~(\ref{eq:I_J_eignsys_a}), in the
$q_1$-representation, are given by
\begin{subequations}
\begin{align}
  &\langle q_1 | \psi_I^{\mathrm{(o,r)}} \rangle = \Theta(-q_1) (-q_1)^{-1/2 + iI/\heff} \, , \label{eq:I_eignfunc_a} \\
  &\langle q_1 | \psi_I^{\mathrm{(o,p)}} \rangle = \Theta(q_1) q_1^{-1/2 + iI/\heff} \, , \label{eq:I_eignfunc_b}
\end{align}
\label{eq:I_eignfunc_out}
\end{subequations}
where $\Theta$ is the Heaviside step function, the superscript ``o''
stands for ``outgoing to'', and ``r'' and ``p'' stand for
``reactants'' and ``products'' respectively. The motivation for this
notation is clear from viewing the solutions given by
Eq.~(\ref{eq:I_J_eignsys}) as Lagrangian states, i.e., by writing
\begin{equation}
  \langle q_1 | \psi_I^{\mathrm{(o,r/p)}} \rangle = A_I^{\mathrm{(o,r/p)}}(q_1) \,
  \exp \left( \frac{i}{\heff} \varphi_I^{\mathrm{(o,r/p)}}(q_1) \right) \, ,
\label{eq:Lagrangian_states}
\end{equation}
where the purely real amplitude and phase functions are given by
\begin{equation}
  A_I^{\mathrm{(o,r/p)}}(q_1) = \Theta(\mp q_1) |q_1|^{-1/2} \, , \qquad
  \varphi_I^{\mathrm{(o,r/p)}}(q_1) = I \ln |q_1| \, ,
\label{eq:Lagrangian_ampl_phase}
\end{equation}
respectively. This way one can associate the one-dimensional Lagrangian
manifolds
\begin{subequations}
\begin{align}
  \Lambda_I^{\mathrm{(o,r)}} &= \bigg\{
  (q_1,p_1)=\bigg(q_1,\frac{\ud}{\ud q_1}
  \varphi_I^{\mathrm{(o,r)}}(q_1)\bigg)=\bigg(q_1,\frac{I}{q_1}\bigg)
  \,:\, q_1<0 \bigg\}\,, \label{eq:lagrman_scatt_a}\\
  \Lambda_I^{\mathrm{(o,p)}} &=\bigg\{ (q_1,p_1)=\bigg(q_1,\frac{\ud
  }{\ud q_1} \varphi_I^{\mathrm{(o,p)}}(q_1)\bigg)
  =\bigg(q_1,\frac{I}{q_1}\bigg)\,:\, q_1>0
  \bigg\} \label{eq:lagrman_scatt_b}
\end{align}
\label{eq:lagrman_scatt}
\end{subequations}
with the states $| \psi_I^{\mathrm{(o,r)}} \rangle$ and $|
\psi_I^{\mathrm{(o,p)}} \rangle$. From the presentation of
$\Lambda_I^{\mathrm{(o,r)}}$ and $\Lambda_I^{\mathrm{(o,p)}}$ in
Fig.~\ref{fig:lagrange_1} one sees that for $q_1 \rightarrow -\infty$
the wave function $\langle q_1 | \psi_I^{\mathrm{(o,r)}} \rangle$
represents a state outgoing to reactants, and for $q_1 \rightarrow
+\infty$ the wave function $\langle q_1 | \psi_I^{\mathrm{(o,p)}}
\rangle$ represents a state outgoing to products.

Another pair of independent solutions of Eq.~(\ref{eq:I_J_eignsys_a}),
$\langle q_1 | \psi_I^{\mathrm{(i,r)}} \rangle$ and $\langle q_1 |
\psi_I^{\mathrm{(i,p)}} \rangle$, corresponding to states ``incoming
from'' reactants and products respectively, is obtained by requiring
their momentum representations to be given by
\begin{equation}
  \langle p_1 | \psi_I^{\mathrm{(i,r)}} \rangle  = \langle \psi_I^{\mathrm{(o,p)}} | q_1 \rangle \big|_{q_1=p_1} \, , \qquad
  \langle p_1 | \psi_I^{\mathrm{(i,p)}} \rangle  = \langle \psi_I^{\mathrm{(o,r)}} | q_1 \rangle \big|_{q_1=p_1} \, .
\label{eq:def_psi_in}
\end{equation}
The corresponding position representations are obtained by way of
Fourier transform,
\begin{equation}
  \langle q_1 | \cdot \rangle = 
  \frac{1}{\sqrt{2\pi\heff}} \int \ue^{i q_1 p_1/\heff}
  \langle p_1 | \cdot \rangle \, \ud p_1 \, ,
\label{eq:Fourier}
\end{equation}
of $\langle p_1 | \psi_I^{\mathrm{(i,r)}} \rangle$ and $\langle p_1 |
\psi_I^{\mathrm{(i,p)}} \rangle$ defined in Eq.~\eqref{eq:def_psi_in},
namely
\begin{subequations}
\begin{align}
  \langle q_1 | \psi_I^{\mathrm{(i,r)}} \rangle &=
  \frac{1}{\sqrt{2\pi\heff}} \int_0^\infty  \ue^{i q_1 p_1/\heff} \, p_1^{-1/2 -i I/\heff} \,
  \ud p_1 \, , \label{eq:psi_in_pm_a}\\
  \langle q_1 | \psi_I^{\mathrm{(i,p)}} \rangle &= 
  \frac{1}{\sqrt{2\pi\heff}} \int_{-\infty}^0  \ue^{i q_1 p_1/\heff} \, (-p_1)^{-1/2 -i I/\heff} \,
  \ud p_1 \, . \label{eq:psi_in_pm_b}
\end{align}
\label{eq:psi_in_pm}
\end{subequations}
The integrals in \eqref{eq:psi_in_pm} are not absolutely convergent,
but can be defined as oscillatory integrals.  The motivation for
defining incoming states according to Eq.~\eqref{eq:def_psi_in}
becomes clear from considering the stationary phase contributions to
the integrals \eqref{eq:psi_in_pm}. These come from the $p_1$
satisfying the stationary phase condition
\begin{equation}
\frac{\ud}{\ud p_1}(-I \ln \abs{p_1} + q_1 p_1)=  0 \,,
\end{equation}
i.e., $p_1=I/q_1$, where $p_1>0$ for $\langle q_1 | \psi_I^{\mathrm{(i,r)}}
\rangle$ and $p_1<0$ for $\langle q_1 | \psi_I^{\mathrm{(i,p)}}
\rangle$. This way one can associate the Lagrangian manifolds
\begin{equation} \label{eq:defLambdairp}
\begin{split}
\Lambda_I^{\mathrm{(i,r)}} &= \bigg\{ (q_1,p_1) = \bigg(q_1,\frac{I}{q_1}\bigg)\,:\, p_1>0 \bigg\}\,,\\
\Lambda_I^{\mathrm{(i,p)}} &= \bigg\{ (q_1,p_1) = \bigg(q_1,\frac{I}{q_1}\bigg)\,:\, p_1<0 \bigg\}\,.
\end{split}
\end{equation}
with the incoming states. These manifolds are also shown in
Fig.~\ref{fig:lagrange_1}. One sees that for $p_1 \rightarrow +\infty$
the wave function $\langle q_1 | \psi_I^{\mathrm{(i,r)}} \rangle$
represents a state incoming from reactants, and for $p_1 \rightarrow
-\infty$ the wave function $\langle q_1 | \psi_I^{\mathrm{(i,p)}}
\rangle$ represents a state incoming from products. Finally,
calculating the Fourier integrals in Eq.~(\ref{eq:psi_in_pm})
\cite{WaalkensSchubertWiggins08}, one obtains expressions
\begin{subequations}
\begin{align}
  &| \psi_I^{\mathrm{(i,r)}} \rangle = \frac{e^{i\left(
        \frac{\pi}{4}-\frac{I}{\heff} \ln\heff \right)}}{\sqrt{2\pi}}
  \Gamma\left( \frac{1}{2}-i\frac{I}{\heff} \right) \! \left(
    e^{\frac{\pi}{2}\frac{I}{\heff}} \, | \psi_I^{\mathrm{(o,p)}} \rangle - i
    e^{-\frac{\pi}{2}\frac{I}{\heff}} \, | \psi_I^{\mathrm{(o,r)}} \rangle \right) , \label{eq:I_eignfunc2_a} \\
  &| \psi_I^{\mathrm{(i,p)}} \rangle = \frac{e^{i\left(
        \frac{\pi}{4}-\frac{I}{\heff} \ln\heff \right)}}{\sqrt{2\pi}}
  \Gamma\left( \frac{1}{2}-i\frac{I}{\heff} \right) \! \left(
    e^{\frac{\pi}{2}\frac{I}{\heff}} \, | \psi_I^{\mathrm{(o,r)}} \rangle - i
    e^{-\frac{\pi}{2}\frac{I}{\heff}} \, | \psi_I^{\mathrm{(o,p)}} \rangle \right) , \label{eq:I_eignfunc2_b}
\end{align}
\label{eq:I_eignfunc_in}
\end{subequations}
that relate the incoming states to the outgoing ones \cite{VerdiereParisse94}.

\begin{figure}[h]
\centerline{\epsfig{figure=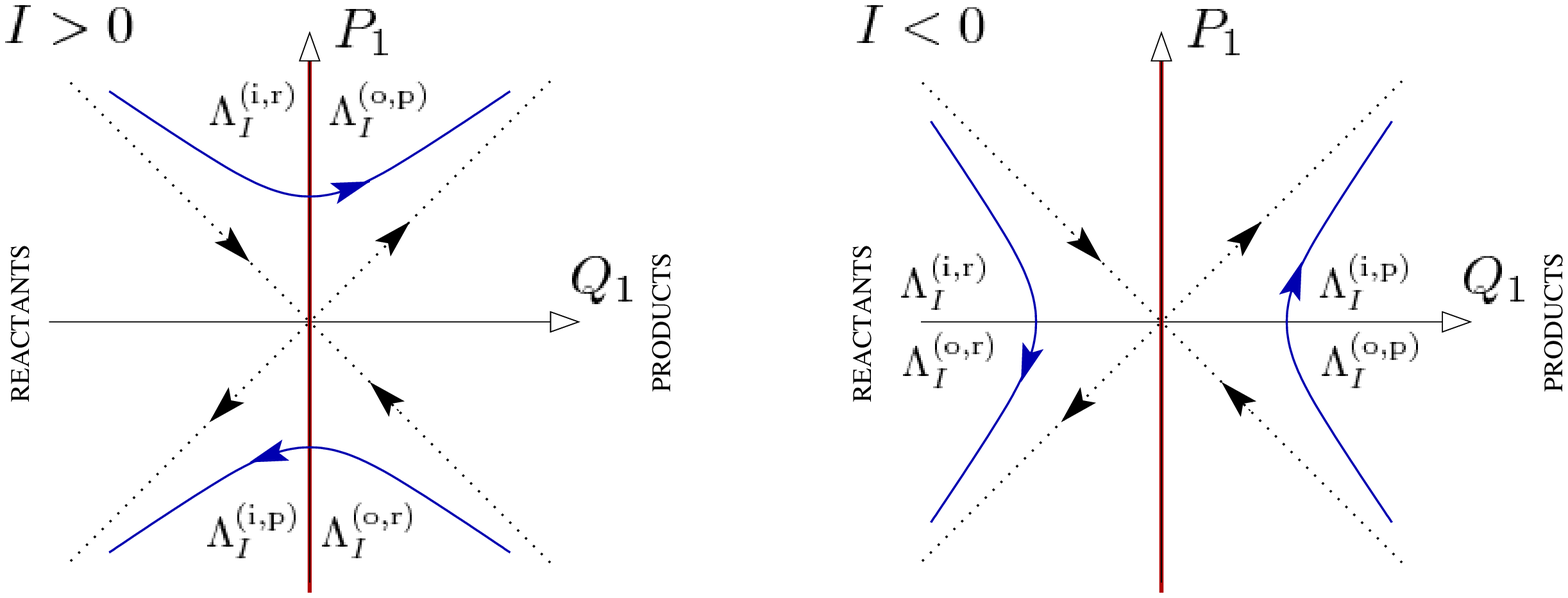,width=5in}}
\caption{The blue lines show the Lagrangian manifolds
  $\Lambda_I^{\mathrm{(i,r)}}$, $\Lambda_I^{\mathrm{(i,p)}}$,
  $\Lambda_I^{\mathrm{(o,r)}}$, and $\Lambda_I^{\mathrm{(o,p)}}$
  associated with the $\hat{I}$-eigenstates $| \psi_I^{\mathrm{(i,r)}}
  \rangle$, $| \psi_I^{\mathrm{(i,p)}} \rangle$, $|
  \psi_I^{\mathrm{(o,r)}} \rangle$, and $| \psi_I^{\mathrm{(o,p)}}
  \rangle$ respectively. The arrows indicate the classical Hamiltonian
  vector fields generated by $I=(P_1^2-Q_1^2)/2$. The red thick line
  corresponds to the dividing surface $s(Q_1,P_1) = Q_1=0$.}
\label{fig:lagrange_2}
\end{figure}

The equation for eigenstates of the action operator $\hat{I}$,
Eq.~(\ref{eq:I_J_eignsys_a}), can also be solved in the
$Q_1$-representation \cite{Chruscinski1,Chruscinski2}. The two
independent solutions are
\begin{subequations}
\begin{align}
  &\langle Q_1 | \psi_I^{\mathrm{(i,r)}} \rangle = \frac{1}{\pi}
  (2\heff)^{-\frac{3}{4}} \, e^{\frac{\pi}{4} \frac{I}{\heff}} \,
  \Gamma \left( \frac{1}{2}-i\frac{I}{\heff} \right)
  D_{-\frac{1}{2}+i\frac{I}{\heff}} \!  \left( e^{-i\frac{\pi}{4}}
    \sqrt{\frac{2}{\heff}} Q_1 \right) , \label{eq:I_eignfunc_Q_a}\\
  &\langle Q_1 | \psi_I^{\mathrm{(i,p)}} \rangle = \frac{1}{\pi}
  (2\heff)^{-\frac{3}{4}} \, e^{\frac{\pi}{4} \frac{I}{\heff}} \,
  \Gamma \left( \frac{1}{2}-i\frac{I}{\heff} \right)
  D_{-\frac{1}{2}+i\frac{I}{\heff}} \!  \left( e^{i\frac{3\pi}{4}}
    \sqrt{\frac{2}{\heff}} Q_1 \right) , \label{eq:I_eignfunc_Q_b}
\end{align}
\label{eq:I_eignfunc_Q}
\end{subequations}
where $D_\nu$ is the parabolic cylinder function of order $\nu$
\cite{AbramovizStegun}. The outgoing states, $|
\psi_I^{\mathrm{(o,r)}} \rangle$ and $| \psi_I^{\mathrm{(o,p)}}
\rangle$, are then obtained from
Eq.~(\ref{eq:I_eignfunc_in}). Fig.~\ref{fig:lagrange_2} shows the
Lagrangian manifolds $\Lambda_I^{\mathrm{(i,r)}}$,
$\Lambda_I^{\mathrm{(i,p)}}$, $\Lambda_I^{\mathrm{(o,r)}}$, and
$\Lambda_I^{\mathrm{(o,p)}}$ in the $(Q_1,P_1)$ coordinates.

Finally, we note that the incoming and outgoing eigenstates of the
$\hat{I}$ operator are orthogonal and satisfy the completeness
relations \cite{Chruscinski1,Chruscinski2}
\begin{subequations}
\begin{align}
  &\int_\R \left(
  | \psi^{\mathrm{(i,r)}}_I \rangle \langle \psi^{\mathrm{(i,r)}}_I | +
  | \psi^{\mathrm{(i,p)}}_I \rangle \langle \psi^{\mathrm{(i,p)}}_I |
  \right) \ud I= \hat{\mathds{1}} \, , \label{eq:completeness_relation_a}\\
  &\int_\R \left(
  | \psi^{\mathrm{(o,r)}}_I \rangle \langle \psi^{\mathrm{(o,r)}}_I | +
  | \psi^{\mathrm{(o,p)}}_I \rangle \langle \psi^{\mathrm{(o,p)}}_I |
  \right) \ud I= \hat{\mathds{1}} \, , \label{eq:completeness_relation_b}
\end{align}
\label{eq:completeness_relation}
\end{subequations}
where $\hat{\mathds{1}}$ stands for the identity operator.

Generally speaking, the $q_1$- and $Q_1$-representations of the
eigenstates of the $\hat{I}$ operator are completely equivalent, and
the choice of a representation is usually dictated by a particular
problem. For example, in order to determine the total probability flux
corresponding to the eigenstates $| \psi_I^{\mathrm{(i,r/p)}} \rangle$
through the dividing surface it is convenient to adopt the
$Q_1$-representation. Indeed, in the $QP$-basis the operator
corresponding to the flux through the dividing surface $s(Q_1,P_1) =
Q_1$ is \cite{WaalkensSchubertWiggins08}
\begin{equation}
  \hat{F} = \frac{i}{\heff} [\hat{I}, \Theta(\hat{Q}_1)]
  = \frac{1}{2} \left( \hat{P}_1 \delta(\hat{Q}_1)
    + \delta(\hat{Q}_1) \hat{P}_1 \right) \, .
\label{eq:dividing_surf_flux_operator}
\end{equation}
Thus, the expectation value of $\hat{F}$ with respect to the incoming
states is given by
\begin{align}
  \langle \psi_I^{\mathrm{(i,r/p)}} | \hat{F} |
  \psi_I^{\mathrm{(i,r/p)}} \rangle &= \heff \Im \left( \langle
    \psi_I^{\mathrm{(i,r/p)}} | Q_1 \rangle
    \frac{\ud}{\ud Q_1} \langle Q_1 | \psi_I^{\mathrm{(i,r/p)}} \rangle \right) \bigg|_{Q_1=0} \nonumber\\
  &= \pm \frac{1}{2\pi\heff} \, \frac{1}{1+\exp( -2\pi I / \heff)} \, .
\label{eq:dividing_surf_flux_expect}
\end{align}

The operators $\hat{J}_k$, defined by Eq.~(\ref{eq:J_op}) and
corresponding to the action integrals of the bath modes, are the
Hamiltonian operators of one-dimensional harmonic oscillators with
unit frequency. Therefore, the wave function representing the
eigenstates in Eq.~(\ref{eq:I_J_eignsys_b}) are given by
\begin{equation}
  \langle q_k | \psi_{n_k} \rangle = \frac{1}{\sqrt{2^{n_k} (n_k!)}}
  \left(\frac{1}{\pi \heff}\right)^{1/4} H_{n_k}\left(
    \frac{q_k}{\sqrt{\heff}} \right) \exp\left( -\frac{q_k^2}{2\heff}
  \right) \, ,
\label{eq:J_eignfunc}
\end{equation}
where $H_n$ denotes the $n$th order Hermite polynomial
\cite{AbramovizStegun}.

Here we note that the quantum normal form Hamiltonian can be
interpreted as representing the activated complex. As it has been
shown above the QNF Hamiltonian has only continuous spectrum,
and so there are no bound states associated with the activated
complex. Physically this corresponds to the fact that the activated
complex has a finite life time. The latter is determined by the
Gamov-Siegert resonances which we discuss in detail in
Section~\ref{sec:resonances}.



\section{The Cumulative Reaction Probability}

\subsection{General Formulation}
\label{sec:smatrix}

The incoming and outgoing eigenstates of $\hat{I}$ defined in
Sec.~\ref{sec:activated_complex} are related to one another by
Eq.~(\ref{eq:I_eignfunc_in}). Therefore, each solution $| \psi_I
\rangle$ of Eq.~\eqref{eq:I_J_eignsys_a} can be written as a linear
combination of $| \psi_I^{\mathrm{(o,r/p)}} \rangle$ or $|
\psi_I^{\mathrm{(i,r/p)}} \rangle$,
\begin{equation}
  | \psi_I \rangle = \alpha_{\mathrm{p}} | \psi_I^{\mathrm{(o,p)}} \rangle
  + \alpha_{\mathrm{r}} | \psi_I^{\mathrm{(o,r)}} \rangle
  = \beta_{\mathrm{p}} | \psi_I^{\mathrm{(i,p)}} \rangle
  + \beta_{\mathrm{r}} | \psi_I^{\mathrm{(i,r)}} \rangle  \, .
\label{eq:psi_I_arbitrary}
\end{equation}
These representations are connected by the (one-dimensional) S-matrix,
\begin{equation}
\begin{pmatrix}\alpha_{\mathrm{p}} \\ \alpha_{\mathrm{r}}\end{pmatrix}
=\mathcal{S}(I)\begin{pmatrix}\beta_{\mathrm{p}}\\ \beta_{\mathrm{r}} \end{pmatrix} \, .
\end{equation}
The entries of the S-matrix can be read of directly from
Eq.~(\ref{eq:I_eignfunc_in}) yielding
\begin{equation}\label{eq:local-S}
\mathcal{S}(I)=\frac{\ue^{i \left( \frac{\pi}{4} - \frac{I}{\heff}\ln\heff \right)}}{\sqrt{2\pi}} \,
\Gamma\bigg(\frac{1}{2}-i\frac{I}{\heff}\bigg) \!
\begin{pmatrix}
-i \ue^{-\frac{\pi}{2}\frac{I}{\heff}} & \ue^{\frac{\pi}{2}\frac{I}{\heff}}  \\
\ue^{\frac{\pi}{2}\frac{I}{\heff}} & -i \ue^{-\frac{\pi}{2}\frac{I}{\heff}}
\end{pmatrix} \, .
\end{equation}
Using the relation $\Gamma(1/2+i y)\Gamma(1/2-i y)=\pi/\cosh(\pi y)$
it is easy to see that $\mathcal{S}(I)^*\mathcal{S}(I)=\mathds{1}$
implying that the S-matrix is unitary.

The transmission coefficient corresponding to the scattering matrix
$\mathcal{S}(I)$ reads
\begin{equation}
  \mathcal{T}(I) =\abs{\mathcal{S}_{12}(I)}^2=\frac{\ue^{\pi\frac{I}{\heff}}}{\ue^{\pi\frac{I}{\heff}}
    +\ue^{-\pi\frac{I}{\heff}}} =\frac{1}{1+\ue^{-2\pi \frac{I}{\heff}}} \, ,
\label{eq:T-coef}
\end{equation}
and the reflection coefficient
\begin{equation}
  \mathcal{R}(I) =\abs{\mathcal{S}_{11}(I)}^2=\frac{\ue^{-\pi\frac{I}{\heff}}}{\ue^{\pi\frac{I}{\heff}}
    +\ue^{-\pi\frac{I}{\heff}}} =\frac{1}{1+\ue^{2\pi \frac{I}{\heff}}} \, .
\label{eq:R-coef}
\end{equation}
As required, one has $\mathcal{T}(I) + \mathcal{R}(I) = 1$. The
characteristic action scale is given by the effective Planck's
constant: $\mathcal{T}$ tends to $1$ if $I \gg \heff$ and to $0$ if $I
\ll -\heff$.

In accordance with Eqs.~(\ref{eq:H_QNF_eignfunc}) and
(\ref{eq:I_J_eignsys}), the incoming and outgoing scattering states of
the $d$-dimensional reactive system at energy $E$ are given by
\begin{subequations}
\begin{align}
  | \Psi^{\mathrm{(i,r/p)}} \rangle &= | \psi_I^{\mathrm{(i,r/p)}}
  \rangle \otimes | \psi_{{\bf n}} \rangle \, , \label{eq:mult_scat_states_a}\\
  | \Psi^{\mathrm{(o,r/p)}} \rangle &= | \psi_I^{\mathrm{(o,r/p)}}
  \rangle \otimes | \psi_{{\bf n}} \rangle \, , \label{eq:mult_scat_states_b}
\end{align}
\label{eq:mult_scat_states}
\end{subequations}
where ${\bf n} = (n_2,\ldots ,n_d) \in \N_0^{d-1}$ is a
$(d-1)$-dimensional vector of scattering quantum numbers, and
\begin{equation}
  | \psi_{{\bf n}} \rangle = | \psi_{n_2} \rangle \otimes \ldots \otimes | \psi_{n_d} \rangle
\label{eq:bath_mode}
\end{equation}
are the eigenstates of a $(d-1)$-dimensional harmonic oscillator
corresponding to the bath modes. Here, $I$ and ${\bf n}$ and not
independent, but related by Eq.~(\ref{eq:H_QNF_eignval}). For energies
$E$ close to the equilibrium point energy $E_0$ this equation can be
resolved for the reactive coordinate action, yielding a single valued
function $I = I_{\bf n}(E)$. Then, the S-matrix connecting the scattering states $|
\Psi^{\mathrm{(i,r/p)}} \rangle$ and $| \Psi^{\mathrm{(o,r/p)}}
\rangle$ is block-diagonal and given by
\begin{equation}
S_{{\bf n},{\bf m}}(E)
=\delta_{{\bf n},{\bf m}} \, \mathcal{S}(I_{{\bf n}}(E)) \, ,
\label{eq:Smatrixfull}
\end{equation}
where $\delta_{{\bf n},{\bf m}}$ is the multi-dimensional Kronecker
symbol, and $\mathcal{S}(I)$ is given by Eq.~\eqref{eq:local-S}.

\rem{The transition matrix $T$ can be defined as the
diagonal sub-block of the S-matrix with the diagonal elements equal to
the $(1,2)$-components of the matrices $\mathcal{S}(I)$ given by
Eq.~\eqref{eq:Smatrixfull}, i.e.,}

The transmission matrix $T$ can be defined as a diagonal matrix with
the diagonal elements equal to the modulus squared of the
$(1,2)$-components of the matrices $\mathcal{S}(I)$ in
Eq.~\eqref{eq:Smatrixfull}, i.e.,
\begin{equation}
  T_{{\bf n},{\bf m}}(E)= \delta_{{\bf n},{\bf m}} \, \big| \mathcal{S}_{1,2}(I_{{\bf n}}(E)) \big|^2
  =\delta_{{\bf n},{\bf m}}
  \bigg[1+\exp\bigg(-2\pi\frac{I_{{\bf n}}(E)}{\heff}\bigg) \bigg]^{-1} .
\label{eq:transmission1D}
\end{equation}
The \emph{cumulative reaction probability} $\crp(E)$ is then defined  as (see,
e.g., \cite{Miller98})
\begin{equation}
  \crp(E) = \text{Tr} \, \big\{ T(E) \big\} \, ,
\label{eq:crp_T_matrix_def}
\end{equation}
where $\text{Tr}$ stands for the trace\rem{, and $\dagger$ denotes
Hermitian conjugation}. Thus, using Eq.~\eqref{eq:transmission1D} one
obtains
\begin{equation}\label{eq:cum_react}
  \crp(E) = \sum_{{\bf n}} T_{{\bf n},{\bf n}}(E)=
  \sum_{{\bf n}} \bigg[1+\exp\bigg(-2\pi\frac{I_{{\bf n}}(E)}{\heff}\bigg)
  \bigg]^{-1} .
\end{equation}

The cumulative reaction probability $\crp(E)$ is the quantum analogue
of the classical flux $\flux(E)$ through the dividing surface or, more
precisely, of the dimensionless quantity $\flux(E)/(2\pi
\heff)^{d-1}$. \rem{that we defined in Equation~\eqref{eq:NWeyl_def}
  in Sec.~\ref{sec:classicalrate}.} To see this one should consider
$\crp(E)$ in the semiclassical limit $\heff\rightarrow 0$, where
\begin{equation}
  \lim_{\heff \rightarrow 0}  \bigg[1+\exp\bigg(-2\pi I/\heff \bigg)
  \bigg]^{-1} = \Theta(I) \, ,
\end{equation}
where $\Theta$ is the Heaviside function. This means that the
transmission coefficients $ T_{{\bf n},{\bf n}}(E)$ in
Eq.~\eqref{eq:cum_react} are characteristic functions, i.e., in the
semiclassical limit, $ T_{{\bf n},{\bf n}}(E)$ is 0 or 1 if $I_{\bf
  n}(E)$ is negative or positive respectively.
This way the cumulative reaction probability can be considered to be a
counting function. For a given energy $E$, it counts how many of the
solutions $I_{{\bf n}}$ of the equations $K_{\text{QNF}}(I_{{\bf
    n}},\heff(n_2+1/2),\ldots,\heff(n_d+1/2))=E$ with scattering
quantum numbers ${\bf n} = (n_2,\ldots,n_d) \in\N_0^{d-1}$ are
positive. \rem{
\begin{equation}
  \crp(E) \rightarrow   \# \{ I_{{\bf n}}>0 \,:\,
  K_{\text{QNF}}(I_{{\bf n}},\heff(n_2+1/2),\ldots,\heff(n_d+1/2))=E,\,
  {\bf n} \in \N_0^{d-1} \}\,,
\end{equation}
as $\heff \rightarrow 0$.} In other words, $\crp(E)$ can be considered
to count the number of open ``transmission channels'', where a
transmission channel with quantum numbers ${\bf n}$ is open if the
corresponding transmission coefficient $T_{{\bf n},{\bf n}}(E)$ is
close to 1.

\def\fighbargrid{%
  (a) Lines $(I,\heff(n_2+1/2),\ldots,\heff(n_d+1/2))$, $I\in\R$,
  $n_k\in\N_0$, $k=2,\ldots,d$, in the space
  $(I,J_2,\ldots,J_d)\in\R\times[0,\infty)^{d-1}$ for $d=3$ and their
  intersections with the surface $K_{\text{QNF}}(I,J_2,J_3)=E$.  (b)
  Grid points $(\heff(n_2+1/2),\ldots,\heff(n_d+1/2))$ in the space
  $(J_2,\ldots,J_d)$ for $d=3$. The blue line marks the contour
  $K_{\text{QNF}}(0,J_2,\ldots,J_d)=E$. In this plot only the
  scattering states for which the quantum numbers $(n_2,n_3)$ have the
  values $(0,0)$, $(0,1)$, $(1,0)$ or $(1,1)$ correspond to ``open
  transmission channels'', see text.} \def\FIGhbargrid{ \centerline{
    \raisebox{6cm}{a)}\includegraphics[width=8.0cm]{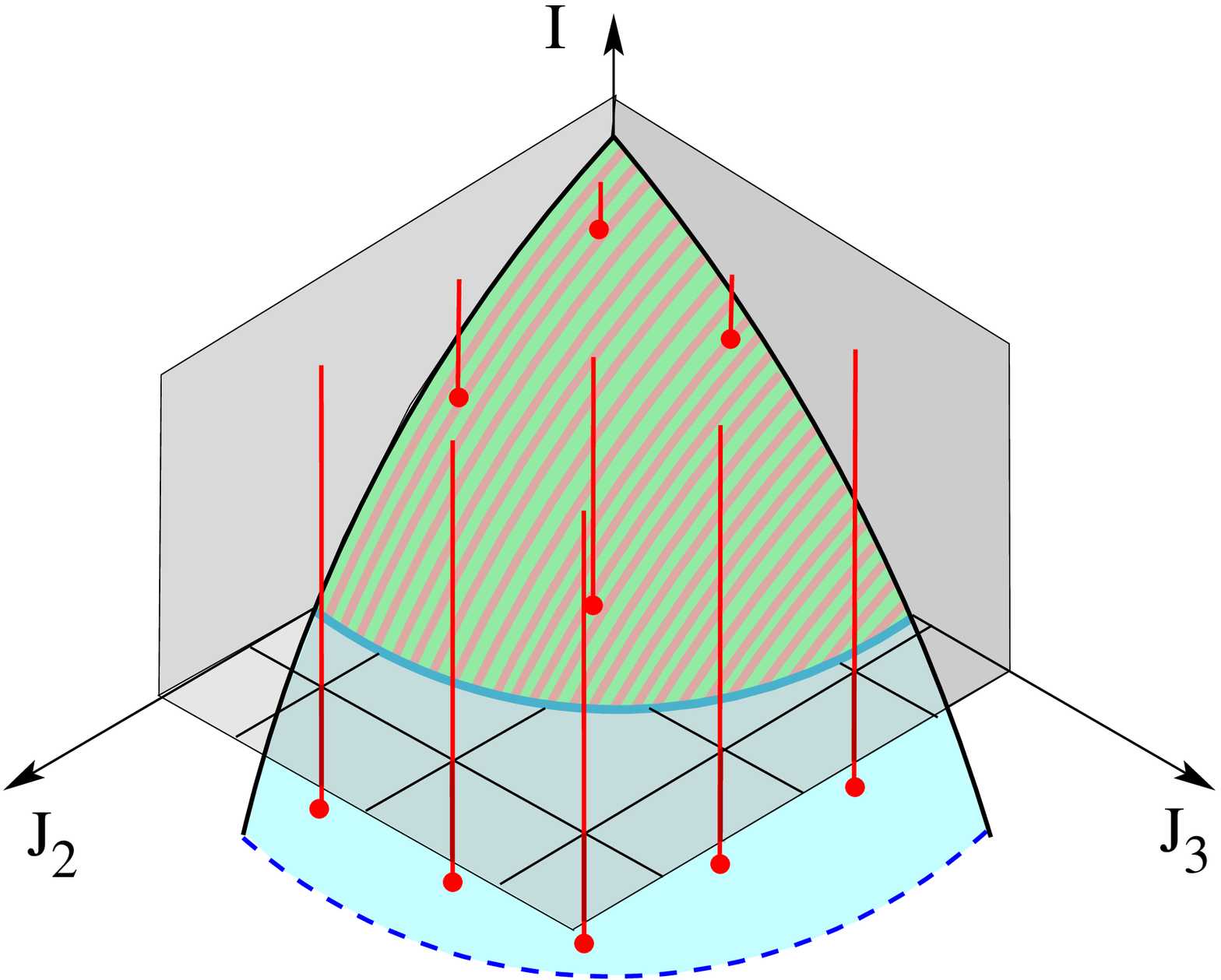}
    \raisebox{6cm}{b)}\includegraphics[width=5.0cm]{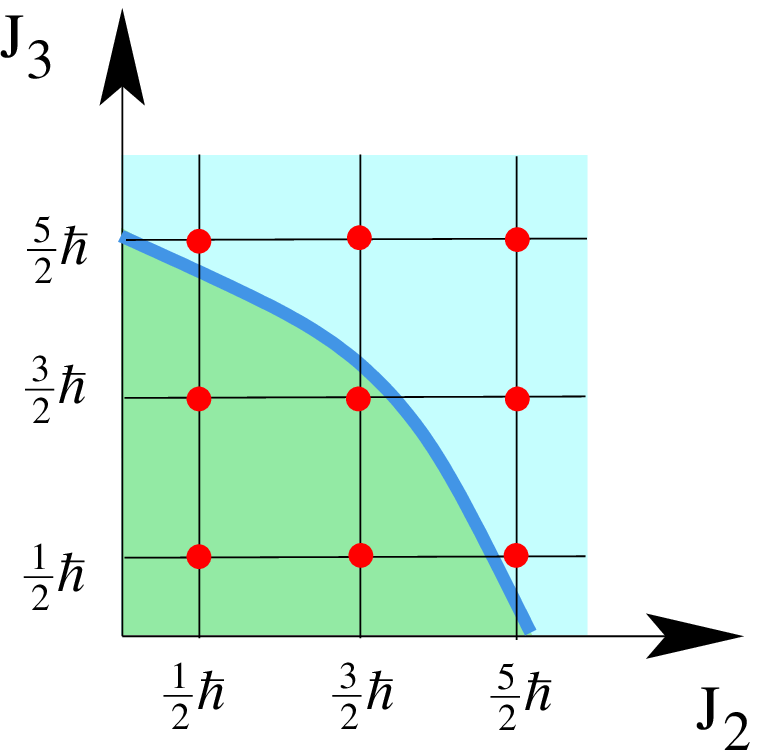}
  } } \FIGo{fig:hbargrid}{\fighbargrid}{\FIGhbargrid}

Graphically $\crp(E)$ can be interpreted as the number of grid points
$(\heff(n_2+1/2),\ldots,\heff(n_d+1/2))$ in the space of action
integrals $(J_2,\ldots,J_d)\in[0,\infty)^{d-1}$ that are enclosed by
the contour $K_{\text{QNF}}(0,J_2,\ldots,J_d)=E$, see
Fig.~\ref{fig:hbargrid}.  The number of grid points is approximately
given by the volume in the space of
$(J_2,\ldots,J_d)\in[0,\infty)^{d-1}$ enclosed by
$K_{\text{QNF}}(0,J_2,\ldots,J_d)=E$ divided by $\heff^{d-1}$. Using
the fact that for $\heff \rightarrow 0$, $K_{\text{QNF}}$ becomes the
function $K_{\text{CNF}}$ which gives the classical energy as a
function of the classical integrals $(I,J_2,\ldots,J_d)$ we find that
the volume in the space of $(J_2,\ldots,J_d)$ enclosed by
$K_{\text{CNF}}(0,J_2,\ldots,J_d)=E$ is given by the classical flux
$\flux(E)$ divided by $(2\pi)^{d-1}$, see
Eq.~\eqref{eq:fluxactions_classical}, and the cumulative reaction
probability $\crp(E)$ is thus approximately given by
$\crp_{\mathrm{Weyl}}(E)=\flux(E)/(2\pi\hbar)^{d-1}$ defined in
Eq.~\eqref{eq:NWeyl_def}. This way we verify our statement in
Sec.~\ref{sec:classicalrate} that $\crp_{\mathrm{Weyl}}(E)$ gives the
mean number of open transmission channels. In fact, as mentioned in
Sec.~\ref{sec:classicalrate}, the classical flux $\flux(E)$ can be
considered to be the phase space volume enclosed by the energy contour
of energy $E$ of the invariant subsystem which has one degree of
freedom less than the full scattering system and which as the so
called activated complex is located between reactants and
products. $\crp_{\mathrm{Weyl}}(E)$ counts how many elementary quantum
cells of volume $(2\pi \heff)^{d-1}$ fit into this phase space volume
and this way gives the Weyl approximation of the cumulative reaction
probability $\crp(E)$.

It is important to note here that like the flux in the classical case
the cumulative reaction probability is determined by local properties
of the Hamilton operator $\hat{H}$ in the neighbourhood of the
equilibrium point only.



\subsection{Nitrogen Exchange Reaction}
\label{sec:nitrogen_exchange}

In this section we focus on the QNF theory as a tool for computation
of the cumulative reaction probability (CRP) and the thermal reaction
rate constant. To demonstrate the efficiency and capability of the QNF
method and to compare it with other existing methods we follow
Ref.~\cite{gsww09} and address the CRP and the thermal rate in a
collinear triatomic reaction, namely the nitrogen exchange reaction
\begin{equation}
  \mathrm{N} + \mathrm{N}_2 \rightarrow \mathrm{N}_2 + \mathrm{N} \, .
\label{eq:nitrogen_reaction}
\end{equation}
The collinear version of Eq.~(\ref{eq:nitrogen_reaction}) corresponds
to a two degree of freedom reactive system with the Hamiltonian
operator of the form
\begin{equation}
  \hat{H} \equiv H(\hat{\tilde{q}}_1, \hat{\tilde{q}}_2, \hat{\tilde{p}}_1, \hat{\tilde{p}}_2)
  = \frac{1}{2} \left( \hat{\tilde{p}}_1^2 + \hat{\tilde{p}}_2^2 \right) + V(\hat{\tilde{q}}_1, \hat{\tilde{q}}_2) \, ,
\label{eq:H_2D}
\end{equation}
where $V(\tilde{q}_1,\tilde{q}_2)$ gives the Born-Oppenheimer
potential energy surface, and $\tilde{q}_1$ and $\tilde{q}_2$ are the
Delves mass-scaled coordinates \cite{delves59}. The effective Planck's
constant is given by $\heff = \mu^{-1/2}$, where $\mu$ is the
dimensionless reduced mass of the triatomic system (given in units of
the electronic mass). For the case of nitrogen exchange reaction $\mu
\approx 1.47 \times 10^{4}$ yielding $\heff \approx 8.2 \times
10^{-3}$. We further adopt the London-Eyring-Polanyi-Sato potential
energy surface \cite{lagana} that possess a single saddle point
governing the reaction from the asymptotic reactants and products
states.

\begin{figure}[htb!]
\centerline{\epsfig{figure=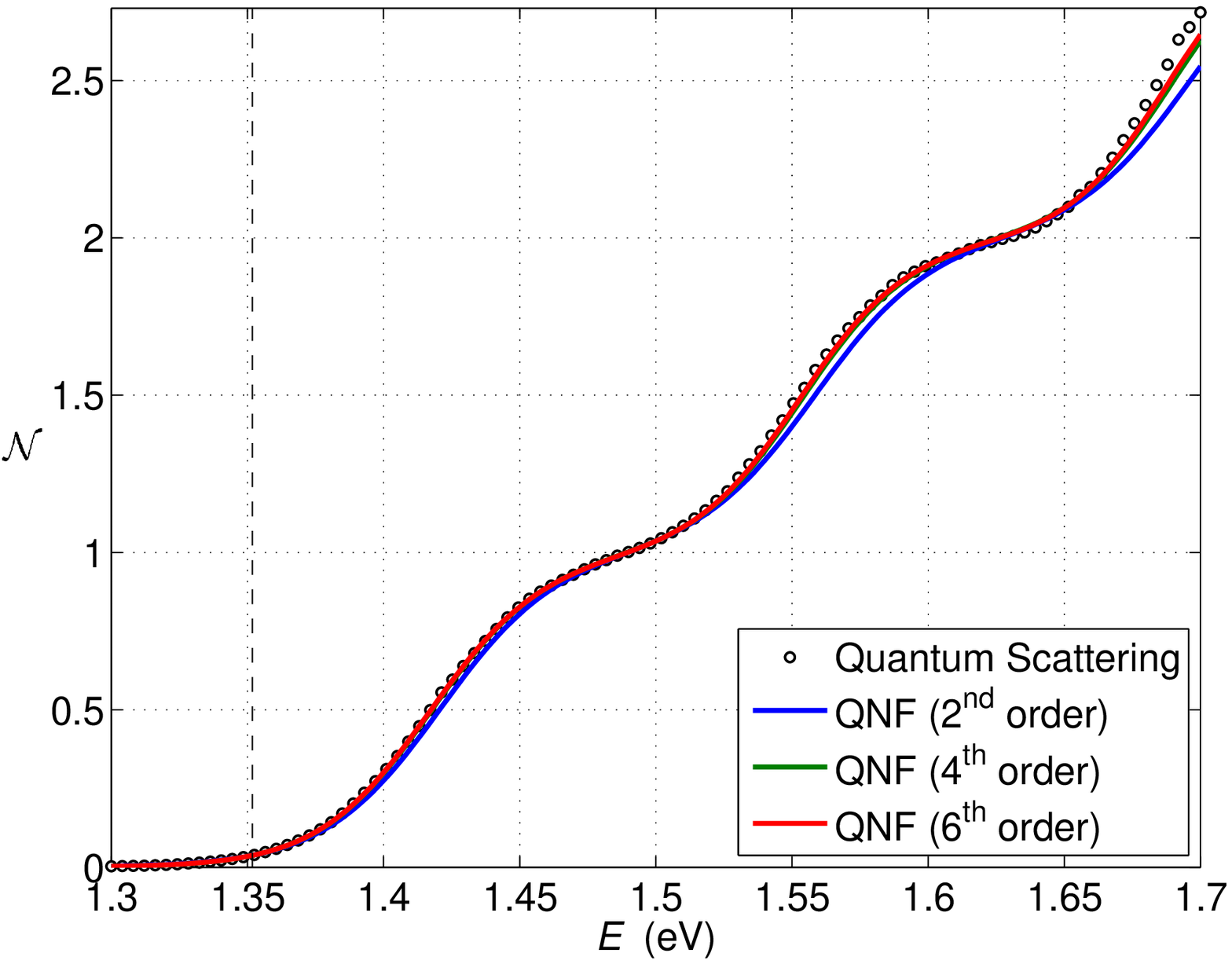,width=4.5in}}
\caption{Cumulative reaction probability as a function of the total
  energy, $\crp(E)$, for the collinear nitrogen exchange reaction
  (\ref{eq:nitrogen_reaction}).  The effective Planck's constant is
  $\heff \approx 8.2 \times 10^{-3}$. The $\crp(E)$ curves obtained
  with the $4^{\mathrm{th}}$ and $6^{\mathrm{th}}$ order QNF are
  essentially indistinguishable for most of the energy range. The
  vertical dashed line shows the saddle point energy, $E_0$.}
\label{fig:CRP}
\end{figure}

Following the algorithmic procedure presented in
Sec.~\ref{sec:activated_complex} and Appendix~\ref{sec:QNF_algorithm}
we compute the $N$th order approximation
\begin{equation}
  \hat{H}_{\mathrm{QNF}}^{(N)} = K_{\mathrm{QNF}}^{(N)}(\hat{I}, \hat{J}_2)
\label{eq:QNF_2D}
\end{equation}
of the original Hamiltonian operator $\hat{H}$. The CRP is then given
by
\begin{equation}
  \crp(E) = \sum_{n_2=0}^{\infty} \left[ 1 + \exp\left( -2\pi \frac{I_{n_2}(E)}{\heff} \right) \right]^{-1}
\label{eq:CRP_2D}
\end{equation}
with $I_{n_2}(E)$ solving
\begin{equation}
  K_{\mathrm{QNF}}^{(N)}\big( I,\heff (n_2+1/2) \big) = E \, .
\label{eq:I_n_2D}
\end{equation}
The resulting CPR-vs-energy curves computed for energies in the range
between $1.3$ and $1.7$ eV are shown in Fig.~\ref{fig:CRP}. The solid
color curves (blue, green, and red) correspond to different orders $N$
of the QNF approximation (2, 4, and 6 respectively). The $\crp(E)$
curves obtained with the $4$th and $6$th order QNF are essentially
indistinguishable for most of the energy range; this fact signals the
rapid convergence of the QNF expansion for the given value of the
effective Planck's constant. The vertical dashed line shows the saddle
point energy, $E_0$, of the London-Eyring-Polanyi-Sato potential
energy surface.

The circular data points in Fig.~\ref{fig:CRP} represent the ``exact''
valued of the CRP obtained through the full {\it reactive quantum
  scattering} calculation \cite{hauke,kupperman}. The latter were
performed by integrating the coupled multichannel Schr\"odinger
equation in hyperspherical coordinates \cite{hauke,kupperman} from the
strong interaction region to the asymptotic reactant and product
configurations. The log-derivative matrix method of Manolopoulos and
Gray \cite{manol} together with the six-step symplectic integrator of
McLachlan and Atela \cite{mclach} was used to integrate the radial
Schr\"odinger equation. It is evident from Fig.~\ref{fig:CRP} that the
quantitative agreement of the exact and QNF values of $\crp(E)$ well
extends up to energies of 1.5 eV. \footnote{It is interesting to note
  that the QNF curves in Fig.~\ref{fig:CRP} seem to converge to a
  result that slightly differs from the exact CRP curve at high
  energies: a small discrepancy starts to show up at energies around
  1.5 eV and becomes more pronounced at energies above 1.65 eV. This
  discrepancy may be attributed to the growing importance of
  corner-cutting tunneling at high energies. We discuss the phenomenon
  of the corner-cutting tunneling in relation to the normal form
  theory in Section~\ref{sec:corner-cutting}.}

We finally note that the QNF calculation of the CRP requires
significantly less computational time than the corresponding full
quantum reactive scattering calculation. For example, the $6$th order
QNF computation of the nitrogen-exchange CRP curve in
Fig.~\ref{fig:CRP} took about 10 minutes on a 2.6 GHz processor, 2 GB
RAM computer, while the corresponding full quantum reactive scattering
computation took more than 12 hours on the same machine. The QNF
approach becomes even more advantageous for treating chemical systems
of atoms heavier than nitrogen: the expense of the full quantum
computations rapidly grows with the number of asymptotic channels
(and, therefore, with mass) \cite{walker}, while the QNF expansion
only becomes more rapidly convergent making the corresponding analysis
computationally cheaper.

\begin{figure}[htb!]
\centerline{\epsfig{figure=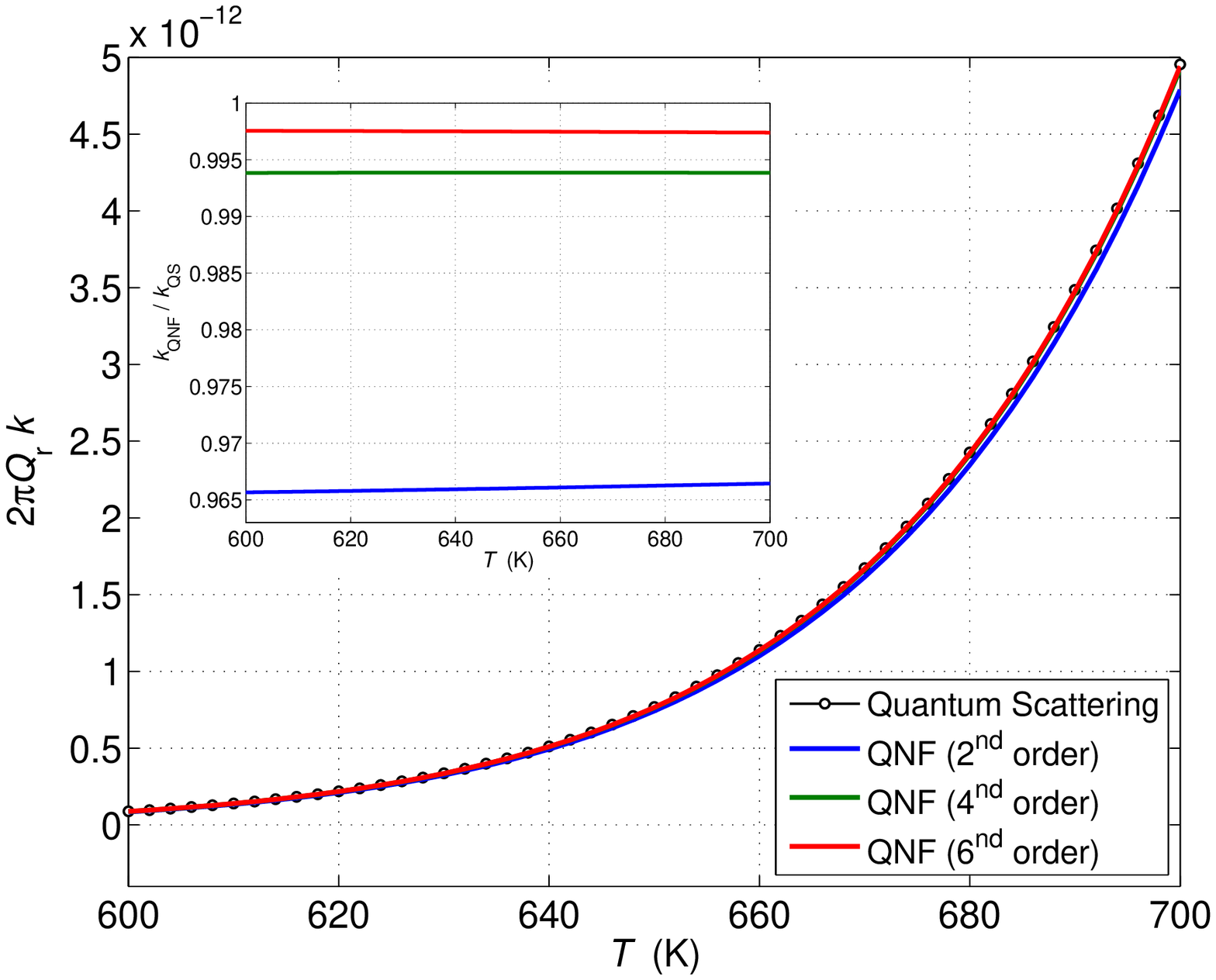,width=4.5in}}
\caption{The thermal rate constant (multiplied by the reactant
  partition function) as a function of temperature for the collinear
  nitrogen exchange reaction (\ref{eq:nitrogen_reaction}). The curves
  obtained with the $4^{\mathrm{th}}$ and $6^{\mathrm{th}}$ order QNF
  are essentially indistinguishable for most of the energy range. The
  inset shows the ratio, $k_{\mathrm{QNF}} / k_{\mathrm{QS}}$, of the
  thermal rate constant computed using the QNF of orders 2, 4, 6, to
  the one obtained from the quantum scattering data.}
\label{fig:thermal_rate}
\end{figure}

We now discuss the straightforward extension of the QNF approach to
computation of the thermal reaction rate constant, $k(T)$, defined as
\cite{seideman}
\begin{equation}
  k(T) = \frac{1}{2\pi Q_{\mathrm{r}}(T)} \int_0^\infty dE
  \exp\left(-\frac{E}{k_{\mathrm{B}}T}\right) \crp(E) \, ,
\label{eq:thermal_rate}
\end{equation}
where $T$ stands for the absolute temperature, $Q_{\mathrm{r}}(T)$ is
the partition function of the reactant, and $k_{\mathrm{B}}$ is the
Boltzmann constant. To this end, we use the CRP data for the nitrogen
exchange reaction, Eq.~(\ref{eq:nitrogen_reaction}), presented in
Fig.~\ref{fig:CRP}. The data allow us to compute $k(T)$ in the range
of temperatures between 600 and 700 K, for which the integrand in the
right hand side of Eq.~(\ref{eq:thermal_rate}) is well localized to
the energy interval between 1.3 and 1.7
eV. Figure~\ref{fig:thermal_rate} provides a comparison of $k(T)$
calculated from $\crp(E)$ obtained using the QNF of orders 2, 4, 6
(color solid lines), and the ``exact'', reactive quantum scattering
technique (black line with circles). The inset in the figure shows the
ratio, $k_{\mathrm{QNF}} / k_{\mathrm{QS}}$, of the thermal rate
computed using the QNF of orders 2, 4, 6, to the one obtained from the
quantum scattering data. One can clearly see that the thermal rate
constant computed with the QNF method rapidly approaches its exact
value as the approximation order $N$ is increased.

\subsection{A 3 \dof Reactive System}
\label{sec:Eckart_Morse_Morse}

In order to illustrate the utility of the QNF technique for
computation of reaction rates in higher dimensional systems we
consider a 3 \dof model system consisting of an Eckart barrier in the
(physical) $\tilde{q}_1$-direction that is coupled to Morse
oscillators in the $\tilde{q}_2$-direction and in the
$\tilde{q}_3$-direction. The Hamiltonian operator is
\begin{equation}
  \hat{H} = \frac{1}{2} \big( \hat{\tilde{p}}_1^2+\hat{\tilde{p}}_2^2+\hat{\tilde{p}}_3^2\big) 
  +V_{\text{E}}(\hat{\tilde{q}}_1) + V_{\text{M};2}(\hat{\tilde{q}}_2) + V_{\text{M};3}(\hat{\tilde{q}}_3) + \epsilon \hat{H}_c\,.
\label{eq:HEckartMorseMorse}
\end{equation}
Here,
\begin{equation}
  V_{\text{E}}(q) = A \frac{\exp((q+q_0)/a)}{1+\exp ((q+q_0)/a)} + B
  \frac{\exp ((q+q_0)/a)}{[1+\exp ((q+q_0)/a)]^2} \, ,
\label{eq:eckartpotential}
\end{equation}
with positive real parameters $a$, $A$, and $B$, and
\begin{equation}
  q_0 = a \ln \frac{B+A}{B-A}\,,
\label{eq:equieckart}
\end{equation}
is the Eckart potential. For $B > A \ge 0$ it possesses a maximum at
$q=0$; the potential value at the maximum is $V_{\text{E}}(0) =
(A+B)^2/(4B)$. The Morse potentials, given by
\begin{equation}
  V_{\text{M};k}(q) = D_{e;k} \left[ \exp\left(-2 a_{\text{M};k} \, q\right) 
    -2\exp\left(-a_{\text{M};k} \, q\right) \right] \, ,
\label{eq:Morsepot}
\end{equation}
are parametrized by $D_{e;k}$ and $a_{\text{M};k}$, for $k=2,3$
respectively. For $\hat{H}_c$ we choose the mutual kinetic coupling
\begin{equation}
  \hat{H}_c = \hat{\tilde{p}}_1 \, \hat{\tilde{p}}_2\, 
  + \hat{\tilde{p}}_2 \, \hat{\tilde{p}}_3\, + \hat{\tilde{p}}_3 \, \hat{\tilde{p}}_1 \, .
\label{eq:kinetic_coupling}
\end{equation}
The strength of the coupling is controlled by the parameter $\epsilon$
in Eq.~\eqref{eq:HEckartMorseMorse}.  The vector field generated by
the corresponding classical Hamilton function has an equilibrium point
at
$(\tilde{q}_1,\tilde{q}_2,\tilde{q}_3,\tilde{p}_1,\tilde{p}_2,\tilde{p}_3)=0$. For
$|\epsilon|$ sufficiently small (for given values of parameters of the
Eckart and Morse potentials), the equilibrium point is of
saddle-centre-centre stability type.

\begin{figure}[htb!]
\centerline{\epsfig{figure=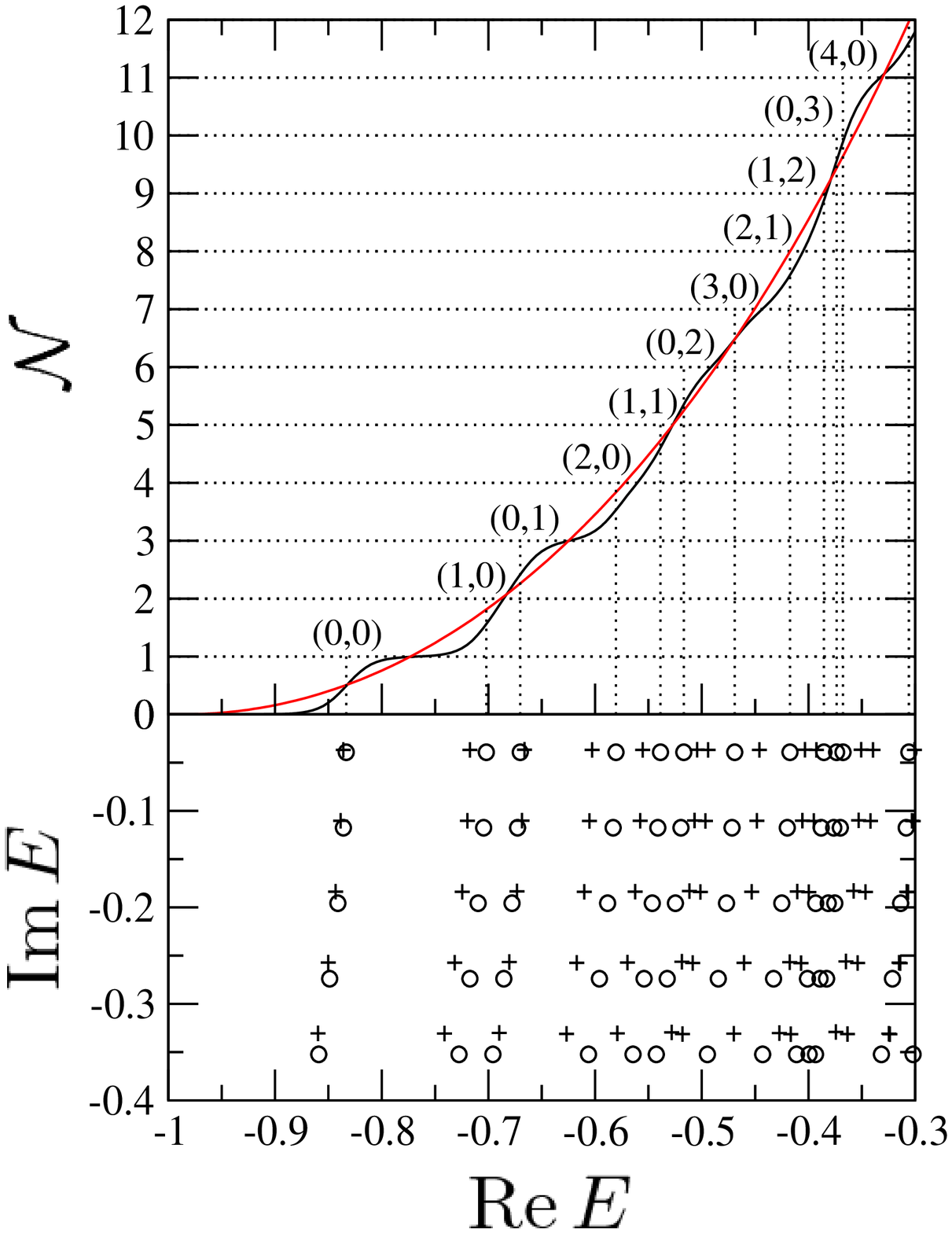,width=3.5in}}
\caption{The top panel shows the cumulative reaction probabilities
  $\crp_{\text{exact}}(E)$ (black oscillatory curve) and
  $\crp_{\text{Weyl}}(E)$ (red smooth curve) for the
  Eckart-Morse-Morse reactive system with the Hamiltonian given by
  Eq.~(\ref{eq:HEckartMorseMorse}) with $\epsilon = 0$. It also shows
  the quantum numbers $(n_2,n_3)$ of the Morse oscillators that
  contribute to the quantization steps. The bottom panel shows the
  resonances in the complex energy plane marked by circles for the
  uncoupled case $\epsilon =0$ and by crosses for the strongly coupled
  case $\epsilon = 0.3$.  The parameters for the Eckart potential are
  $a=1$, $A=0.5$, and $B=5$. The parameters for the Morse potential
  are $D_{e;2}=1$, $D_{e;3}=1.5$, and $a_{M;2}=a_{M;3}=1$. Also,
  $\heff=0.1$.}
\label{fig:eckartmorsemorsecoupled_cum}
\end{figure}

In the uncoupled case, $\epsilon = 0$, it is easy to calculate the
exact CRP, $\crp_{\mathrm{exact}}(E)$, of the Eckart-Morse-Morse
system analytically. Indeed, in accordance with
Eq.~(\ref{eq:cum_react}),
\begin{equation}
  \crp_{\mathrm{exact}}(E) = \sum_{n_2=0}^{\infty} \sum_{n_3=0}^{\infty}
  T_{\mathrm{exact}} \left(E - E_{\mathrm{M};2,n_2} - E_{\mathrm{M};3,n_3}\right) \, ,
\label{eq:exact_CRP_3DoF}
\end{equation}
where $T_{\mathrm{exact}}$ denotes the exact transmission coefficient
for the Eckart barrier, and is given by \cite{Eckart30}
\begin{equation}
  T_{\text{exact}}(E) = 1 - \frac{ \cosh ( 2\pi(\alpha-\beta) ) 
    + \cosh (2\pi \delta) }{\cosh ( 2\pi(\alpha+\beta) ) + \cosh (2\pi \delta)}
\label{eq:TEexact}
\end{equation}
with
\begin{equation}
  \alpha = \frac12 \sqrt{\frac{E}{C} } \,,\quad
  \beta = \frac12 \sqrt{\frac{E-A}{C} } \,,\quad
  \delta = \frac12 \sqrt{\frac{B-C}{C} } \,,\quad
  C= \frac{\heff^2}{8 a^2}\,.
\label{eq:TExact_parameters}
\end{equation}
\rem{Note that $T_{\text{exact}}(E)\rightarrow 0$ when the energy $E$
  approaches the limiting value $A$ of the potential from above.} In
Eq.~(\ref{eq:exact_CRP_3DoF}), $E_{\mathrm{M};k,n_k}$ (with $k=2,3$)
are the energy levels of the one-dimensional Morse oscillators,
\begin{equation}
  E_{\text{M};k,n_k} = -\frac{1}{2} a_{\text{M};k}^2 \, \heff^2 \left( n_k+\frac{1}{2} - 
    \frac{\sqrt{2 D_{e;k}}}{a_{\text{M};k} \, \heff} \right)^2 , \qquad n_k=0,1,2,\dots\,.
\label{eq:Morse_spectrum}
\end{equation}
The graph of $\crp_{\text{exact}}(E)$, obtained from
Eqs.~(\ref{eq:exact_CRP_3DoF}-\ref{eq:Morse_spectrum}), is shown by
the black oscillatory curve in the top panel of
Fig.~\ref{fig:eckartmorsemorsecoupled_cum}. Here, the parameters for
the Eckart potential are $a=1$, $A=0.5$, and $B=5$, the parameters for
the Morse potential are $D_{e;2}=1$, $D_{e;3}=1.5$, and
$a_{M;2}=a_{M;3}=1$, and the effective Planck's constant $\heff=0.1$.

\begin{figure}[htb!]
\centerline{\epsfig{figure=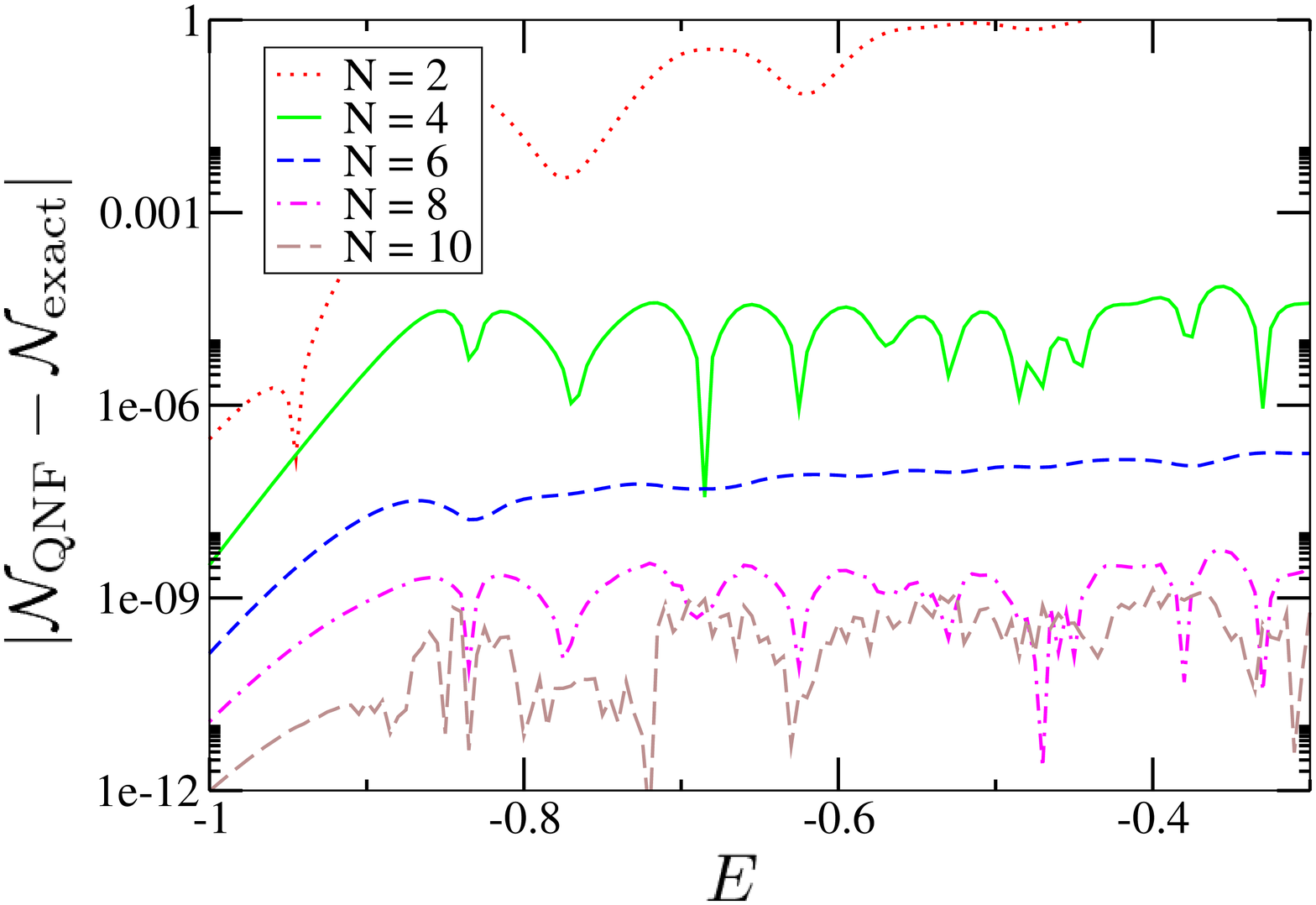,width=5in}}
\caption{Errors $|\crp_{\mathrm{QNF}} - \crp_{\mathrm{exact}}|$ for
  the cumulative reaction probability as a function of energy $E$ of
  the Eckart-Morse-Morse system for different order $N$ of the quantum
  normal form.}
\label{fig:CRP_error}
\end{figure}

The quantum normal form computation of the CRP for the
Eckart-Morse-Morse system gives
\begin{equation}
  \crp_{\text{QNF}}(E) = \sum_{n_2=0}^{\infty} \sum_{n_3=0}^{\infty} 
  \left[ 1 + \exp\left( -2\pi \frac{I_{n_2,n_3}(E)}{\heff} \right) \right]^{-1}
\label{eq:CRP_3D}
\end{equation}
with $I_{n_2,n_3}(E)$ solving
\begin{equation}
  K_{\mathrm{QNF}}^{(N)}\big( I, \heff (n_2+1/2), \heff (n_3+1/2) \big) = E \, .
\label{eq:I_n_3D}
\end{equation}
The high quality of the quantum normal form approximation of the
cumulative reaction probability is illustrated in
Fig.~\ref{fig:CRP_error}, which shows $|\crp_{\mathrm{QNF}} -
\crp_{\mathrm{exact}}|$ as a function of the energy $E$ for different
order $N$ of the quantum normal form. It is evident that the CRP
computed with the QNF method rapidly approaches its exact value as the
approximation order is increased.



\section{Gamov-Siegert Resonances}
\label{sec:resonances}

As we have discussed in Section~\ref{sec:CNF} in classical mechanics
the transition state is represented by a lower dimensional invariant
subsystem, the NHIM. In the quantum world, due to Heisenberg's
Uncertainty Principle, we cannot localise quantum states entirely on
the NHIM, so there can not be any invariant quantum subsystem
representing the transition states. Instead we expect a finite
lifetime for the transition state. The lifetime of the transition
state is determined by the Gamov-Siegert resonances, whose importance
in the theory of reaction rates has been emphasized in the literature
\cite{SeidemanMiller91, Zare06}.

Intuitively, a resonance is a complex eigenvalue of a Hamiltonian, 
\begin{equation}
\hat{H}|\psi_{res}\rangle=E|\psi_{res}\rangle
\end{equation}
where the imaginary part has to satisfy $\Im E \leq 0$.  The time
evolution of a resonance state is
\begin{equation*}
|\psi_{res}(t)\rangle=\ue^{-i E t/\heff}|\psi_{res}\rangle=\ue^{ -\Im E t/ \heff}\ue^{-i \Re Et /\heff}|\psi_{res}\rangle  
\end{equation*}
and so we see that a resonance state decays in time with lifetime given by 
\begin{equation*}
T=\frac{\heff}{\abs{\Im E}}\,\, .
\end{equation*}
Notice that since $\hat H$ is self-adjoint, a resonant state with complex energy cannot be square integrable. 

  In order to see how the resonances can be computed from the QNF Hamiltonian, consider first  the example of a one-dimensional system described by
the Hamilton operator
\begin{equation}\label{eq:basic-op}
 \hat{I}= \frac{1}{2} (\hat{q} \hat{p} + \hat{p} \hat{q}) \, .
\end{equation}
We easily see that $|n\ra$ defined by $\langle q| n\rangle =q^n$ is a resonance eigenstate
\begin{equation}\label{eq:I-res}
\hat{I} |n\ra=-i\heff (n+1/2)|n\rangle\,\, .
\end{equation}
with complex eigenvalue
\begin{equation}
E_n =-i \heff  (n+1/2)\,\, ,\quad  n=0,\ldots, N\,\, .
\end{equation}
These are the resonances of the operator $\hat I$. Notice that
compared to the scattering states $|I\rangle$ which depended on a
continuous parameter $I$, the resonance states are quantized and
depend on a discrete parameter $n\in\N_0$.  Using this simple example
we can now directly determine the resonances and the corresponding
resonance states for a Hamilton operator in quantum normal form
$\hat{H}_{\mathrm{QNF}}=K_{\mathrm{QNF}}(\hat{I} ,\hat{J}_2,\dots ,\hat{J}_{d})$.  Let $\psi_{n_k}$
denote the $n_k^{\text{th}}$ harmonic oscillator eigenfunction (see
\eqref{eq:J_eignfunc}).
For $n=(n_1,\ldots ,n_d)\in \N_0^d$, set 
\begin{equation}
|n_1,n_2,\cdots,n_d\rangle=|n_1\rangle\otimes| \psi_{n_2}\rangle\otimes \cdots \otimes|\psi_{n_{d}}\rangle 
\, ,
\label{eq:multiresstate}
\end{equation}
then we have 
\begin{align}
  &\hat{H}_{\mathrm{QNF}}|n_1,n_2,\cdots ,n_d\rangle \nonumber\\
  &=K_{\mathrm{QNF}}\big(-i\heff(n_1+1/2),\heff(n_{2}+1/2), \ldots
  ,\heff(n_d+1/2)\big) |n_1,n_2,\cdots ,n_d\rangle \,,
\label{eq:mult-dim-eigen}
\end{align}
and so the resonances of $\hat{H}_{\mathrm{QNF}}$ are given by 
\begin{equation}\label{eq:resonances-ndim}
E_{n_1,n_2,\ldots, n_d} =K_{\mathrm{QNF}}\big(-i \heff(n_1+1/2), \heff(n_{2}+1/2), \ldots ,\heff(n_d+1/2)\big)\,\, .
\end{equation}
The bottom panel of Fig.~\ref{fig:eckartmorsemorsecoupled_cum} shows
the resonances in the complex energy plane that have been computed
using Eq.~\eqref{eq:resonances-ndim} for the case of the coupled
Eckart-Morse-Morse reactive system discussed in
Section~\ref{sec:Eckart_Morse_Morse}.


The resonances computed from the QNF describe the lifetime of the activated complex. To see this in more detail consider a state localized at time $t=0$ on the NHIM, 
i.e.,  the dependence on the local normal form reaction coordinate $q_1$ is of the form 
\begin{equation}
\langle q_1|\varphi_0 \rangle=\frac{1}{(\pi\heff)^{1/4}}\ue^{-\frac{1}{\heff}\frac{q_1^2}{2}}\,\, .
\end{equation}
which is a minimal uncertainty state. In the bath coordinates we take the state 
to be given by  harmonic oscillator eigenfunctions $\psi_{n_k}$, so that 
\begin{equation}\label{eq:NHIM-state}
|\Psi\ra=|\varphi_0\ra\otimes|\psi_{n_2}\ra\otimes \cdots\otimes |\psi_{n_d}\ra
\end{equation}
for some fixed quantum numbers $n_2,\dots, n_d\in\N_0$. Then,
expanding $|\Psi\ra$ into the basis of Eq.~\eqref{eq:multiresstate} we
obtain the time evolved state
\begin{equation}
| \Psi(t) \ra = \frac{1}{(\pi\heff)^{1/4}} \sum_{n=0}^{\infty} \frac{1}{n!} \frac{(-1)^n}{(2\heff)^n}
\ue^{-i E_{2n,n_2,\ldots,n_d}t/\heff} | 2n, n_2, \ldots, n_d \ra \, .
\label{eq:NHIM-state_time_evolved}
\end{equation}

A suitable quantity for measuring the lifetime of a state like
\eqref{eq:NHIM-state} is the decay of the autocorrelation function
\begin{equation}
\abs{\la \Psi| \Psi(t)\ra}^2\,.
\end{equation}
\rem{
which is constant if the state $\psi$ is invariant and decays otherwise. 
}
For the Hamiltonian in quantum normal form the overlap reads
\rem{
We will compute the autocorrelation function for the case that the Hamiltonian is in quantum normal form. 
Inserting the expression \eqref{eq:NHIM-state} for $|\psi\ra$ and expanding the Gaussian into a Taylor series gives}
\begin{equation}
\la \Psi|\Psi(t) \ra 
= \bigg(\frac{2}{\pi}\bigg)^{1/2}\sum_{n=0}^{\infty} \frac{\Gamma(n+1/2)}{n!}(-1)^n
\ue^{-iE_{2n,n_2,\ldots,n_d}t / \heff} \, .
\end{equation}
The leading term in this sum for $t\to\infty$ is given by the smallest resonance with $n=0$. Hence, 
\begin{equation}
\abs{\la \Psi |\Psi(t) \ra}^2 \sim 2\ue^{2\Im E_{0,n_2,\ldots,n_d}t / \heff}\,,
\end{equation}
and this determines the maximal lifetime  
of a quantum state of the activated complex, i.e. a 
state initially localised near the classically  invariant subsystem given by the NHIM.

For small $\heff$ the quantum normal form is dominated by its quadratic part and that gives 
\begin{equation}
2 \Im E_{0,n_2,\ldots,n_d}/\heff \approx-\lambda \, ,
\end{equation}
and therefore for small $\heff$ 
\begin{equation}
\abs{\la \Psi | \Psi(t) \ra}^2 \sim 2\ue^{-t\lambda}\,\, .
\end{equation}
Thus, the quantum lifetime of the activated complex is in leading
order for small $\heff$ given by the reciprocal value of the classical
Lyapunov exponent associated with the saddle equilibrium point.



\section{Further Challenges}
\label{sec:challenges}

\subsection{Corner-Cutting Tunneling}
\label{sec:corner-cutting}

The approach based on normal forms is designed to give an accurate description 
of the dynamics near the saddle, or near the transition state. If the reaction 
is influenced by processes which avoid the neighborhood of the saddle 
then the normal form is unlikely to capture them. One such process is 
corner  cutting tunneling. We want to point out that the  results derived 
from the quantum normal form, like the computation of the $S$-matrix and 
cumulative reaction probability, contain tunneling contributions, but these 
are related to tunneling paths which are close to the saddle. One strength of our  
approach is that it describes uniformly the transition from energies 
below the saddle energy, where only tunneling is possible, to 
energies above the saddle energy.  

But there are situations where other tunneling contributions have a
large effect.  If the reaction path (e.g., the minimum energy path) in
configuration space is strongly curved in the regions where it
connects reactants and products, then the reactant valley and the
product valley can have a small angle relative to each other.  In this
situation there can be quite short tunneling paths which connect
reactants and products and which don't go near the saddle region.  So
these tunneling paths ``cut the corner''. The contribution from such
trajectories play an important role, in particular for hydrogen
exchange reactions at room temperature and lower
\cite{Miller_science86}.

In these situations the normal form approach has to be combined with 
multidimensional semiclassical tunneling calculations.

\subsection{State-to-State Reaction Rates}

It is important to emphasize again that, so far, our approach to
quantum reaction dynamics has been local, i.e., it is derived
completely from the properties of the quantum normal form that is
valid in the neighborhood of the saddle-centre-$\cdots$-centre equilibrium point. 
The property of the resulting S-matrix in \eqref{eq:Smatrixfull} being block-diagonal reflects the fact that the  quantum normal form is integrable
in the sense that the
basis of scattering states can be chosen in the product form \eqref{eq:mult_scat_states}. In a different basis the
matrix will lose this feature, and phenomena like mode mixing are related to how other incoming
and outgoing scattering states are related to this special basis. It is natural to embed  
the study of this phenomenon in a study of the global dynamics which we will describe in this section.
The global formalism is in particular required in order to compute general state-to-state
reaction rates.

Let us start by
describing the scattering or reaction process in classical
mechanics by using Poincar{\'e} sections. Recall that a
Poincar{\'e} section at energy $E$ is given by a smooth
hypersurface $\Sigma(E)$ of the energy surface with energy $E$
which is transversal to the flow ($\Sigma(E)$ is allowed to have
several components). If we have two such Poincar{\'e} sections
$\Sigma_1(E)$ and $\Sigma_2(E)$ such that all the flow lines
intersecting $\Sigma_1(E)$ intersect at a later time
$\Sigma_2(E)$, too, then moving along the flow from $\Sigma_1(E)$
to $\Sigma_2(E)$ defines a Poincar{\'e} map 
\begin{equation}
P^{(2,1)}(E):\Sigma_1(E)\to \Sigma_2(E)\,\, .
\end{equation}
Such Poincar{\'e} maps can be composed. If $\Sigma_3(E)$ is another Poincar{\'e} section which lies behind
$\Sigma_2(E)$ in the sense that the flow lines that intersect $\Sigma_2(E)$ also intersect $\Sigma_3(E)$ at a later time,
and if $P^{(3,2)}(E):\Sigma_2(E)\to \Sigma_3(E)$ is the corresponding Poincar{\'e} map, then the Poincar{\'e} map
\begin{equation}
P^{(3,1)}(E):\Sigma_1(E)\to \Sigma_3(E)\,\,
\end{equation}
is given by
\begin{equation}
P^{(3,1)}(E)=P^{(3,2)}(E) \circ P^{(2,1)}(E)\,\, .
\end{equation}

Using this construction we can describe transport through phase space regions by a sequence of maps.
Given some Poincar{\'e} section  $\Sigma_{\text{initial}}(E)$ located in the area of initial points
in the reactants region
where we prepare the system and a Poincar{\'e} section $\Sigma_{\text{final}}(E)$ in the products region
where we measure the outcome, a succession
of Poincar{\'e} maps
\begin{equation}\label{eq:Poincare_sect_succession}
\Sigma_{\text{initial}}(E)\to \Sigma_{1}(E)\to \Sigma_{2}(E)\to \cdots \to \Sigma_{\text{final}}(E)
\end{equation}
tells us how the initial points are transported through the
system.\footnote{ We here ignore the difficulties involved in
constructing global Poincar{\'e} sections (see, e.g.,
\cite{DullinWittek95}); we assume that the sequence of
Poincar{\'e} sections \eqref{eq:Poincare_sect_succession} is
intersected transversally by the trajectories with initial points
from a suitable open subset in the reactants region.}

The advantage of subdividing the flow into a sequence of maps lies
in the fact that different regions in phase space might need
different techniques to compute the flow. In our case of interest
Poincar{\'e} sections can be constructed to the products and
reactants side of a saddle-centre-$\cdots$-centre equilibrium point. The dynamics `across'
this equilibrium point can then be described by the normal form
while the dynamics between neighbourhoods of different saddle
points can be obtained from integrating the original equations of
motions \cite{Creagh04,Creagh05,WaalkensBurbanksWiggins05b}.
Moreover, the phase space structures obtained from the local
normal form can be ``globalized'' following the discussion in
Section \ref{sec:global}.

A similar procedure can be developed in the quantum case. The Poincar{\'e} maps
\begin{equation}
P^{(j,i)}(E):\Sigma_i(E)\to \Sigma_j(E)\,\,
\end{equation}
are symplectic maps, and as such can be quantised using the theory of Fourier integral operators.
The quantisations will be unitary operators which we interpret as local S-matrices,
\begin{equation}
S^{(j,i)}(E):L^2_{\Sigma_i(E)}\to L^2_{\Sigma_j(E)}\,\, ,
\end{equation}
where $L^2_{\Sigma(E)}$ is a Hilbert space obtained by geometric quantisation of $\Sigma(E)$, see, e.g., \cite{Kir01}.
This is similar to the quantisation developed in \cite{Bogomolny92}.
As in classical dynamics we can  compose these matrices to obtain a global S-matrix
\begin{equation}
S^{(\text{final}, \text{initial})}(E)=S^{(\text{final},n)}(E)S^{(n,n-1)}(E)\cdots S^{(1, \text{initial})}(E)
\end{equation}
which tells us how initial states in
$L^2_{\Sigma_{\text{initial}}(E)}$ are transformed into final
states in $L^2_{\Sigma_{\text{final}}(E)}$. The reasons for
introducing this splitting of the S-matrix are the same as in the
classical case. We can employ different techniques for computing
the S-matrices according to different  local properties of the system.
Near equilibrium  points the dynamics can be described by the quantum
normal form we developed in this paper. 
Notice that the neighbourhoods of the saddle-centre-$\cdots$-centre
equilibriuml points are the regions where we expect quantum effects to
be of most importance due to partial reflection at and tunnelling
through the barriers associated with saddle points. 
The quantum transport between neighbourhoods of
different equilibrium points can be described by a standard van Vleck
type formalisms, using, e.g, {\em initial value representations} (IVRs) 
which are very common in theoretical chemistry
(see, e.g., \cite{Miller98,Miller1} for references).

\subsection{Flux-Flux Autocorrelation Function Formalism}
\label{sec:flux-flux}


In this section we discuss the flux-flux autocorrelation function
approach to computation of reaction rates in both classical and
quantum theories.

\paragraph{Classical formulation:} The directional flux through the
dividing surface that determines the classical reaction rate, see
Sec.~\ref{sec:classicalrate}, can be written as \cite{Yamamoto60,
  MillerSchwartzTromp83, Miller98}
\begin{equation}
  \flux(E) = \int_{\R^{2d}} \delta(E-H(z)) F(z) P_{\mathrm{r}}(z) \, \ud z \, .
\label{eq:dir_flux}
\end{equation}
The dynamical meaning of Eq.~(\ref{eq:dir_flux}) is as follows. First,
$z = (q_1,p_1,\ldots,q_d,p_d)$ is a point in the system's phase space,
and the function $\delta(E-H(z))$ restricts the integration to the
energy surface of energy $E$ under consideration. The flux factor $F$
is given by
\begin{equation}
  F(z) = \frac{\ud}{\ud t} \Theta \big( s(z_t) \big) \bigg|_{t=0} \, ,
\label{eq:flux_factor}
\end{equation}
where $z_t$ is the Hamiltonian flow satisfying the initial condition
$z_{t=0} = z$. The dividing surface is defined as the zero level set
of the function $s$, i.e., $\{ z\in\R^{2d} \, : \, s(z)=0 \}$. It is
assumed that this surface divides the phase space into two components:
the region given by $s(z)<0$ corresponds to reactants, while the
region given by $s(z)>0$ to products. Finally, the projection function
$P_{\mathrm{r}}$ is defined as
\begin{equation}
  P_{\mathrm{r}}(z) = \lim_{t \rightarrow \infty} \Theta \big( s(z_t) \big) \, ,
\label{eq:projection}
\end{equation}
and equals unity (zero) if a trajectory starting at $z$ ends up in the
product (reactant) region for infinitely long times.

In order to explain the relation of Eq.~\eqref{eq:dir_flux} to
Eq.~\eqref{eq:fluxactions_classical} we first rewrite the projection
function as
\begin{equation}
  P_{\mathrm{r}}(z) = \Theta \big( s(z_{t_0}) \big) + 
  \int_{t_0}^{\infty} \frac{\ud}{\ud t} \Theta \big( s(z_t) \big) \, \ud t \, ,
\label{eq:projection_integral}
\end{equation}
where we choose $t_0 = -\epsilon$ with $\epsilon \rightarrow
0^{+}$. At this stage it is important to note that since the flux
factor $F(z)$ is proportional to $\delta\big( s(z) \big)$ the integral
in Eq.~\eqref{eq:dir_flux} is effectively restricted to the dividing
surface\footnote{More precisely, since in addition to the flux factor
  $F$ one also has to take into account the function $\delta(E-H)$ the
  integral in Eq.~\eqref{eq:dir_flux} is a $(2d-2)$-dimensional
  integral over the intersection of the dividing surface with the
  energy surface of energy $E$.}. But, using such properties of the
dividing surface that it is (i) recrossing-free and (ii) transverse to
the Hamiltonian flow, we have
\begin{equation}
  \lim_{\epsilon \rightarrow 0^{+}} \Theta \big( s(z_{-\epsilon}) \big) 
  = 1 - P_{\mathrm{r}}(z)
\label{eq:projection_identity}
\end{equation}
for $z$ on the dividing surface. Then, substituting
Eqs.~\eqref{eq:flux_factor} and \eqref{eq:projection_identity} into
Eq.~\eqref{eq:projection_integral}, and solving for the projection
function $P_{\mathrm{r}}$ we obtain
\begin{equation}
  P_{\mathrm{r}}(z) = \frac{1}{2} + \lim_{\epsilon \rightarrow 0^{+}} \frac{1}{2}
  \int_{-\epsilon}^{\infty} F(z_t) \, \ud t
\label{eq:projection_func}
\end{equation}
for $z$ on the dividing surface. Finally, substituting
Eq.~\eqref{eq:projection_func} into Eq.~\eqref{eq:dir_flux}, and
taking into account the fact that the total flux through the dividing
surface is zero, i.e.,
\begin{equation}
  \int_{\R^{2d}} \delta(E-H(z)) F(z) \, \ud z = 0 \, ,
\end{equation}
we arrive at the expression for the directional flux,
\begin{equation}
  \flux(E) = \lim_{\epsilon \rightarrow 0^{+}}  \frac{1}{2} \int_{-\epsilon}^{\infty} C_F(t) \, \ud t \,,
\label{eq:dir_flux_autocor}
\end{equation}
as a time integral of the {\it flux-flux autocorrelation function}
\begin{equation}
  C_F(t) = \int_{\R^{2d}} \delta(E-H(z)) F(z) F(z_t) \, \ud z \, .
\label{eq:flux-flux_autocor}
\end{equation}

We now explicitly calculate the time dependence of the function $C_F$
using the classical normal form theory. To this end we first express
the flux factor $F$ in the normal form coordinates $(Q_1, P_1, q_2,
p_2, \ldots, q_d, p_d)$, see Sec.~\ref{sec:normal_form}, in which the
dividing surface is given by $s(z) = Q_1$:
\begin{equation}
  F(z) = \frac{\ud}{\ud t} \Theta\big( Q_1(t) \big) \bigg|_{t=0}
  = \delta(Q_1) \frac{\ud}{\ud t} Q_1(t) \bigg|_{t=0} = \delta(Q_1) \Lambda P_1 \, ,
\label{eq:flux_factor_NF}
\end{equation}
where
\begin{equation}
  \Lambda = \frac{\partial}{\partial I} \Hcnf(I,J_2,\ldots,J_d)
\label{eq:Lambda}
\end{equation}
and $\Hcnf$ is the normal form Hamiltonian as a function of the action
integrals $I = (P_1^2-Q_1^2)/2$ and $J_k = (p_k^2+q_k^2)/2$,
$k=2,\ldots,d$. Equation~\eqref{eq:flux-flux_autocor} then reads
\begin{equation}
  C_F(t) = \int_{\R^{2d}} \delta(E-\Hcnf) \delta(Q_1) 
  \delta\big( Q_1(t) \big) \Lambda^2 P_1 P_1(t) \, \ud z \,.
\label{eq:flux-flux_autocor_CNF}
\end{equation}
The product of the $\delta$-functions of $Q_1$ and the corresponding
time evolved coordinate $Q_1(t)$ signals that only the infinitesimally
short time scales $t \rightarrow 0$ give a non-vanishing contribution
to the integral. Linearization the time evolution of $z_t$ around
$t=0$ yields
\begin{align}
  C_F(t) &= \delta(t) \int_{\R^{2d}} \delta(E-\Hcnf) \delta(Q_1) \Lambda \abs{P_1} \, \ud z \nonumber\\
  &= 2 \delta(t) (2\pi)^{d-1} \int_{\R_+^d} \delta(E-H_{\mathrm{CNF}})
  \frac{\partial \Hcnf}{\partial I} \, \ud I \ud J_2 \ldots
  \ud J_d \nonumber\\
  &= 2\delta(t) (2\pi)^{d-1} \int_{I(E,J_2,\ldots,J_d)>0} \ud J_2
  \ldots \ud J_d \,.
\label{eq:flux-flux_autocor_CNF-2}
\end{align}
The last integral in Eq.~\eqref{eq:flux-flux_autocor_CNF-2} is nothing
but the volume $\mathcal{V}(E)$ in the action space $(J_2,\dots,J_d)$
enclosed by the contour $\Hcnf(0,J_2,\dots,J_d)=E$, and illustrated in
Fig.~\ref{fig:flux}, that defines the directional flux through the
dividing surface, see Eq.~\eqref{eq:fluxactions_classical}. Thus, we
obtain
\begin{equation}
  C_F(t) = 2\delta(t) (2\pi)^{d-1} \mathcal{V}(E) = 2\delta(t) \flux(E) \,,
\label{eq:flux-flux_autocor-final}
\end{equation}
which is in agreement with Eq.~\eqref{eq:dir_flux_autocor}. Finally,
we note that in view of Eq.~\eqref{eq:flux-flux_autocor-final} one can
rewrite Eq.~\eqref{eq:dir_flux_autocor} as
\begin{equation}
  \flux(E) = \frac{1}{2} \int_{-\infty}^{+\infty} C_F(t) \, \ud t \,.
\label{eq:dir_flux_autocor_new}
\end{equation}

\paragraph{Quantum formulation:} In the quantum version of the
flux-flux autocorrelation function approach Eq.~\eqref{eq:dir_flux}
for the directional flux is replaced by a corresponding equation for
the cumulative reaction probability,
\begin{equation}
  \crp(E) = 2\pi\heff \, \mathrm{Tr} \left\{ \delta(E-\hat{H}) \hat{F} \hat{P}_{\mathrm{r}} \right\} \, .
\label{eq:flux-flux_CRP}
\end{equation}
Here, the flux factor operator $\hat{F}$ is given by
\begin{equation}
  \hat{F} = \frac{\ud}{\ud t} \left( \ue^{i\hat{H}t/\heff} \widehat{\Theta(s)} 
    \ue^{-i\hat{H}t/\heff} \right) \bigg|_{t=0} = \frac{i}{\heff} [\hat{H}, \widehat{\Theta(s)}] \, ,
\label{eq:flux_op_quant}
\end{equation}
where $\widehat{\Theta(s)}$ is a quantization of the composition of
the Heaviside function with a function $s$ defining the dividing
surface. The projection operator $\hat{P}_{\mathrm{r}}$ is defined as
\begin{equation}
  \hat{P}_{\mathrm{r}} = \lim_{t \rightarrow \infty} \ue^{i\hat{H}t/\heff} \widehat{\Theta(s)} 
    \ue^{-i\hat{H}t/\heff} \,.
\label{eq:project_quant}
\end{equation}
Then, as shown in Ref.~\cite{Miller98}, Eq.~\eqref{eq:flux-flux_CRP}
can be written in the close analogy with its classical counterpart
Eq.~\eqref{eq:dir_flux_autocor}:
\begin{equation}
  \crp(E) = (2\pi\heff) \, \frac{1}{2} \int_{-\infty}^{+\infty} \mathcal{C}_F(t) \, \ud t \,,
\label{eq:crp_vs_flux-flux}
\end{equation}
where
\begin{equation}
  \mathcal{C}_F(t) = \mathrm{Tr} \left\{ \delta(E-\hat{H}) \hat{F} \ue^{i\hat{H}t/\heff}
    \hat{F} \ue^{-i\hat{H}t/\heff} \right\}
\label{eq:flux-flux_autocor_quant}
\end{equation}
is the quantum flux-flux autocorrelation function.

\bigskip

As we have shown above the classical flux-flux autocorrelation
function $C_F$ can be explicitly analyzed within the framework of the
classical normal form theory. The following natural question arises:
can one obtain the time dependence of the quantum flux-flux
autocorrelation function $\mathcal{C}_F$ using the methods of the
quantum normal form theory? Of course, the QNF technique provides one
with an approximation of the original Hamiltonian operator. This
approximation is only accurate in the vicinity of the equilibrium
saddle point in phase space, so one should not expect a perfect
agreement between the QNF flux-flux autocorrelation function and the
exact one to hold up to infinitely long times. Instead, the QNF theory
will provide an approximation of the exact flux-flux autocorrelation
function in a certain time interval whose length will depend on the
effective Planck's constant among other parameters.


\subsection{Convergence of Quantum Normal Form}
\label{sec_convergence}

Both the classical and quantum normal forms have the form of  power series where each successive term is constructed via an iterative technique. Consequently, in general, we will only be able to compute a finite number of terms of the normal form.  Therefore the obvious question that arises is ÔÕhow many terms of the normal form are required in order that the quantities derived from the normal form are accurateÕÕ?   A discussion, as well as some references with specific examples, was given in Section \ref{sec:trunc_error} for the classical normal form.  In this section we are concerned with the behavior of the quantum normal form, for which there has been essentially no work from this point of view.

First, we begin with a brief background discussion. Since the normal forms, both classical and quantum, are given as series representations of the Hamiltonian function and Hamilton operator, respectively, a natural question to ask is ÔÕdo these series convergeÕÕ? In the classical setting there has been previous work on this question. However, practically speaking, from the point of view of using the results of the theory, there are three related  questions: 1)convergence of the series representing the normal form Hamiltonian, 2) convergence of the transformation from the original coordinate to the normal form coordinates, and 3) convergence of any integrals of the motion that arise from the normalization procedure.  A recent discussion of these issues, as well as a discussion of earlier results for the classical Hamiltonian normal form setting can be found in \cite{pm03}.   Briefly, the situation with respect to convergence for Hamiltonian systems with three or more degrees-of-freedom is not optimistic. Generically,  divergence is the expected behavior in normal form theory for classical Hamiltonian systems. However, the situation is not so pessimistic as one my initially believe.  There are many examples of  divergent series which still yield useful information in an asymptotic sense. Resummation techniques and Pade approximation techniques can be used to find an optimal number of terms that yield a desired accuracy. For classical Hamiltonian normal forms these issues have been  examined in \cite{kr92, robnik93,ceg03,egc04}.  These issues have yet to be explored in the quantum normal form setting.

It is useful to note that the case of two degree-of-freedom Hamiltonian  systems is special.  In this case a classical result of \cite{moser} (see also \cite{giorgilli})  gives convergence results for the classical Hamiltonian normal form in the neighborhood of a saddle-center equilibrium point.  Recently, the first results on convergence of the quantum normal form have appeared. In \cite{anikin} convergence results for a one and a half degree-of-freedom system (i.e. time-periodically forced one degree-of-freedom Hamiltonian system) have been given. It is not unreasonable that these results can be extended to the quantum normal form in the neighborhood of a saddle-center equilibrium point of a two degree-of-freedom system.

In the following we provide a qualitative discussion of the convergence of the QNF based on our calculations performed for the triatomic collinear reactions. In this situation the QNF approximates the Hamiltonian of the reacting system in a
phase-space neighborhood of the saddle-center equilibrium point. Thus, for
instance, in computing the CRP one only expects this approximation to
render reliable results in a certain energy range around the saddle
point energy $E_0$ of the potential energy surface under consideration. The energy difference $(E-E_0)$ may therefore be considered as one small
parameter in the QNF expansion. The role of the other small parameter
is played by the effective Planck's constant, $\heff$. It is the
convergence of the QNF with respect to this second small parameter
that we focus on in this section.

We proceed by considering the right hand side of Eq.~(\ref{eq:qnf_H}),
i.e., the QNF, at $I=0$, corresponding to no ``energy'' in the reaction
coordinate, and $n_2=0$, giving the zero-point `vibrational energy' of
the transverse degree of freedom. Then, Eq.~(\ref{eq:qnf_H}) becomes

\begin{equation}
  E = E_0 + \sum_{n=1}^{\lfloor N/2 \rfloor} c_n \heff^n \, .
\label{eq:hbar_expansion}
\end{equation}

\noindent
For the case of the collinear hydrogen exchange reaction, $\mathrm{H}
+ \mathrm{H}_2 \rightarrow \mathrm{H}_2 + \mathrm{H}$, on the
Porter-Karplus potential energy surface the first five expansion
coefficients were obtained in Ref.~\cite{gsww09}: $c_1 = 0.161982$,
$c_2 = 1.193254$, $c_3 = 14.90023$, $c_4 = 378.7950$, and $c_5 =
1227.035$. As $N \rightarrow \infty$ the radius of convergence
$\heff^{(0)}$ of the sum in Eq.~(\ref{eq:hbar_expansion}) is given by

\begin{equation}
  \heff^{(0)} = \lim_{n\to \infty} \frac{ c_n}{c_{n+1} } \, .
\label{4-02}
\end{equation}

\noindent
Here, we make a crude estimate of $\heff^{(0)}$ by only considering
the first five expansion coefficients in Eq.~(\ref{4-02}), i.e., $c_n$
with $n=1,\ldots,5$; then, the radius of convergence is given by
$\heff^{(0)} \sim 0.04$.

\begin{figure}[htb!]
\centerline{\epsfig{figure=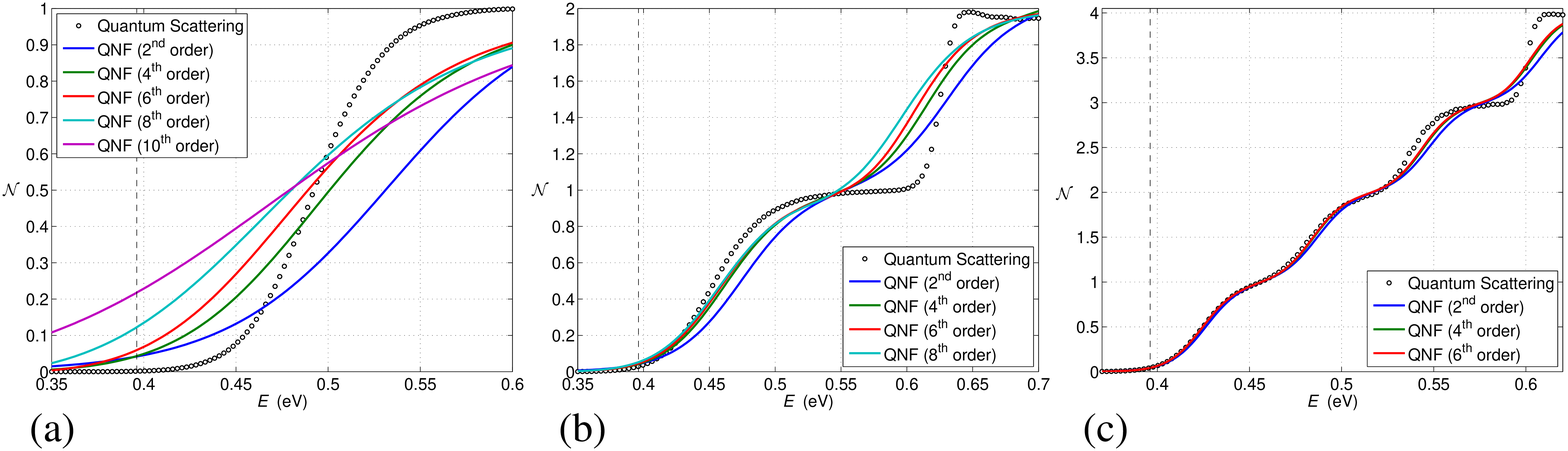,width=6in}}
\caption{Cumulative reaction probability as a function of the total
  energy, $\crp(E)$, for the collinear hydrogen exchange
  reaction. Figure (a) corresponds to $^1\mathrm{H}$ isotope
  characterized by $\heff = 3.07 \times 10^{-2}$, (b) corresponds to
  $^3\mathrm{H}$ isotope characterized by $\heff = 1.77 \times
  10^{-2}$, and (c) corresponds to a hypothetical $^{20}\mathrm{H}$
  ``isotope'' characterized by $\heff = 6.9 \times 10^{-3}$. The
  vertical dashed line shows the saddle point energy, $E_0$.}
\label{fig:convergence}
\end{figure}

The estimated value of $\heff^{(0)}$ sheds light on the seeming
inefficiency of the QNF theory for CRP computations in light atom
reactions. Indeed, the $^1$H exchange reaction, see
Fig.~\ref{fig:convergence}a, is characterized by $\heff = 3.07 \times
10^{-2}$. This value being close to $\heff^{(0)}$ signals that the
corresponding QNF expansion converges very slowly, if at all, and,
possibly, terms of orders far beyond $N=10$ are needed for a reliable
CRP prediction in Fig.~\ref{fig:convergence}a.

In the case of the $^3$H exchange reaction the effective Planck's
constant is $\heff = 1.77 \times 10^{-2}$ and is thus smaller than
$\heff^{(0)}$. This fact is in agreement with the apparent speed-up of
the convergence of the CRP values, see Fig.~\ref{fig:convergence}b, in
comparison with the $^1$H case. Finally, the convergence is very fast
and pronounced for the case of the heavy (hypothetical) $^{20}$H
atoms, see Fig.~\ref{fig:convergence}c, for which $\heff = 6.9 \times
10^{-3}$ which is much smaller that the estimated convergence radius.

Clearly, there is a great deal of scope for further studies of convergence aspects  of the quantum normal form as well as related optimal truncation and resummation techniques.



\section{Conclusions}
\label{sec:conclusions}

We have reviewed a new dynamical systems, phase space approach to quantum reactive scattering. The quantum theory arises from an underlying  classical theory that reveals the geometrical structures in phase space  that govern trajectories evolution from reactants to products in the reaction region.   The classical theory is valid for any number of degrees-of-freedom, and the new concept which has led to this is the introduction of the notion of a {\em normally hyperbolic invariant manifold}, or NHIM, into classical reaction dynamics. The NHIM is the key phase space structure that leads to the construction of a surface dividing reactants from products having the no (local) re-crossing and minimal flux properties. Moreover, the stable and unstable manifolds of the NHIMs have one less dimension than the energy surface (i.e. they are ``co-dimension one'') and they bound regions of the energy surface (being invariant manifolds, or ``impenetrable barriers'') which contain {\em all} reacting trajectories.  These phase space structures area realized in specific systems through the use of the  classical Poincar\'e-Birkhoff normal form. Quantization is carried out through an analogous quantum normal form theory and the Weyl quantization procedure. The related structure of the classical and quantum normal form theories makes the quantum manifestation of the classical phase space structures transparent.   The phase space structures in the classical case  define the ``landscape'' in the {\em energy surface} which constrain the location and evolution of reacting trajectories, and this renders the need for the  calculation of trajectories to evaluate, e.g. classical flux across the dividing surface, unnecessary. This characteristic is inherited in the quantum setting in the sense that quantum expressions governing reaction  are expressed in terms of quantities that can be  computed  from the quantum normal form without the need to compute classical trajectories. Thus this  phase space approach to quantum reactive scattering provides a completely new approach to the computation of quantities describing quantum reactive scattering which may prove to be fruitful in the study of ``large'' quantum systems.



\section*{Appendix}

\appendix

\section{The  Normal Form Algorithm: Classical and Quantum}
\label{sec:QNF_algorithm}

In Sec.~\ref{sec:CNF} we mentioned that the crucial point to realize the phase space structures governing reaction dynamics is  to choose a suitable set of phase space coordinates in terms of which the Hamilton function assumes a very simple form. Similarly, the description of the activated complex  in Sec.~\ref{sec:activated_complex} leading to the explicit expressions for the cumulative reaction probabilities and the Gamov-Siegert resonances as explained in Sections~\ref{sec:smatrix} and \ref{sec:resonances}, respectively, relied on expressing the Hamilton operator in a suitable basis. Both the canonical (or equivalently \emph{symplectic}) transformation of the phase space coordinates and the corresponding transformation of the Hamilton function in the classical case, and the unitary transformation of the basis set together with the corresponding transformation of the Hamilton operator  in the quantum case can be constructed in an algorithmic fashion. The algorithms are based on a (classical) \emph{normal form} (CNF) in the classical case and on an analogous \emph{quantum normal form} (QNF) in the quantum mechanical case. In the following we give a brief description of the algorithms to compute these normal forms. We describe the algorithms in such a way that the similarities between both algorithms become apparent.  For rigorous mathematical statements, proofs and further
details  we refer to Ref.~\cite{SchubertWaalkensWiggins06,WaalkensSchubertWiggins08}. 

\subsection{The classical normal form algorithm}
\label{sec:cnf_algorithm}

We begin by considering a Hamiltonian system with phase space $\R^{d}\times \R^d$ with phase space coordinates
$\tilde{\vecq} = (\tilde{q}_1, \tilde{q}_2, \ldots,\tilde{q}_d)$ and $\tilde{\vecp} = (\tilde{p}_1,\tilde{p}_2, \ldots, \tilde{p}_d)$ which we group in the $2d$-dimensional vector $\tilde{\vecz}=(\tilde{\vecq},\tilde{\vecp})$, and a Hamilton function $H(\tilde{\vecq},\tilde{\vecp})$. 
For convenience, we can use atomic units. This imposes no restriction.  However, it has the advantage that  the phase space coordinates become dimensionless. 
We will now assume that Hamilton's equations associated with $H(\vecq,\vecp)$ 
have a (single) equilibrium point, $\tilde{\vecz}_0 \equiv
(\tilde{\vecq}_0, \tilde{\vecp}_0)$, of saddle-center-$\ldots$-center stability
type. By this we mean that the matrix associated with the
linearization of Hamilton's equations about this equilibrium point has
two real eigenvalues, $\pm \lambda$, of equal magnitude and opposite
sign, and $d-1$ purely imaginary complex conjugate pairs of
eigenvalues $\pm \ui \omega_k$, $k=2,\ldots,d$.
If the  Hamilton function is of the form kinetic energy plus
potential energy then this type of equilibrium point of Hamilton's
equations correspond to an index one saddle point of the potential
energy.

The CNF procedure consists of a sequence of canonical (or equivalently  \emph{symplectic}) transformations changing
the phase space coordinates $\tilde{\vecz} = (\tilde{\vecq}, \tilde{\vecp})$ 
in a  phase space neighborhood of the equilibrium point
$\tilde{\vecz}_0$.
This is accomplished order by order according to the sequence
\begin{equation}
\tilde{\vecz}\equiv \vecz^{(0)} \rightarrow \vecz^{(1)} \rightarrow \vecz^{(2)} \rightarrow \vecz^{(3)} 
\rightarrow \ldots \rightarrow \vecz^{(N)}  \,,
\label{eq:seq_trafos_z}
\end{equation}
where  $\vecz^{(n)} $ is obtained from $\vecz^{(n-1)} $ 
by means of a symplectic transformation 
\begin{equation}
\vecz^{(n-1)} \mapsto \vecz^{(n)} =  \phi_{W_n} \vecz^{(n-1)} \,.
\label{eq:zn-1twozn}
\end{equation}
generated by a homogenous polynomial $W_n(\vecz)$ of order $n$, i.e.
\begin{eqnarray}
W_n \in  \mathcal{W}_{\text{cl}}^n := \mathrm{span} \left\{ q_1^{\alpha_1} \ldots q_d^{\alpha_d}
    p_1^{\beta_1} \ldots p_d^{\beta_d}  :   |\alpha|+|\beta|= n \right\} \, .
\label{2-09_cl}
\end{eqnarray}
More precisely, the  $\phi_{W_n}$ in \eqref{eq:zn-1twozn} denote the time-one maps of the flows generated by the Hamiltonian vector fields corresponding to the polynomials $W_n$ (see \cite{WaalkensSchubertWiggins08} for the details). 
The maximum order $N$ in \eqref{eq:seq_trafos_z} is the desired order of accuracy at which the expansion will be terminated and truncated. 


Expressing the Hamilton function $H$ in the coordinates  $\vecz^{(n)}$, $n=1,\ldots,N$, 
we get a sequence of Hamilton functions $H^{(n)}$,
\begin{equation}
H\equiv H^{(0)} \rightarrow H^{(1)} \rightarrow H^{(2)} \rightarrow H^{(3)}
\rightarrow \ldots \rightarrow H^{(N)}\,, 
\label{eq:seq_trafos_cl}
\end{equation}
where for $n=1,\ldots,N$, and
$
H^{(n)} (\vecz^{(n)}) = H^{(n-1)} (\vecz^{(n-1)}  ) = H^{(n-1)} ( \phi_{W_n}^{-1} \vecz^{(n)})     
$, i.e. 
\begin{equation}
H^{(n)}  = H^{(n-1)} \circ  \phi_{W_n}^{-1}\,.
\end{equation}
To avoid a proliferation of notation we will in the following neglect the superscripts $(n)$ for the phase space coordinates.

In the first transformation in \eqref{eq:seq_trafos_z} we
shift the equilibrium point
$\tilde{\vecz}_0$ to the origin, i.e. $\vecz \mapsto  \phi_{W_1} (\vecz) := \vecz - \tilde{\vecz}_0$. This gives  
\begin{equation}
  H^{(1)}(\vecz) = H^{(0)} (\vecz+\tilde{\vecz}_0) \, .
\label{2-02_cl}
\end{equation}
The next steps of the CNF algorithm rely on  the power series
expansions of $H^{(n)}$, 
\begin{equation}
  H^{(n)}(\vecz) = E_0 + \sum_{s=2}^\infty H_s^{(n)}(\vecz) \, ,
\label{2-03_cl}
\end{equation}
where the $H_s^{(n)}$ are homogenous polynomials in $ \mathcal{W}_{\text{cl}}^s $ given by
\begin{equation}
  H_s^{(n)}(\vecz)  =  \displaystyle \sum_{|\alpha|+|\beta|=s}
    \frac{H_{\alpha_1,\ldots,\alpha_d,\beta_1,\ldots,\beta_d}^{(n)}}{\alpha_{1}!
      \ldots \alpha_{d}! \beta_{1}! \ldots \beta_{d}! }   
    \, q_1^{\alpha_1} \ldots q_d^{\alpha_d} p_1^{\beta_1} \ldots
    p_d^{\beta_d}  
 \label{2-04_cl}
\end{equation}
with $\alpha_k,\beta_k\in \mathbb{N}_0$, $|\alpha| = \sum_k
\alpha_k$, $|\beta| = \sum_k \beta_k$.
For $n=1$, the coefficients in \eqref{2-04_cl} are given by the Taylor expansion of $H^{(1)}$ about the origin,
\begin{equation}
   H_{\alpha_1,\ldots,\alpha_d,\beta_1,\ldots,\beta_d}^{(1)} 
  = \displaystyle \left.
  \prod_{k,l=1}^d \frac{\partial^{\alpha_k}}{\partial
    q_k^{\alpha_k}} \frac{\partial^{\beta_l}}{\partial
    p_l^{\beta_l}} 
  H^{(1)}(\vecz) \right|_{\vecz={\bf 0}} \!\!\! .
  \label{2-04.1_cl}
\end{equation}
For $n\ge 3$, the coefficients in \eqref{2-04_cl}   are obtained recursively. For $n=2$, i.e. 
the second step in the sequence of transformations \eqref{eq:seq_trafos_z}, the coefficients in  \eqref{2-04_cl} are determined by
 a linear transformation of the phase space coordinates according to
\begin{equation}
\vecz \mapsto  \phi_{W_2} (\vecz) :=  M\, \vecz\,.
\end{equation}
Here,  $M$ is a symplectic $2d \times 2d$ matrix which is chosen in such a way that
the second order term of the transformed  Hamilton function
\begin{equation}
  H^{(2)}(\vecz) = H^{(1)}(M^{-1} \vecz)
\label{2-04.2_cl}
\end{equation}
assumes the particularly simple form
\begin{equation}
  H_2^{(2)}(\vecz) = \lambda q_1 p_1 + \sum_{k=2}^d \frac{\omega_k}{2}
  (q_k^2 + p_k^2) \,.
\label{2-05_cl}
\end{equation}
Section 2.3 of Ref.~\cite{WaalkensSchubertWiggins08} provides an explicit procedure for
constructing the transformation matrix $M$.

For the first two steps in the sequence \eqref{eq:seq_trafos_z}, we actually did not give explicit expressions for the generating functions $W_1$ and $W_2$. For conceptual reasons (and to justify the notation) we mention that such expressions can be determined (see \cite{WaalkensSchubertWiggins08}) but since it is not necessary for the computation we do not discuss these generating functions here. The situation is different though for the next  steps in  \eqref{eq:seq_trafos_z} which rely on the explicit computation of the generating functions $W_n$ with $n\ge3$.
In order to deal  with these higher order transformations
we introduce
the  Poisson bracket of two functions $A(\vecz)$ and
$B(\vecz)$ which for convenience, we write as
\begin{equation}
 \{ A, B \} =   A
 \sum_{j=1}^d \left( \frac{\overleftarrow{\partial}}{\partial q_j}
    \frac{\overrightarrow{\partial}}{\partial p_j} - \frac{\overleftarrow{\partial}}{\partial p_j}
    \frac{\overrightarrow{\partial}}{\partial q_j} \right)  B\,.
\label{2-06_cl}
\end{equation}
In this notation 
the arrows 
indicate whether the
partial differentiation acts to the left (on $A$) or to the right (on $B$).
With the Poisson bracket we associate the adjoint operator
\begin{equation}
  \mathrm{ad}_A : B \mapsto \mathrm{ad}_A B \equiv \{ A, B \}
  \, .
\label{2-08_cl}
\end{equation}
The transformation \eqref{eq:zn-1twozn} then
leads to a transformation of the Hamilton function $H^{(n-1)}$ to $H^{(n)}$ with $n \ge 3$ which in terms of the adjoint operator 
\eqref{2-08_cl} reads
\begin{equation}
  H^{(n)} = \sum_{k=0}^\infty \frac{1}{k!} \left[ \mathrm{ad}_{W_n} \right]^k
  H^{(n-1)} \, .
\label{2-10_cl}
\end{equation}
Inserting the power series for  $H^{(n-1)}$ (see Equations~\eqref{2-03_cl} and \eqref{2-04_cl}) in \eqref{2-10_cl} and 
ordering terms one finds for the $s^{\text{th}}$ order term of the power series of $H^{(n)}$:
\begin{equation}
  H^{(n)}_s = \sum_{k=0}^{\left\lfloor \frac{s}{n-2} \right\rfloor}
  \frac{1}{k!} \left[ \mathrm{ad}_{W_n} \right]^k H^{(n-1)}_{s-k(n-2)} \, ,
\label{2-11_cl}
\end{equation}
where $\lfloor \cdot \rfloor$ gives the integer part of a number,
i.e., the `floor'-function. 

Using Eq.~(\ref{2-11_cl}) one can show that
the transformation defined by Eq.~(\ref{2-10_cl}) satisfies the following
important properties for $n \ge 3$. Firstly, at step $n$, $n\ge3$, the terms of order less than $n$ in the power series of the Hamilton function are unchanged, i.e.
\begin{equation}
  H_s^{(n)} = H_s^{(n-1)} \, , \;\;\; \mathrm{for} \;\;\; s < n \, ,
\label{2-12_cl}
\end{equation}
so that, in particular, $H_2^{(n)} = H_2^{(2)}$.
Defining
\begin{equation}
  \mathcal{D} \equiv \mathrm{ad}_{H_2^{(2)}} = \{ H_2^{(2)} , \cdot \} \: .
\label{2-14_cl}
\end{equation}
we get for the terms of order $n$,
\begin{equation}
  H_n^{(n)} = H_n^{(n-1)} - \mathcal{D} W_n \, .
\label{2-13_cl}
\end{equation}
This is the so-called {\it homological  equation} which will determine the generating functions $W_n$ for $n\ge3$ from 
requiring   $\mathcal{D} H_n^{(n)} = 0$, or equivalently $H_n^{(n)}$ to be in the kernel of the
restriction of $\mathcal{D}$ to $\mathcal{W}_{\text{cl}}^n$. 
In view of Eq.~(\ref{2-13_cl}) this condition yields
\begin{equation}
  H_n^{(n-1)} - \mathcal{D} W_n \in \mathrm{Ker} \,\mathcal{D} |_{\mathcal{W}_{\text{cl}}^n} \, .
\label{2-15_cl}
\end{equation}
Section 3.4.1 of Ref.~\cite{WaalkensSchubertWiggins08} provides the explicit
procedure of finding the solution of Eq.~\eqref{2-15_cl}.  In the generic situation where the
linear frequencies $\omega_2,\ldots,\omega_d$ in \eqref{2-05_cl} are
rationally independent, i.e. $m_2\omega_2+\ldots+m_d \omega_d=0$
implies $ m_2=\ldots=m_d=0$ for all integers $m_2,\ldots,m_d$, it
follows that for odd $n$, $H_n^{(n)} =0$, and for even $n$,
\begin{equation}
  H_n^{(n)} \in \mathrm{span} \left\{ I^{\alpha_1} J_2^{\alpha_2} J_3^{\alpha_3}
    \ldots J_d^{\alpha_d}  : |\alpha|=n/2 \right\} \, ,
\label{2-15.2_cl}
\end{equation}
where $I = q_1 p_1$ and $J_k = (q_k^2 + p_k^2)/2$, with $k = 2, \ldots, d$.

Applying the transformation (\ref{2-10_cl}), with the generating function
defined by \eqref{2-15_cl}, for $n=3,\ldots,N$, and truncating the
resulting power series at order $N$ one
arrives at 
the $N^{\mathrm{th}}$ order {\it classical normal form} (CNF) 
\begin{equation}
  H_{\mathrm{CNF}}^{(N)}(\vecz) = E_0 + \sum_{s=2}^N H_s^{(N)}(\vecz) \, .
\label{2-15.5_cl}
\end{equation}
We stress that $H_{\mathrm{CNF}}^{(N)}$ represents an
$N^\mathrm{th}$ order approximation of the original Hamiltonian $H$ obtained from
expressing $H$ in terms of the phase space coordinates
\begin{equation}
\vecz^{(N)} = \phi (\tilde{\vecz}) :=  (\phi_{W_N} \circ \phi_{W_{N-1}}    \circ   \cdots  \circ \phi_{W_{2}}    \circ  \phi_{W_{1}})(\tilde{\vecz})  
\label{eq:def_phi}
\end{equation}
(see \eqref{eq:seq_trafos_z}).  Truncating the powers series expansion of $\vecz^{(N)}$ at order $N$ we get the
transformation to the \emph{normal form coordinates} $\vecz=(\vecq,\vecp) =T(\tilde{q},\tilde{p})$ which we introduced in Eq.~\eqref{eq:NFtransform}. 
The neighborhood $\mathcal{L}$ of validity of the CNF is now defined as a neighborhood of the equilibrium $\tilde{\vecz}_0$ in which
the difference between  $H^{(N)}$ and $H_{\text{CNF}}^{(N)}$, i.e. the remainder term consisting  of the nonnormalized tail of terms of order greater than $N$,  and also the difference between 
the normal form coordinates $\vecz$ and their untruncated version $\vecz^{(N)}$ can be considered to be sufficiently small for the application under consideration (see the comments in Sec.~\ref{sec:trunc_error}).

By construction the CNF Hamilton function $ H_{\mathrm{CNF}}^{(N)}$ is a
polynomial of order $N/2$ in the functions $I$ and $J_k$, see \eqref{nf_int}.
We thus have 
\begin{equation}
\dot{I} = \{I,   H_{\mathrm{CNF}}^{(N)} \}=0\,,\quad \dot{J}_k = \{J_k,  H_{\mathrm{CNF}}^{(N)}  \}=0\,, \quad k=2,\ldots,d\,,
\label{eq:IJinvariance_cl}
\end{equation}
i.e.  $I$, and $J_k$, $k = 2, \ldots, d$, are constants of the motion or \emph{integrals} for Hamilton's equation
with Hamiltonian function $H_{\mathrm{CNF}}^{(N)}$.
As discussed in Sec.~\ref{sec:CNF} it is the conservation  of these integrals which allows one to construct the phase space 
structures governing reaction dynamics in the neighborhood of the saddle equilibrium point $\tilde{\vecz}_0$.
We emphasize that the full algorithm to compute $H_{\mathrm{CNF}}^{(N)} $ and the corresponding coordinate transformation is algebraic
in nature, and can be implemented on a computer.

\subsection{The quantum normal form algorithm}
\label{sec:qnf_algorithm}

In the quantum mechanical case
we start with a  Hamilton operator $\hat{H}$ which we
assume to be obtained from the Weyl quantization of a classical
Hamilton function $H(\tilde{\vecq},\tilde{\vecp})$. Like in the previous subsection,  $\tilde{\vecq} = (\tilde{q}_1, \tilde{q}_2, \ldots,
\tilde{q}_d)$ and $\tilde{\vecp} = (\tilde{p}_1,\tilde{p}_2, \ldots, \tilde{p}_d)$ denote the canonical
coordinates and momenta, respectively, of a Hamiltonian system with
$d$ degrees of freedom.  
For convenience, we again choose atomic
units, so that $\tilde{\vecq}$ and $\tilde{\vecp}$ are dimensionless.  
We denote
the corresponding operators by $\hat{\tilde{\vecq}} = (\hat{\tilde{q}}_1, \hat{\tilde{q}}_2,
\ldots, \hat{\tilde{q}}_d)$ and $\hat{\tilde{\vecp}} = (\hat{\tilde{p}}_1, \hat{\tilde{p}}_2, \ldots,
\hat{\tilde{p}}_d)$. In the coordinate representation their components
correspond to multiplication by $\tilde{q}_j$ and the differential operators
$\hat{\tilde{p}}_j = -\ui \heff \, \partial / \partial \tilde{q}_j$. Here $\heff$ is a
dimensionless parameter which corresponds to a scaled, effective
Planck's constant (see the discussion in Sec.~\ref{sec:activated_complex}).  

Like in the classical case 
we will now assume that $H(\vecq,\vecp;\heff)$  (i.e. the classical Hamilton function)
has an equilibrium point, $\tilde{\vecz}_0 \equiv
(\tilde{\vecq}_0, \tilde{\vecp}_0)$, of saddle-center-$\ldots$-center stability
type.  
Similarly to the classical case
the idea of the QNF procedure is to approximate the original Hamilton
operator $\hat{H}$   by a simpler Hamilton operator where the approximation is valid in a 'neighborhood of the equilibrium $\tilde{\vecz}_0$'.
In the classical case this simplification is achieved by a symplectic transformation leading to a specific choice of the phase space coordinates.
In the quantum mechanical case the simplification will be achieved by a unitary transformation to a suitable choice of a 
basis for the eigenstates. 
The basis for computing such unitary transformation is the
Wigner-Weyl symbol calculus. 
The crucial point of the symbol calculus is that it allows one to identify an operator with its Weyl symbol which is a
(in general $\heff$ dependent) phase space functions.  
This way the QNF theory   provides  a systematic
procedure to obtain a local approximation, $\hat{H}_{\mathrm{QNF}}$,
of the Hamiltonian $\hat{H}$ in a phase space neighborhood of the
equilibrium point $\tilde{\vecz}_0$ 
in order to facilitate further computation
of various quantities, such as the CRP, of the reaction system under
consideration, which only depend on properties of  $\hat{H}$ near the equilibrium $\tilde{\vecz}_0$.
From a computational point of view the symbol calculus is extremely beneficial since
the resulting phase space functions can  be dealt with in a similar way as in the classical case described in the previous subsection.

The Weyl symbol of an operator $\hat{H}$ is defined as
\begin{equation}
  H(\tilde{\vecq},\tilde{\vecp};\heff) = \int \! \ud\vecx \, \langle \tilde{\vecq}-\vecx/2 |
  \hat{H} | \tilde{\vecq}+\vecx/2 \rangle \, e^{i \tilde{\vecp} \vecx / \heff} .
\label{2-01}
\end{equation}
The map $\hat{H} \mapsto H(\tilde{\vecq},\tilde{\vecp};\heff)$
leading to \eqref{2-01} is also called the Wigner map. It is the
inverse of the transformation which yields a Hamilton operator
$\hat{H}$ from the Weyl quantization, $\mathrm{Op}[H] $, of a phase
space function $H$ (the Weyl map) which, using Dirac notation, is
given by
\begin{equation}
\hat{H}=  \mathrm{Op}[H] = \iint \frac{\ud\tilde{\vecq} \ud\tilde{\vecp}}{\left( 2\pi\heff \right)^d}
  H(\tilde{\vecq},\tilde{\vecp};\heff)  \displaystyle \int
  \ud\vecx | \tilde{\vecq}-\vecx/2 \rangle e^{-\ui \tilde{\vecp}\vecx / \heff}
  \langle \tilde{\vecq}+\vecx/2 | \,.  
\label{eq:def_Weyl_map}
\end{equation}
Accordingly, $H(\tilde{\vecq},\tilde{\vecp};\heff)$ in \eqref{2-01} agrees with
the classical Hamilton function $H(\tilde{\vecq},\tilde{\vecp})$ in our case. The
argument $\heff$ is introduced for convenience since the Weyl symbol
of the unitarily transformed Hamilton operator we will deal with below will in general
explicitly depend on $\heff$.

Analogously to the classical case we will now construct a sequence of unitary transformations which will simply the Hamilton operator order by order in the neighborhood of the equilibrium point $z_0$.  
These unitary transformation will be of the form
\begin{equation}
\hat{U}_n = \ue^{-\ui \hat{W}_n}\,,
\label{eq:def_Un}
\end{equation}
where $ \hat{W}_n$ is the Weyl quantization  $  \mathrm{Op}\left[ {W}_n \right]$ of a symbol $W_n$ which is a homogenous polynomial in the space
\begin{equation}
   \mathcal{W}_{\text{qm}}^n := \mathrm{span} \left\{ q_1^{\alpha_1} \ldots q_d^{\alpha_d}
    p_1^{\beta_1} \ldots p_d^{\beta_d} \heff^{j} :  |\alpha|+|\beta|+2j = n \right\} \, .
\label{2-09}
\end{equation}
The main difference between the symbols $W_n$  and  the generating functions defining the symplectic transformations in the classical case in \eqref{eq:zn-1twozn} is that  the $W_n$ in \eqref{eq:def_Un}  are polynomials in the phase space coordinates 
$({\vecq},{\vecp})$ \emph{and} $\hbar$ where for the definition of the order, the power of $\hbar$ counts double.

Using unitary transformations of the form \eqref{eq:def_Un} we get the sequence of Hamilton operators
\begin{equation}
\hat{H} \equiv \hat{H}^{(0)} \rightarrow \hat{H}^{(1)} \rightarrow \hat{H}^{(2)} \rightarrow \hat{H}^{(3)}
\rightarrow \ldots \rightarrow \hat{H}^{(N)}\,,
\label{eq:seq_trafos_operators}
\end{equation}
where
\begin{equation}
\hat{H}^{(n)} = \hat{U}_n \hat{H}^{(n-1)} \hat{U}^*_n \,.
\label{eq:conjugation_Hn}
\end{equation}
Expressing the sequence \eqref{eq:seq_trafos_operators} in terms of symbols we get a sequence
\begin{equation}
H \equiv H^{(0)} \rightarrow H^{(1)} \rightarrow H^{(2)} \rightarrow H^{(3)}
\rightarrow \ldots \rightarrow H^{(N)}
\label{eq:seq_trafos}
\end{equation}
which  will be the direct analogue of the classical sequence \eqref{eq:seq_trafos_cl}.
As we will see below, in terms of the symbols the  unitary transformations \eqref{eq:def_Un} can be viewed as $\hbar$ dependent transformations of the phase space coordinates. 

The first two steps of  the transformations \eqref{eq:seq_trafos} are identical to the classical case. 
The first step serves  to shift the equilibrium point
$\tilde{\vecz}_0$ to the origin according to 
\begin{equation}
  H^{(1)}(\vecz;\heff) = H^{(0)}(\vecz+\tilde{\vecz}_0;\heff) \, ,
\label{2-02}
\end{equation}
The higher order transformations then work on power series expansions of the symbol of the form
\begin{equation}
  H^{(n)}(\vecz;\heff) = E_0 + \sum_{s=2}^\infty H_s^{(n)}(\vecz;\heff) \, ,
\label{2-03}
\end{equation}
where
\begin{equation}
  H_s^{(n)}(\vecz;\heff)  =   \sum_{|\alpha|+|\beta|+2j=s}
    \frac{H_{\alpha_1,\ldots,\alpha_d,\beta_1,\ldots,\beta_d,j}^{(n)}}{\alpha_{1}!
      \ldots \alpha_{d}! \beta_{1}! \ldots \beta_{d}! j!}  
     q_1^{\alpha_1} \ldots q_d^{\alpha_d} p_1^{\beta_1} \ldots
    p_d^{\beta_d} \, \heff^{j}  \in  \mathcal{W}_{\text{qm}}^s   \, .
 \label{2-04}
\end{equation}
For $n=1$, the coefficients in \eqref{2-04} are again given by the Taylor expansion of $H^{(1)}$ about the origin 
\begin{eqnarray}
   H_{\alpha_1,\ldots,\alpha_d,\beta_1,\ldots,\beta_d,j}^{(1)}   =  \left(
  \prod_{k,l=1}^d \frac{\partial^{\alpha_k}}{\partial
    q_k^{\alpha_k}} \frac{\partial^{\beta_l}}{\partial
    p_l^{\beta_l}}   \right)  \frac{\partial^{j}}{\partial \varepsilon^{j}}
  H^{(n)}(\vecz; \varepsilon) |_{ (\vecz;\epsilon) =({\bf 0};0)} \!\!\! \,,
  \label{2-04.1}
\end{eqnarray}
where we included powers of $\hbar$ only for completeness (i.e. for the type of starting classical Hamilton functions in this paper  $H^{(1)}$ does not have terms with nonzero powers of $\hbar$). 
Like in the classical case the second step of the sequence of transformations \eqref{eq:seq_trafos}
consists of a linear transformation of the phase space coordinates by a symplectic $2d\times2d$ matrix $M$ such that the second order term of the symbol
\begin{equation}
  H^{(2)}(\vecz; \heff) = H^{(1)}(M^{-1} \vecz; \heff)
\label{2-04.2}
\end{equation}
takes the form
\begin{equation}
  H_2^{(2)}(\vecz; \heff) = \lambda q_1 p_1 + \sum_{k=2}^d \frac{\omega_k}{2}
  (q_k^2 + p_k^2) \,.
\label{2-05}
\end{equation}
The required matrix $M$ is identical to the classical case.

The reason for the equality of the classical transformations and the quantum transformations to second order is the commutativity of the Weyl quantization and affine linear symplectic transformations. 
For the nonlinear transformations corresponding to steps $n\ge 3$ the commutativity ceases to exist and the symbol calculus develops its full power. 

In order to simplify the terms of order  $n\ge 3$ we have to introduce the notion of
the {\it Moyal bracket}. 
Given two symbols $A(\vecz;\heff)$ and
$B(\vecz;\heff)$, corresponding to operators $\hat{A} $ and $\hat{B}$
respectively, the Moyal bracket is defined as
\begin{equation}
\begin{split}
 \{ A, B \}_{\mathrm{M}}  =  \frac{2}{\heff} A
\sin\left[ \frac{\heff}{2} \sum_{j=1}^d \left( \frac{\overleftarrow{\partial}}{\partial q_j}
    \frac{\overrightarrow{\partial}}{\partial p_j} - \frac{\overleftarrow{\partial}}{\partial p_j}
    \frac{\overrightarrow{\partial}}{\partial q_j} \right) \right] B\,.
\end{split} \label{2-06}
\end{equation}
The Moyal bracket gives the Weyl symbol of the operator $i[\hat{A},\hat{B}]/\heff$, where
$[\cdot,\cdot]$ denotes the commutator. Like in Eq.~\eqref{2-06_cl} the arrows in \eqref{2-06}
indicate whether the
partial differentiation acts to the left (on $A$) or to the right (on $B$).
Equation~(\ref{2-06}) implies that for $\heff \rightarrow 0$,
\begin{equation}
  \{ A, B \}_{\mathrm{M}} = \{ A, B \} + \mathcal{O}(\heff^2) \, ,
\label{2-07}
\end{equation}
i.e. for $\heff = 0$, we recover the classical Poisson bracket.
Moreover, if at 
least one of the functions $A$, $B$ is a second order polynomial in
the variables $(\vecq, \vecp)$ then $\{ A, B \}_{\mathrm{M}} = \{ A, B\}$. 
Analogously to the adjoint operator associated with the Poisson bracket in the classical case  (see Eq.~\eqref{2-08_cl})
we now define a Moyal-adjoint operator associated with the Moyal bracket:
\begin{equation}
  \mathrm{Mad}_A : B \mapsto \mathrm{Mad}_A B \equiv \{ A, B \}_{\mathrm{M}}  \, .
\label{2-08}
\end{equation}
The symbol of $\hat{H}^{(n)}$ obtained from a unitary transformation of $\hat{H}^{(n-1)}$ according to Eq.~\eqref{eq:conjugation_Hn}
is then given by 
\begin{equation}
  H^{(n)} = \sum_{k=0}^\infty \frac{1}{k!} \left[ \mathrm{Mad}_{W_n} \right]^k
  H^{(n-1)} \, .
\label{2-10}
\end{equation}
Inserting the power series of $H^{(n-1)}$ and ordering terms one finds for the $s^{\text{th}}$ order  terms of
the powers series of  $H^{(n)}$
\begin{equation}
  H^{(n)}_s = \sum_{k=0}^{\left\lfloor \frac{s}{n-2} \right\rfloor}
  \frac{1}{k!} \left[ \mathrm{Mad}_{W_n} \right]^k H^{(n-1)}_{s-k(n-2)} \, ,
\label{2-11}
\end{equation}
where $\lfloor \cdot \rfloor$ again denotes the floor-function. 
Equation~\eqref{2-11} looks formally the same as  Eq.~\eqref{2-11_cl} in the classical case. The only difference is the occurrence of the Moyal adjoint operator instead of the adjoint operator associated with the Poisson bracket. This difference however is significant, as the Moyal bracket will in general introduce $\hbar$ dependent terms although the original symbol we started  with (i.e. the classical Hamilton function) had no $\hbar$ dependence.  
Still, the Moyal bracket has similiar properties as the Poisson bracket. 
Using Eq.~(\ref{2-11}) we see  analogously the classical case   that 
at each step $n \ge 3$, the terms of order of order less than $n$ remain unchanged:
\begin{equation}
  H_s^{(n)} = H_s^{(n-1)} \, , \;\;\; \mathrm{for} \;\;\; s < n \, ,
\label{2-12}
\end{equation}
so that we in particular again have $H_2^{(n)} = H_2^{(2)}$ like in the classical case. 
Defining 
\begin{equation}
  \mathcal{D} \equiv \mathrm{Mad}_{H_2^{(2)}} = \{ H_2^{(2)} , \cdot \} 
\label{2-14}
\end{equation}
we find for the terms of order $n$
\begin{equation}
  H_n^{(n)} = H_n^{(n-1)} - \mathcal{D} W_n \, .
\label{2-13}
\end{equation}
This is the {\it quantum homological equation} which analogously to the classical case is used to determine the symbols $W_n$
that lead to the unitary transformations $\hat{U}_n$ defined in \eqref{eq:def_Un}.
Similar to the classical case we determine the $W_n$ by requiring $\mathcal{D}
H_n^{(n)} = 0$, or equivalently $H_n^{(n)}$ to be in the kernel of the
restriction of $\mathcal{D}$ to $\mathcal{W}_{\text{qm}}^n$. In view of
Eq.~(\ref{2-13}) this condition yields
\begin{equation}
  H_n^{(n-1)} - \mathcal{D} W_n \in \mathrm{Ker} \,\mathcal{D} |_{\mathcal{W}_{\text{qm}}^n} \, .
\label{2-15}
\end{equation}
Section 3.4.1 of Ref.~\cite{WaalkensSchubertWiggins08} provides the explicit
procedure of finding the solution of Eq.~(\ref{2-15}).  Provided the
linear frequencies $\omega_2,\ldots,\omega_d$ in \eqref{2-05} are
rationally independent it again
follows that for odd $n$, $H_n^{(n)} =0$, and for even $n$,
\begin{equation}
  H_n^{(n)} \in \mathrm{span} \left\{ I^{\alpha_1} J_2^{\alpha_2} J_3^{\alpha_3}
    \ldots J_d^{\alpha_d} \heff^j  : |\alpha|+j=n/2 \right\} \, ,
\label{2-15.2}
\end{equation}
where $I = q_1 p_1$ and $J_k = (q_k^2 + p_k^2)/2$, with $k = 2, \ldots, d$,
are the analogues of the classical integrals.

Applying the transformation (\ref{2-10}), with the symbol
defined by Eq.~(\ref{2-15}), for $n=3,\ldots,N$, and truncating the
resulting power series (\ref{2-03}) at the $N^\mathrm{th}$ order one
arrives at the Weyl symbol $H_{\mathrm{QNF}}^{(N)}$ corresponding to
the $N^{\mathrm{th}}$ order {\it quantum normal form} (QNF) of the
Hamiltonian $\hat{H}$,
\begin{equation}
  H_{\mathrm{QNF}}^{(N)}(\vecz;\heff) = E_0 + \sum_{s=2}^N H_s^{(N)}(\vecz;\heff) \, .
\label{2-15.5}
\end{equation}

We thus see that the computation of the symbol of the QNF operator is very similar to the classical case. 
The major difference is the use of the Moyal bracket in the quantum case which is more complicated than the Poisson bracket in the classical case. 
What remains to be done to get the  $N^{\text{th}}$ order QNF operator is to compute the Weyl quantization of the symbol $H_{\mathrm{QNF}}^{(N)}(\vecz;\heff) $:
\begin{equation}
  \hat{H}_{\mathrm{QNF}}^{(N)} = \mathrm{Op}\left[ H_{\mathrm{QNF}}^{(N)}
  \right] \, .
\label{2-16}
\end{equation}

The Weyl  quantization of the classical integrals $I$ and $J_k$, $k=2,\ldots,d$, are
\begin{eqnarray}
  \hat{I} &\equiv& \mathrm{Op}[I] = \displaystyle \frac{1}{2} (\hat{q} \hat{p}
  + \hat{p} \hat{q}) \, , \label{2-20a}\\
  \hat{J}_k &\equiv& \mathrm{Op}[J_k] = \displaystyle \frac{1}{2} (\hat{q}_k^2 +
  \hat{p}_k^2),\quad k=2,\ldots,d.
  \label{2-20b}
\end{eqnarray}
Using Eq.~(\ref{2-05}) and the linearity of the Weyl quantization we get
\begin{equation}
  \hat{H}_2^{(2)} = \lambda \hat{I} + \sum_{k=2}^d \omega_k \hat{J}_k  \, .
\label{2-19}
\end{equation}
Since the higher order terms in \eqref{2-15.5} are polynomials in $I$
and $J_k$, $k=2,\ldots,d$ (see \eqref{2-15.2}), we need to know how to
quantize powers of $I$ and $J_k$.  As shown in \cite{WaalkensSchubertWiggins08} this can
be accomplished using the recurrence relations
\begin{equation}
\mathrm{Op}\left[ I^{n+1} \right] = \hat{I}    \mathrm{Op}\left[ I^{n} \right] - \left(\frac{\hbar}{2} \right)^2 n^2  \mathrm{Op}\left[ I^{n-1} \right]
\label{eq:recurrence_I}
\end{equation}
and
\begin{equation}
\mathrm{Op}\left[ J_k^{n+1} \right] = \hat{J}_k    \mathrm{Op}\left[ J_k^{n} \right] + \left(\frac{\hbar}{2} \right)^2 n^2  \mathrm{Op}\left[ J_k^{n-1} \right]
\label{eq:recurrence_J}
\end{equation}
for $k=2,\ldots,d$. Hence, $ \hat{H}_{\mathrm{QNF}}^{(N)}$ is a
polynomial function of the operators $\hat{I}$ and $\hat{J}_k$ of the form Eq.~\eqref{eq:qnf_H}. The coefficients  $\kappa_{n,\alpha,j}$
are systematically obtained by the QNF procedure to compute the symbol
$H_{\mathrm{QNF}}^{(N)} $, as desribed above, and the recurrence
relations \eqref{eq:recurrence_I} and \eqref{eq:recurrence_J}.  So like in the case of  the classical normal form
the full procedure to compute $\hat{H}_{\mathrm{QNF}}^{(N)} $ is algebraic
in nature, and can be implemented on a computer. Our software for
computing the quantum normal form as well as the classical normal form
which is recovered for $\heff=0$ is publicly available at
\url{http://lacms.maths.bris.ac.uk/publications/software/index.html}.

In summary the  $N^{\text{th}}$ order quantum normal form operator $\hat{H}_{\mathrm{QNF}}^{(N)}$ is  obtained from
conjugating the original Hamiltonian $\hat{H}$ by the unitary
transformation
\begin{equation}
\hat{U} =  \ue^{-\ui\hat{W}_1/\heff}  \ue^{-\ui\hat{W}_2/\heff}   \cdots   \ue^{-\ui\hat{W}_N/\heff} \,,
\label{eq:def_U}
\end{equation} 
where we used the fact that the first two steps in the sequence
\eqref{eq:seq_trafos} can also be implemented using suitable
generators $\hat{W}_1$ and $\hat{W}_2$ (see \cite{WaalkensSchubertWiggins08} for
more details). 
It is an $N^{\text{th}}$ order approximation in the sense that the symbol corresponding to the conjugated operator is truncated at 
order $N$ where the order is defined according to Eq.~\eqref{2-09}. 
Using arguments based on the symbol calculus 
the remainder term consisting of the Weyl quantization of the unnormalized tail $H^{(N)}-H_{\text{QNF}}^{(N)}$ 
is small 'in a neighborhood of the equilbrium point  $\tilde{\vecz}_0$'.  
The local approximation given by the QNF  is ideally suited  to compute the cumulative reaction probability which 
is directly related to properties of the Hamilton operator in the neighborhood of $\tilde{\vecz}_0$ because of the simple structure of
$ \hat{H}_{\mathrm{QNF}}^{(N)}$. Since  $ \hat{H}_{\mathrm{QNF}}^{(N)}$ is  a polynomial function of $\hat{I}$ and $\hat{J}_k$, $k=2,\ldots,d$, we have 
\begin{equation}
[\hat{I}, \hat{H}_{\mathrm{QNF}}^{(N)}] =0\,,\quad 
[\hat{J}_k, \hat{H}_{\mathrm{QNF}}^{(N)}] =0\,,\quad k=2,\ldots,d
\label{eq:commutativities}
\end{equation}
which is the quantum analog of the invariance of the classical integrals in Eq.~\eqref{eq:IJinvariance_cl}.
The commutativities \eqref{eq:commutativities} imply that the eigenstates of $\hat{H}_{\mathrm{QNF}}^{(N)}$ are product states of the one dimensional eigenstates of $\hat{I}$ and the  $\hat{J}_k$. As we demonstrated in Sections~\ref{sec:smatrix} and \ref{sec:resonances}
the simplicity of the spectral properties of $\hat{I}$ and the  $\hat{J}_k$ upon which the QNF is built can be exploited to give an efficient procedure to compute the cumulative reaction probability and the Gamov-Siegert resonances associated with the equilibrium $\tilde{\vecz}_0$.



\section*{Acknowledgments}

This work was supported by the EPSRC under grant no.~EP/E024629/1 and
ONR under grant no.~N00014-01-1-0769.

\small
\bibliographystyle{unsrt}

\def\cprime{$'$}

\end{document}